\documentclass[twocolumn]{aa}
\usepackage{graphicx}
\usepackage{txfonts}
\newcommand{\mtot}{\relax \ifmmode M_{\rm tot}\else $M_{\rm tot}$\fi}
\newcommand{\Reff}{\relax \ifmmode R_{\rm e}\else $R_{\rm e}$\fi}
\newcommand{\SBe}{\relax \ifmmode \langle SB_{\rm e}\rangle \else $\langle SB_{\rm e}\rangle$\fi}
\newcommand{\mB}{\relax \ifmmode M_{\rm B}\else $M_{\rm B}$\fi}
\newcommand{\ReB}{\relax \ifmmode R_{\rm e,B}\else $R_{\rm e,B}$\fi}
\newcommand{\mueB}{\relax \ifmmode \mu_{\rm e,B} \else $\mu_{\rm e,B}$\fi}
\newcommand{\mD}{\relax \ifmmode M_{\rm D}\else $M_{\rm D}$\fi}
\newcommand{\muo}{\relax \ifmmode \mu_{\rm 0}\else $\mu_{\rm 0}$\fi}
\newcommand{\rd}{\relax \ifmmode h\else $h$\fi}
\newcommand{\db}{\relax \ifmmode D/B\else $D/B$\fi}
\newcommand{\nb}{\relax \ifmmode n\else $n$\fi}
\newcommand{\inc}{\relax \ifmmode i\else $h$\fi}
\newcommand{\ellip}{\relax \ifmmode \epsilon\else $\epsilon$\fi}

\newcommand{\vmax}{\relax \ifmmode V_{\rm max}\else $V_{\rm max}$\fi}

\def\kms{\relax \ifmmode {\,\rm km\,s}^{-1}\else \,km\,s$^{-1}$\fi}
\def\ks{\relax \ifmmode  K_{\rm s}\else $K_{\rm s}$\fi}
\def\ha{\relax \ifmmode {\rm H}\alpha\else H$\alpha$\fi}
\def\hb{\relax \ifmmode {\rm H}\beta\else H$\beta$\fi}
\def\hi{\relax \ifmmode {\rm H\,{\sc i}}\else H\,{\sc i}\fi}
\def\hii{\relax \ifmmode {\rm H\,{\sc ii}}\else H\,{\sc ii}\fi}
\def\h2{\relax \ifmmode {\rm H}_2\else H$_2$\fi}
\def\lha{\relax \ifmmode L_{{\rm H}\alpha}\else $L_{{\rm H}\alpha}$\fi}
\def\shi{\relax \ifmmode \sigma_{{\rm HI}}\else $\sigma_{\rm HI}$\fi}
\def\sh2{\relax \ifmmode \sigma_{{\rm H}_2}\else $\sigma_{{\rm H}_2}$\fi}
\def\degr{\hbox{$^\circ$}}
\def\arcmin{\hbox{$^\prime$}}
\def\arcsec{\hbox{$^{\prime\prime}$}}
\def\deg{\hbox{$^\circ$}}

\def\sec{\hbox{$^{\prime\prime}$}}
\def\fdg{\hbox{$.\!\!^\circ$}}
\def\fs{\hbox{$.\!\!^{\rm s}$}}
\def\farcm{\hbox{$.\mkern-4mu^\prime$}}
\def\farcs{\hbox{$.\!\!^{\prime\prime}$}}
\def\degd#1.#2{ #1\fdg#2 }                 % degrees over decimal point
\def\mind#1.#2{ #1\farcm#2 }               % minutes over decimal poixxxnt
\def\secd#1.#2{ #1\farcs#2 }               % seconds over decimal point
\def\hhh{\ifmmode {\rm ^h}              % hours symbol
         \else {${\rm ^h}$}
         \fi}
\def\sss{\ifmmode {\rm ^s}              % seconds symbol
         \else {${\rm ^s}$}
         \fi}
\def\hms#1h#2m#3s{                      % hms format (for RA)
                  \relax
                  \ifmmode #1^{\rm h}\,#2^{\rm m}\,#3^{\rm s}
                  \else \hbox{$#1^{\rm h}\,#2^{\rm m}\,#3^{\rm s}$}
                  \fi
                 }
\def\dms#1d#2m#3s{                      % dms format (for Dec)
                  \relax
                  #1\degr\,#2\arcmin\,#3\arcsec 
                 }
\def\hmsd#1h#2m#3.#4s{                  % hms format with decimal point (RA)
                      \relax
                      \ifmmode #1^{\rm h}\,#2^{\rm m}\,#3\fs#4
                      \else \hbox{$#1^{\rm h}\,#2^{\rm m}\,#3\fs#4$}
                      \fi
                     }
\def\dmsd#1d#2m#3.#4s{                  % dms format with decimal point (Dec)
                      \relax
                      #1\degr\,#2\arcmin\,#3\farcs#4
                     }
\def\mag{\relax                          % magnitudes symbol
        \ifmmode ^{\rm m}
        \else $^{\rm m}$
        \fi
       }
\def\magd#1.#2{                          % magnitudes over decimal point
              \relax
              \ifmmode #1^{\rm m}
                       \hskip-0.55em.\hskip0.22em#2
              \else \hbox{#1$^{\rm m}
                    \hskip-0.55em.\hskip0.22em$#2}
              \fi
             }
\input psfig.sty
\begin{document}
\title{A composite \hii\ region luminosity function in H$\alpha$ of
unprecedented statistical weight}
\author{T.~R.~Bradley\inst{1,2}
\and J.~H.~Knapen\inst{2}
\and J.~E.~Beckman\inst{3,4}
\and S.~L.~Folkes\inst{2}
}

\offprints{J. H. Knapen}  

\institute{Centre for Astrophysics, University of Central Lancashire, 
Preston PR1 2HE, UK
\and Centre for Astrophysics Research,
University of Hertfordshire, Hatfield, Herts AL10 9AB, UK\\
\email{j.knapen@star.herts.ac.uk}
\and Instituto de Astrof\'\i sica de Canarias, E-38200 La Laguna, 
Spain
\and Consejo Superior de Investigaciones Cient\'\i ficas, Spain
}

\date{Received ; accepted 15 Sep. 2006}

\abstract
{Statistical properties of \hii\ region populations in disk
	galaxies yield important clues to the physics of massive star
	formation.}
{We present a set of \hii\ region catalogues and luminosity
functions for a sample of 56 spiral galaxies in order to derive the
most general form of their luminosity function.}
{\hii\ region luminosity functions are derived for
individual galaxies which, after photometric calibration, are summed
to form a total luminosity function comprising 17,797 \hii\ regions
from 53 galaxies.}
{The total luminosity function, above its lower limit of
completeness, is clearly best fitted by a double power law with a
significantly steeper slope for the high luminosity portion of the
function. This change of slope has been reported in the literature for
individual galaxies, and occurs at a luminosity of $\log L =
38.6\pm0.1$ ($L$ in erg\,s$^{-1}$) which has been termed the
Str\"omgren luminosity. A steep fall off in the luminosity function
above $\log L = 40$ is also noted, and is related to an upper limit to
the luminosities of underlying massive stellar clusters. Detailed data
are presented for the individual sample galaxies.}
{The luminosity functions of \hii\ regions in spiral
galaxies show a two slope power law behaviour, with a significantly
steeper slope for the high luminosity branch. This can be modelled by
assuming that the high luminosity regions are density bounded, though
the scenario is complicated by the inhomogeneity of the ionized
interstellar medium. The break, irrespective of its origin, is of
potential use as a distance indicator for disc galaxies.}

\keywords{galaxies: spiral -- galaxies: structure -- ISM: H{\sc ii}
regions} 

\maketitle

%
%________________________________________________________________

\section{Introduction}

In their paper giving a major overview of the luminosity functions
(LFs) in \ha\ emission of the \hii\ regions in disk galaxies
Kennicutt, Edgar \& Hodge (1989; hereafter KEH) noted that an
important subset of their objects showed LFs with a clear break in
slope at a luminosity of $L(\ha) = 38.7$\,dex (units of
erg\,s$^{-1}$). At lower luminosities the power law slope is flatter,
whereas above the break it is steeper. KEH called those LFs showing a
break ``type II", suggesting that any galaxy will exhibit type II
behaviour if it has sufficient high luminosity \hii\ regions. A galaxy
with sufficient luminous \hii\ regions to show a clear break, M51, was
observed by Rand (1992), who measured it at $L(\ha) =
38.6$\,dex. Rand's work, using CCD data, was especially precise, and
revealed, as well as the break, a narrow peak in the LF close to the
break luminosity. In the LF of M33, Hodge et al. (1999) found no
break, which is not surprising since the function is defined only up
to $L(\ha) = 38.4$\,dex, with only a dozen regions above $L(\ha) =
38$\,dex. Further examples of this type are NGC~6822 and the dwarf
galaxy Holmberg~II (Hodge, Strobel \& Kennicutt 1994), smaller
galaxies with small \hii\ region numbers at $L(\ha) > 38$\,dex.  On
the other hand, M101 has some 60 \hii\ regions with $L(\ha) >
38.5$\,dex and shows a clear break at $L(\ha) ~ 38.6$\,dex (Scowen,
Dufour \& Hester 1992).  Our group has published LFs for galaxies
selected to have large numbers of high luminosity \hii\ regions to
test the suggestion that the break luminosity shows low scatter. In
three articles (Rozas, Beckman \& Knapen 1996; Rozas et al. 1999; and
Rozas, Zurita \& Beckman 2000), eight galaxies were measured and all
showed type II LFs, most strikingly NGC~7479, which has the largest
number of luminous \hii\ regions.

In spite of the general result reported in KEH, and in spite of the
type II LFs found in individual objects, the question has been raised
persistently whether the break is a real feature or an artifact. Two
types of doubts can be found in the literature. One claims that
although breaks can be found, they are not a universal physical
feature, as the break luminosity varies over quite a wide range
between objects. One example here is by Thilker, Braun \& Walterbos
(2000) who derived an LF for M51 using an automated method, finding a
break at 38.9\,dex, compared with the value of 38.6 by Rand (1992) who
used an interactive, region by region method. The point raised is
whether for statistical reasons connected with the chosen binning
parameters a clear break luminosity cannot be derived, given the low
number of regions per luminosity bin. Others have not found consistent
evidence for breaks, and certainly not at a specific luminosity, among
these being Gonz\'alez Delgado \& P\'erez (1997) in an \ha\ survey of
27 galaxies selected for their nuclear activity. Their sample
comprised some 2000 \hii\ regions, i.e., $\approx75$ per galaxy in a
luminosity range from $L(\ha) = 37$\,dex to 39.5\,dex. The numbers per
bin above 38.5\,dex are on average quite small for each galaxy which
would lead to difficulty in detecting breaks.

On completing a recent imaging survey in \ha\ of a set of nearby
galaxies (Knapen et al. 2004) we saw that we could make a significant
test of the suggestion that a dual slope LF with a well defined break
gives a better general description than a single power law slope, and
that the original result of KEH, with its clean break at a specific
luminosity, is present in a data set with a sufficient statistical
base. Our observations have allowed us to isolate and catalogue almost
18,000 \hii\ regions in 56 galaxies. Here we give the key results of
this study and also offer on-line catalogues for the full set of
objects.

\section{Sample selection and \ha\ imaging}

\subsection{Sample selection and observations}

We have used the data set published by Knapen et al. (2004), which
contains a full set of continuum-subtracted \ha\ images of 57
relatively face-on ($i<$50\deg), nearby, Northern, spiral
galaxies. Most of the images were obtained with the 1\,m Jacobus
Kapteyn Telescope (JKT), but some with other telescopes or from the
literature.  Most of the \ha\ images were taken through $\approx
50$\AA\ wide filters matched to the recessional velocity of each
galaxy.  Full details on the observations are given in Knapen et
al. (2004), whereas the sample galaxies are listed, and their
distances, plate scales, and resolutions given, in Table~1 ({\bf
online only}).

\subsection{The effects of seeing on \hii\ region selection}

The FWHM seeing in our \ha\ images varies from 0.8 to 3.7\,arcsec
(1.53\, arcsec in the median), which corresponds to a range in spatial
resolution of 30 to 295\,pc (137\,pc in the median). We thus need to
consider possible effects of blending on the resulting \hii\ region
catalogues. The two major effects are that fainter \hii\ regions may
be spatially coincident with larger, brighter, \hii\ regions, reducing
the number of faint \hii\ regions observed, and overlapping of \hii\
regions, which may lead to a measured luminosity higher than the true
luminosity. KEH showed that whereas degrading images up to resolutions
of 200\,pc had little effect on the luminosities of interest here, and
that degrading them to a resolution of 200-400\,pc will affect the
faint end of the luminosity function while preserving the shape of the
upper LF, blending seriously affects the entire LF and causes
increases in the luminosities of first ranked \hii\ regions in images
with a resolution of 300-500 pc.  Rand (1992) modelled the main
effects of blending for the arm LF of M51, and found that they did not
significantly affect the shape of the LF.

More recently, \ha\ LFs have been derived from {\it HST} imaging
(Pleuss, Heller \& Fricke 2000; Scoville et al. 2001),
which has significantly higher spatial resolution than the typical
size of an \hii\ region in the luminosity range of interest
here. Under these circumstances, it is possible to classify a single
\hii\ region, due to a single coeval OB association, artificially as a
set of aggregated \hii\ regions, each of a lower luminosity.

We thus conclude that blending should not have serious consequences
for the main results presented in this letter, although for individual
galaxies at larger distances blending may introduce uncertainties in
the LFs.

\section{Calibration}

The continuum subtraction and photometric calibration of the \ha\
images has been described in detail by Knapen et al. (2004).  We
estimated the uncertainties in the calibration to be $L=0.1$\,dex by
assuming errors of 3\% in the distance to the galaxy, and of one
standard deviation of the adopted background value of the calibration
star.

Since the final continuum-subtracted images of most galaxies contain
emission from both the \ha\ line and the [N{\sc ii}] lines at
6548\,{\AA} and 6583\,{\AA}, we need to correct for the latter before
determining the star formation rates. We do this by assuming a fixed
ratio between the [N{\sc ii}]\,6583 and 6548 lines of three, and a
value $\eta$ for the [N{\sc ii}]\,6583/\ha\ ratio of 0.25 for galaxies
that do not contain any or only a few ``giant'' \hii\ regions (e.g.,
early-type galaxies) and of 0.16 for the later Hubble types, based on
descriptions in Osterbrock (1989). We compute the final correction
factor for each galaxy by additionally determining the relative filter
transmissions in the \ha\ and [N{\sc ii}] lines, for which we
used the filter transmission characteristics as measured by the Isaac
Newton Group (ING), which operated the JKT.

We found that the [N{\sc ii}] contribution varied due to filters used,
recessional velocity, and galaxy type, and ranged from 9\% to 36{\%}
with a mean of 17{\%}. NGC~3631 and NGC~4321 were both observed with
\ha\ filters of width 15\,{\AA} which was narrow enough not to include
any [N{\sc ii}] emission. 

Thilker et al.  (2002) use values of $\eta $ which vary from 0.1 to
0.4, but only one galaxy is common to both their and our sample.  This
is NGC 5457, for which Thilker et al.'s value is 0.2, as opposed to
ours of 0.16. Using the former value would lead to a difference of
3{\%} in the [N{\sc ii}] contribution to our image, which is not a
significant factor in the analysis of the LF. No other such
comparisons could be made from the literature.

\section{\hii\ region catalogues}

We used a semi-automatic method to derive the \hii\ region catalogues
for all sample galaxies which uses the {\sc region} programme
(C.~H. Heller, private communication; see Rozas et al. 1999; Pleuss et
al. 2000). Using a tree algorithm, {\sc region} catalogues each \hii\
region, yielding the position of its centre, its area in pixels and
its background-subtracted flux. An advantage of {\sc region} over any
fully automated programme is that it allows for for defining local
backgrounds, and for manual editing of the resulting catalogue to
remove features like cosmic rays or remnants of imperfect continuum
subtraction of foreground stars.  We chose not to catalogue \hii\
regions in the central 1\,kpc region of each galaxy for two
reasons. Firstly, crowding and blending is especially severe there,
and secondly, especially in the AGN hosts in our sample, the \ha\ can
be severely contaminated by emission resulting from shocks or AGN
emission.  The \hii\ region catalogues are all available
electronically through the Centre de Donn\'ees Stellaires
(CDS)\footnote{Via anonymous ftp to cdsarc.u-strasbg.fr (130.79.128.5)
or via http://cdsweb.u-strasbg.fr/cgi-bin/qcat?J/A+A/}. The catalogues
list for all galaxies a number identifying the \hii\ region (col.~1),
the position of the pixel of maximum intensity of the \hii\ region
relative to the centre of the galaxy, in RA and dec (pixel scale is
0.241\,arcsec\,pixel$^{-1}$; cols.~2 and 3), the logarithm of its
integrated luminosity in erg\,s$^{-1}$ (col.~4), the area of the \hii\
region in units of pc$^2$ (col.~5), and the value of the local
background (col.~6).

\section{Luminosity functions}

\subsection{Individual galaxies}

On the basis of our \hii\ region catalogues, we constructed luminosity
functions (LFs) for all our galaxies, which we present in Fig.~1 ({\bf
online only}). The bin widths for plotting and fitting were
determined following Scott (1979), and vary from 0.2 to 0.7\,dex as a
function of the number of \hii\ regions (see Table~\ref{table1}). We
fit slopes to the LFs, following the equation $N(L)dL=AL^{-a}dL$,
where $N(L)dL$ is the number of \hii\ regions with luminosities in the
range $L$ to $L+dL$, and $a$ is the slope, as introduced by
KEH. Consistent with KEH and most of the more recent work, we made an
adjustment of $-$1 to the slopes of the plotted LFs, which is due to
the fact that we show differential LFs as determined with logarithmic
binning, whereas the slopes refer to a differential LF with linear
binning. The slopes, as determined with a weighted fit, have been
plotted on the LFs in Fig.~1, and are listed, along with their formal
errors and coefficients of determination $r^2$, in Table~1.

\setcounter{figure}{1}
\begin{figure}
\psfig{figure=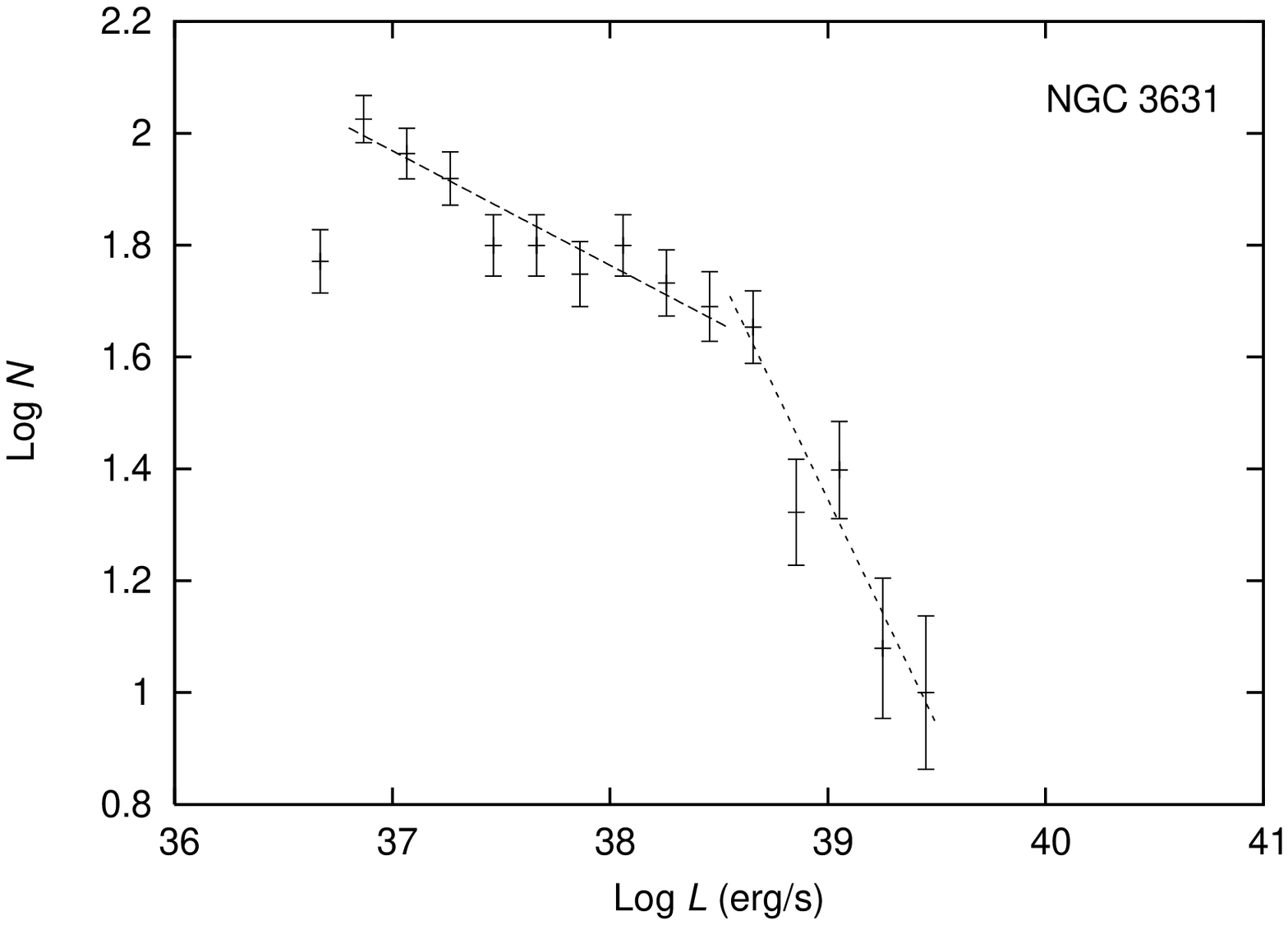,angle=-90,width=9cm}
\psfig{figure=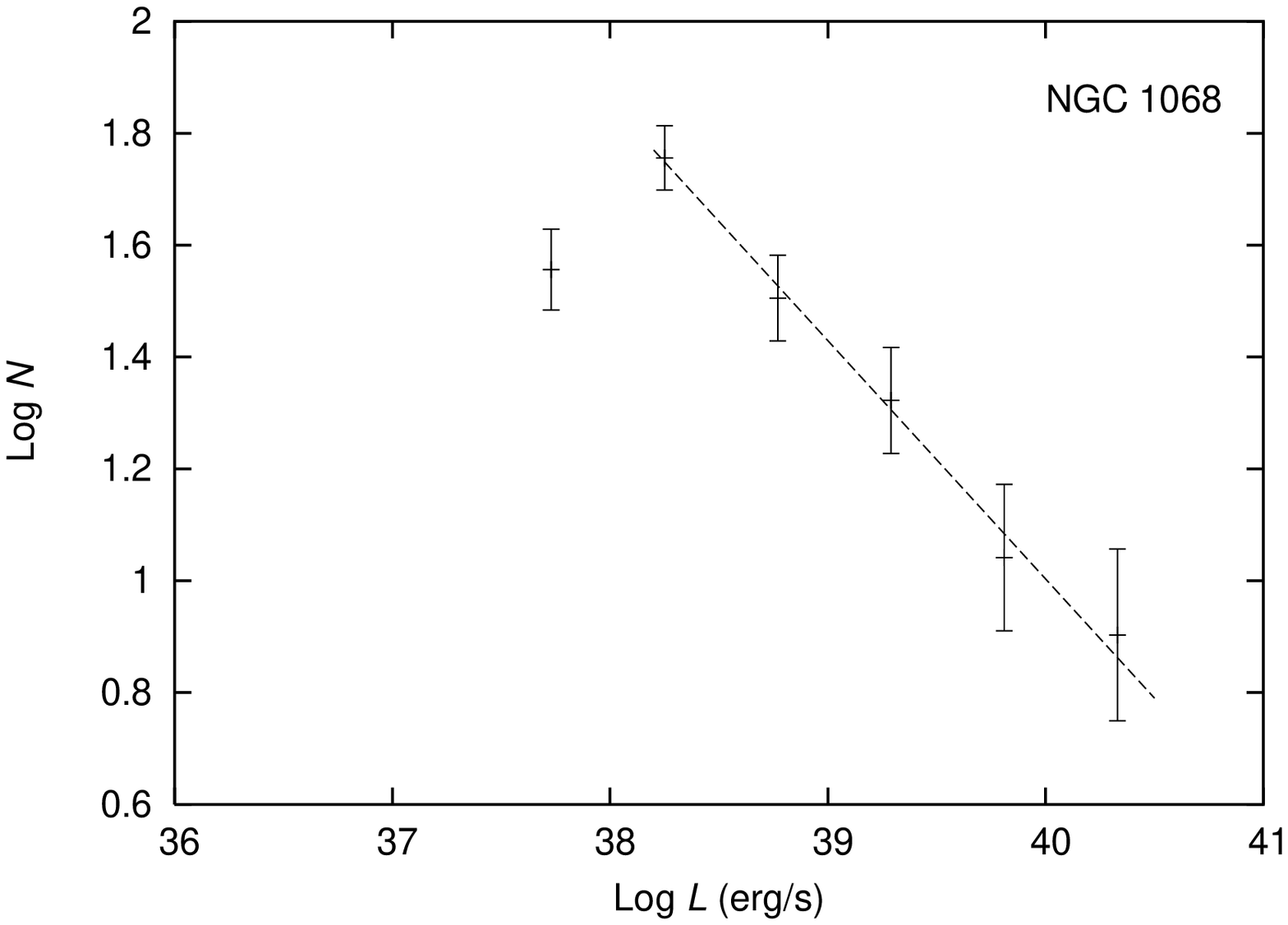,angle=-90,width=9cm}
\caption{\hii\ region LFs of NGC~3631 ({\it top panel}), whose
  catalogue contains 801 \hii\ regions, fitted with a double power
  law, and of NGC~1068 ({\it lower panel}), with 166 \hii\ regions,
  and fitted with a single power law. The LFs have been plotted
  with bin widths of 0.2 and 0.5\, dex, repectively.}
\label{lf-examples}
\end{figure}

KEH pointed out (see also, e.g., Rand 1992; Beckman et al. 2000) that
for those galaxies which have large enough numbers of \hii\ regions
with $L>39$\,dex, the LFs were, in general, best fitted with a
double power law, with the fit to the higher-$L$ portion of the LF
steeper-sloped than that of the lower-$L$ portion. This is borne out
in the present sample, where only eight of our galaxies satisfy the
requirement of sufficient high-$L$ \hii\ regions to entail a
statistically significant double power law fit. Fitting the LFs of
these galaxies with a single power law gives a demonstrably inferior
fit. Parameters of the single power law fits are listed in Table~1 for
all those galaxies in the sample with sufficient \hii\ regions to do
this reliably, whereas the results of the double fits are listed in
Table~2 ({\bf online only}). Fig.~2 shows representative examples
of a galaxy where a double fit is superior ({\it top panel}) and of
one with insufficient high-$L$ \hii\ regions ({\it lower panel}).

\subsection{Composite luminosity functions}

\begin{figure}
\psfig{figure=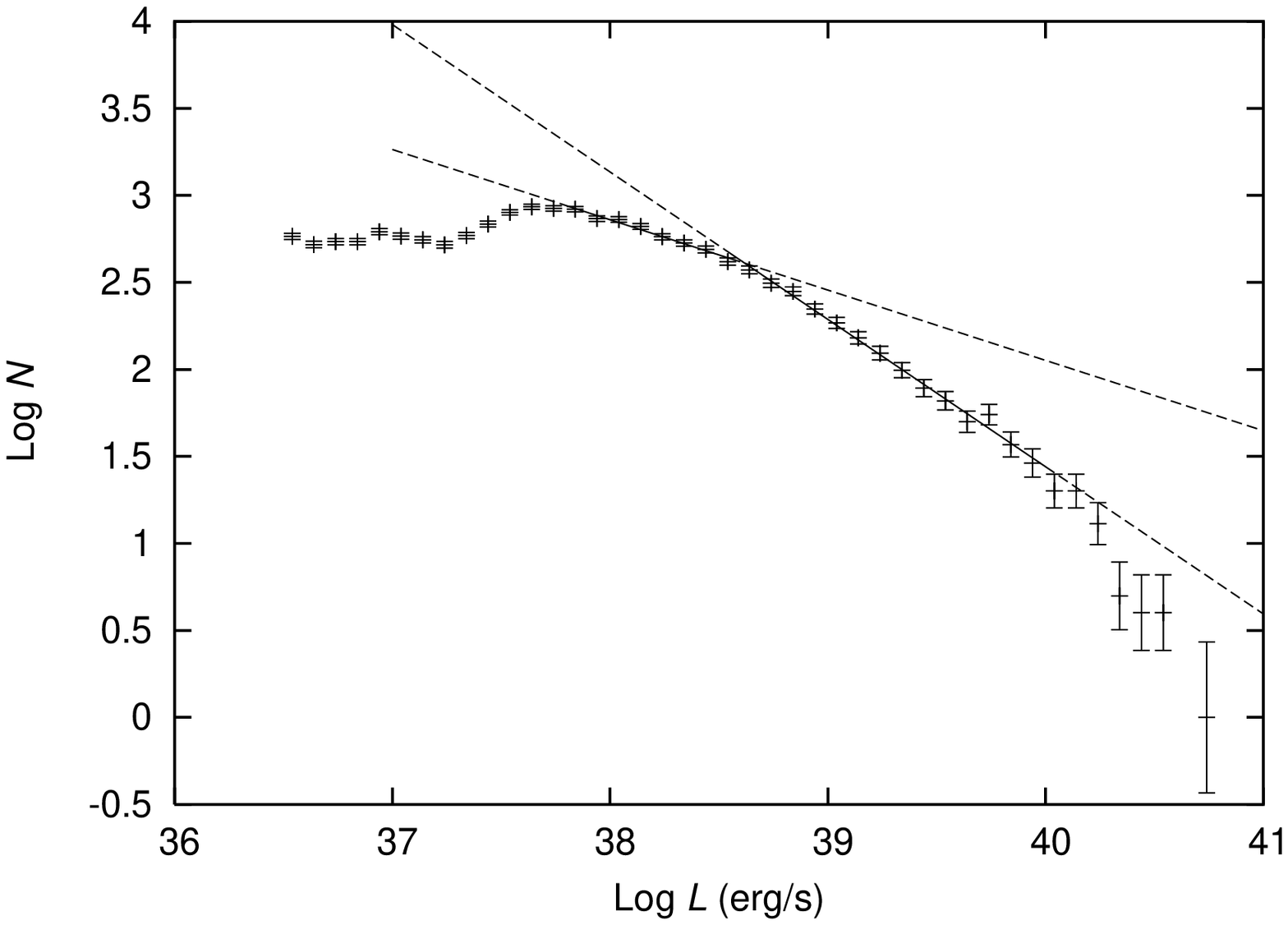,angle=-90,width=9cm}
\caption{LF made from the combined \hii\ region catalogues of 53 of
  our sample galaxies. The large number of \hii\ regions allows
  plotting the LF in steps of 0.1\,dex. The best double-slope fit is
  indicated by the lines, where dashes show extrapolations beyond the
  range over which the best fit has been made.}
\label{lfbig}
\end{figure}

In the present {\it Letter}, our purpose is to discuss the combined LF
for the complete sample (rather than to use separately the individual
LFs shown in Fig.~1). We can produce a combined LF because all our
data have been uniformly calibrated photometrically, and because,
separating the sample into three basic morphological blocks: early,
intermediate and late types, we did not find significant differences
in the slopes of the low or high luminosity portions of the LF.  The
result, which contains 17,797 \hii\ regions, is shown in
Fig.~\ref{lfbig} (bin width 0.1\,dex). Overlaid on that Figure
is the best double power law fit ($a_{\rm upper}=-1.36\pm0.02,
r^2=0.98; a_{\rm lower}=-1.86\pm0.03, r^2=1.00$).  A single power law
can be fitted ($a_{\rm single}=-1.58\pm0.01, r^2=0.94$), but has a
significantly lower coefficient of determination. The low-$L$ limit to
the fitted range occurs at a value of $L=37.6$\,dex, which is a
completeness limit imposed by a combination of signal-to-noise ratio
and spatial resolution in the \ha\ images. As we will see below,
this completeness limit occurs well below the break luminosity and
allows fits to both the low and high-$L$ slopes.

\begin{figure}
\psfig{figure=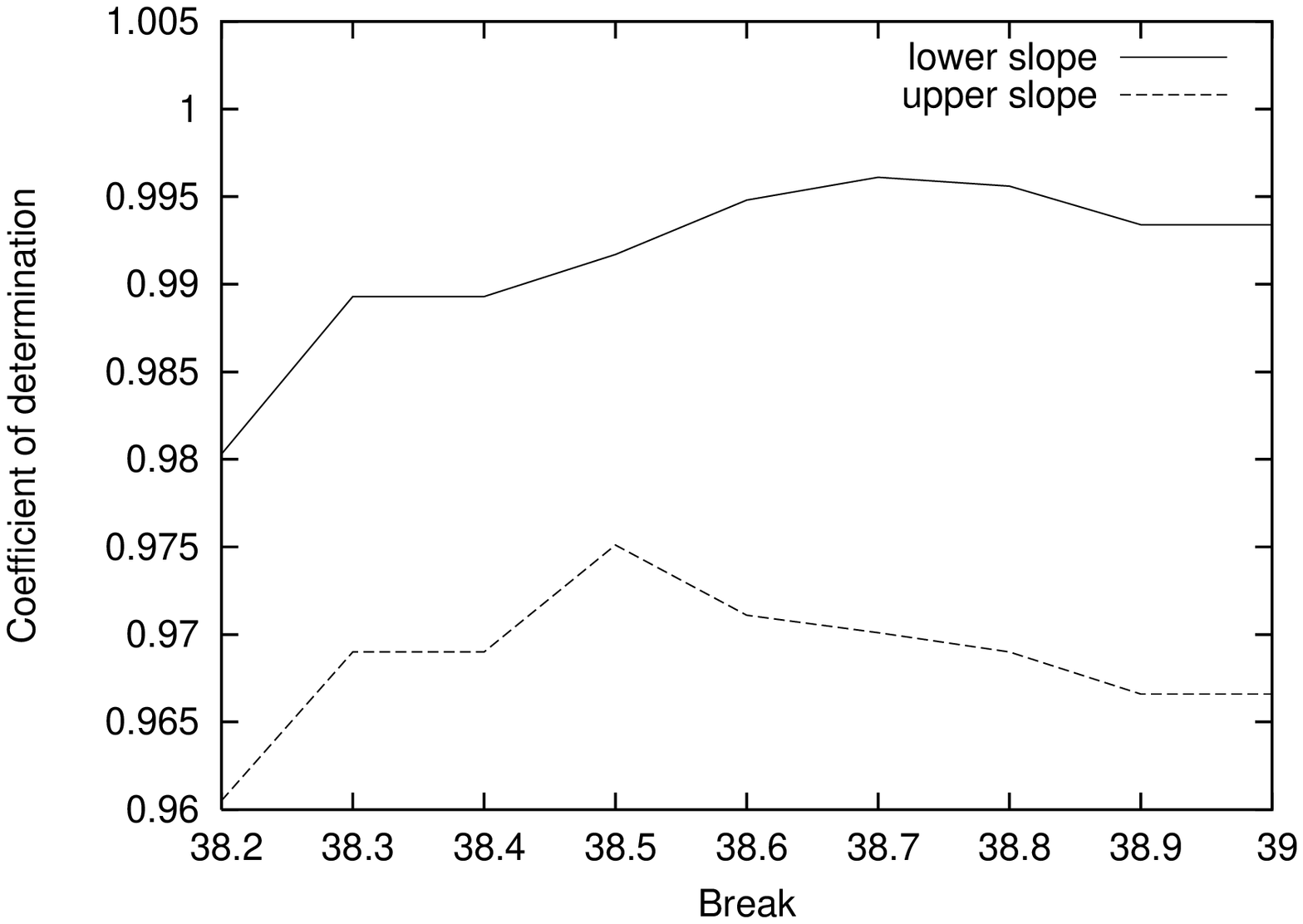,angle=-90,width=9cm}
\caption{Coefficient of determination, $r^2$, for the
  upper and lower slopes of the double-slope fit to the combined LF of
  Fig.~\ref{lfbig}, as a function of the $\log\,L$ at which the break
  between the two slopes occurs. Better fits are closer to $r^2=1$, so
  the best fit is achieved for a break at $\log L {\rm
  (erg\,s}^{-1})=38.6\pm0.1$, the average of the best values for
  the low and high-$L$ slopes.}
\label{lffitquality}
\end{figure}

Not only does the statistical strength of our combined LF of almost
18,000 \hii\ regions allow us to confirm so clearly {\it that} a break
occurs in the LF slope, but the quality of our data is good enough for
us to establish exactly {\it where} that break occurs. We made
successive fits to the two slopes, varying the break luminosity
between $L=37.7$\,dex and 39.1\,dex. Fig.~\ref{lffitquality} shows the
parameter which best indicates the quality of each fit, the
coefficient of determination, $r^2$, as a function of $\log\,L$ for
both the lower and upper slopes. Before interpreting the results, we
must note that $r^2$ will move closer to unity (indicating better
fits) as more points are fitted, which will occur for higher break-$L$
for the lower-$L$ slope, and vice versa for the upper slope. Even so,
Fig.~\ref{lffitquality} clearly shows a maximum for each slope,
indicating a distinct and restricted range of $\log\,L$ where the best
fit quality is obtained. The resulting value for the break, or
Str\"omgren, luminosity is $\log\,L ({\rm erg\,s}^{-1}) = 38.6\pm
0.1$.

Theoretical explanations for the break have been discussed previously,
notably by Beckman et al. (2000) who suggested that it might be caused
by the transition from ionisation bounding to density bounding as the
\hii\ regions become more luminous, and who made semi-quantitative
predictions of the relations between the slopes of the two power laws
based on this scenario. Pleuss et al. (2000) argued that the effect
might be due to a clustering property in which \hii\ regions of
increasing luminosity have an increasing probability of absorbing
smaller \hii\ regions by overlap, and that this may be subject to a
scaling law related to the separation between star formation regions.
Any quantitative attempt to implement the latter or similar proposals,
however, leads to an LF in which the upper luminosity slope is
shallower, not steeper, than that in the lower luminosity range. So
the density bounding scenario retains its basic plausibility, although
it must be carefully quantified to take into account the fact that
\hii\ regions are highly clumpy (see, e.g., Giammanco et al. 2004 for
a study of how clumpiness affects the propagation of ionising
radiation in \hii\ regions), rather than being idealised Str\"omgren
spheres of uniform density. An alternative explanation is the one 
proposed by Oey \& Clarke (1998), in which the break in the LF is due
to the evolution of the \ha\ luminosities of the \hii\ regions.

It is interesting to note that at the high luminosity end of the LF,
above $L = 40$\,dex, the curve falls away quite sharply. It is true
that the number of \hii\ regions per bin in this range is small, and
this is shown clearly in Fig.~\ref{lfbig} in the increasing amplitude
of the error bars as $L$ increases. The LF would take this form if we
were seeing a true cut-off in the luminosity of \hii\ regions due to a
physical limit on the masses of high mass stars modulated by a
statistical effect due to the variation of the total stellar mass in
young clusters. Before reaching a firm conclusion on this, however, it
would be useful to be able to increase the statistical base in this
range by further observations.

\section{Summary and concluding remarks}

We have analysed continuum-subtracted \ha\ images of 57 nearby
galaxies, and here present the LFs and catalogues of their populations
of \hii\ regions. Using the combined data set we have produced a
single LF covering the range in \ha\ luminosity from $L = 37.5$\,dex
to just over $L = 40$\,dex. The total number of \hii\ regions
contributing to this LF is a little under 18,000, which is almost an
order of magnitude greater than that of any previously published LF of
this type.  The best functional fit of the LF, above the completeness
limit, is by a double power law, having a break at $L = 38.6$\,dex,
with the steeper slope to higher luminosity and the shallower slope to
lower luminosity. The existence of this break suggests a change in
physical regime going from lower to higher \ha\ luminosity. If
confirmed, and by virtue of its occurrence at high \hii\ region
luminosities, the break is of potential use as a distance indicator
for star-forming disk galaxies.

The number of \hii\ regions in our sample is large enough to suggest
that the steep fall off in the LF above $L = 40$\,dex indicates that
we are sampling the range where there is a cut-off in the luminosities
of the \hii\ regions, implying a drop-off in the LF of the underlying
massive stellar clusters. This is giving us potentially important
information about a physical limit to the masses of stars in
young clusters of varying total stellar mass.

\begin{acknowledgements}

We thanks the anonymous referee for a number of excellent suggestions
which helped improve this Letter. J.H.K. acknowledges support of the
Leverhulme Trust in the form of a Leverhulme Research Fellowship,
while S.L.F. thanks the Royal Astronomical Society for the award of a
summer student bursary. This research was partly supported by Project
AYA-2004-08251-CO2-01 of the Spanish Ministry for Education and
Science, and by Project P3/86 of the Instituto de Astrof\'\i sica de
Canarias. The JKT has been operated on the island of La Palma by the
ING in the Spanish Observatorio del Roque de los Muchachos of the
Instituto de Astrof\'\i sica de Canarias.  This research has made use
of the NASA/IPAC Extragalactic Database (NED) which is operated by the
Jet Propulsion Laboratory, California Institute of Technology, under
contract with the National Aeronautics and Space Administration.

\end{acknowledgements}

\newpage

\Online

\begin{table*}
\centering
\begin{tabular}{rccrrcrccccccccc}
\hline
NGC & Cal. & Compl. & $N$ & Scale & $D$ & Res. & max& min& bin width & slope &  error & $r^2$\\
\hline
& log($L$) & log($L$) & & (pc/\sec) & (Mpc) & (pc) & log($L$) & log($L$) & (dex)  & & &\\
\hline
\hline
%\end{tabular}
%\hline
210 & 38.49 & 38.40 & 396 & 98 & 20.3 & 151 & 39.95& 37.64& 0.3& $-$1.54& 0.13& 0.725\\
337A & 37.94 & 37.07 & 189 & 66 & 13.7 & 94 & 39.44& 36.78& 0.3& $-$1.63& 0.07& 0.965\\
488 & 38.46 & 38.37 & 35 & 142 & 29.3 & 209 & 39.45& 38.28& 0.4& \multicolumn{3}{c}{Too few \hii\ regions}\\
628 & N/A & N/A & 2027 & 47 & 9.7 & 54 & \multicolumn{2}{c}{No photometric data} & 0.2& $-$1.25& 0.03& 0.899\\
864 & 38.35 & 38.15 & 227 & 97 & 20 & 133 & 40.32& 37.84& 0.3& $-$1.40& 0.09& 0.726\\
1042 & 38.21 & 37.65 & 158 & 81 & 16.7 & 133 & 39.35& 37.36& 0.3& $-$1.57& 0.09& 0.985\\
1068 & 38.41 & 38.16 & 166 & 70 & 14.4 & 157 & 40.59& 37.48& 0.5& $-$1.43& 0.06& 0.994\\
1073 & 38.08 & 37.41 & 170 & 74 & 15.2 & 100 & 39.37& 37.10& 0.5& $-$1.49& 0.07& 0.970\\
1169 & 36.65 & PQ & 26 & 163 & 33.7 & 295 & 36.52& 35.74& 0.4& \multicolumn{3}{c}{Too few \hii\ regions}\\
1179 & 37.47 & PQ & 18 & 103 & 21.2 & 228 & 38.46& 37.11& 0.5& \multicolumn{3}{c}{Too few \hii\ regions}\\
1300 & 38.42 & 38.03 & 84 & 91 & 18.8 & 206 & 39.60& 37.93& 0.3& $-$1.68& 0.16& 0.883\\
2775 & 38.17 & 37.53 & 66 & 82 & 17 & 156 & 38.70& 37.44& 0.3& \multicolumn{3}{c}{Too few \hii\ regions}\\
2805 & 38.23 & 38.02 & 94 & 136 & 28 & 229 & 39.76& 37.65& 0.4& $-$1.47& 0.14& 0.924\\
2985 & 38.37 & 37.92 & 338 & 109 & 22.4 & 194 & 39.16& 37.40& 0.2& $-$1.54& 0.09& 0.931\\
3184 & 37.85 & 37.92 & 576 & 42 & 8.7 & 73 & 39.25& 36.63& 0.2& $-$1.35& 0.10& 0.733\\
3227 & 38.71 & 37.34 & 190 & 100 & 20.6 & 142 & 39.39& 37.05& 0.4& \multicolumn{3}{c}{Too few \hii\ regions}\\
3344 & 37.22 & 36.66 & 669 & 30 & 6.1 & 41 & 38.56& 36.38& 0.2& $-$1.52& 0.05& 0.920\\
3351 & 37.53 & 37.42 & 55 & 39 & 8.1 & 62 & 38.66& 37.33& 0.6& \multicolumn{3}{c}{Too few \hii\ regions}\\
3368 & 37.44 & 36.98 & 77 & 39 & 8.1 & 68 & 38.68& 36.89& 0.3& $-$1.72& 0.18& 0.945\\
3486 & 37.27 & 36.55 & 612 & 36 & 7.4 & 46 & 38.83& 36.30& 0.2& $-$1.43& 0.04& 0.895\\
3631 & 33.88 & 36.87 & 801 & 105 & 21.6 & 137 & 39.54& 36.57& 0.2& $-$1.21& 0.03& 0.896\\
3726 & 38.28 & 37.17 & 614 & 82 & 17 & 115 & 39.48& 36.82& 0.2& $-$1.45& 0.05& 0.915 \\
3810 & 37.19 & 36.23 & 400 & 82 & 16.9 & 107 & 38.66& 35.96& 0.3& $-$1.14& 0.04& 0.725\\
4030 & 38.89 & 38.03 & 276 & 126 & 25.9 & 194 & 39.96& 37.74& 0.2& $-$1.08& 0.07& 0.107\\
4051 & 38.12 & 37.60 & 232 & 82 & 17 & 124 & 39.49& 37.28& 0.3& $-$1.28& 0.10& 0.933\\
4123 & 38.88 & 37.56 & 250 & 123 & 25.3 & 166 & 39.89 & 37.47 & 0.3& $-$1.40 & 0.06 & 0.941\\
4145 & N/A & PQ & 35 & 100 & 20.7 & 270 & \multicolumn{2}{c}{No photometric data} & 0.4& \multicolumn{3}{c}{Too few \hii\ regions}\\
4151 & 38.69 & 36.00 & 263 & 98 & 20.3 & 146 & 39.48& 36.98& 0.3& $-$1.17& 0.086& 0.972\\
4242 & N/A & PQ & 44 & 36 & 7.5 & 114 & \multicolumn{2}{c}{No photometric data}& 0.4& \multicolumn{3}{c}{Too few \hii\ regions}\\
4254 & 38.09 & 37.80 & 626 & 81 & 16.8 & 123 & 39.92& 37.51& 0.2& $-$1.22& 0.08& 0.833\\
4303 & 38.13 & 37.81 & 873 & 74 & 15.2 & 105 & 40.10& 37.32& 0.2& $-$1.44& 0.04& 0.952\\
4314 & 37.93 & NDR & 14 & 47 & 9.7 & 84 & \multicolumn{2}{c}{No photometric data} & \multicolumn{3}{c}{No disk regions}\\
4321 & 34.26 & 37.20 & 2647 & 81 & 16.8 & 80 & 40.15& 36.93& 0.2& $-$1.72& 0.10& 0.957\\
4395 & 35.78 & 35.90 & 498 & 18 & 3.6 & 32 & 38.66& 35.60& 0.3& $-$1.52& 0.05& 0.979\\
4450 & N/A & PQ & 32 & 81 & 16.8 & 70 & \multicolumn{2}{c}{No photometric data} & 0.6& \multicolumn{3}{c}{Too few \hii\ regions}\\
4487 & 38.38 & 37.64 & 146 & 97 & 19.9 & 192 & 39.84& 37.46& 0.3& \multicolumn{3}{c}{Too few \hii\ regions}\\
4535 & 38.20 & 7.64 & 518 & 81 & 16.8 & 152 & 39.82& 37.34& 0.2& $-$1.51& 0.05& 0.924\\
4548 & 38.12 & 37.48 & 74 & 81 & 16.8 & 141 & 38.90& 37.20& 0.3& $-$1.83& 0.16& 0.961\\
4579 & 38.04 & 37.38 & 121 & 81 & 16.8 & 127 & 39.75& 37.07& 0.4& $-$1.60& 0.15& 0.973\\
4618 & 37.67 & 36.44 & 290 & 35 & 7.3 & 46 & 38.89& 36.15& 0.4& \multicolumn{3}{c}{Too few \hii\ regions}\\
4689 & 38.26 & 37.50 & 160 & 81 & 16.8 & 146 & 39.20& 37.17& 0.3& $-$1.40& 0.14& 0.786\\
4725 & 37.97 & 37.82 & 134 & 60 & 12.4 & 140 & 39.51& 37.32& 0.3& $-$1.90& 0.11& 0.969\\
4736 & 36.91 & 36.39 & 294 & 21 & 4.3 & 30 & 37.82& 36.09& 0.4& $-$1.55& 0.13& 0.972\\
5247 & 38.79 & 38.23 & 157 & 108 & 22.2 & 247 & 40.27& 38.14& 0.3& $-$1.20& 0.13& 0.525\\
5248 & 38.48 & 37.72 & 381 & 110 & 22.7 & 156 & 40.18& 37.63& 0.3& $-$1.42& 0.04& 0.945\\
5334 & 38.75 & 37.78 & 106 & 120 & 24.7 & 217 & 39.39& 37.50& 0.3& $-$1.93& 0.15& 0.969\\
5371 & 38.76 & 38.28 & 264 & 183 & 37.8 & 243 & 40.28& 38.00& 0.3& $-$1.43& 0.11& 0.939\\
5457 & 36.71 & 36.60 & 978 & 26 & 5.4 & 62 & 39.03& 35.40& 0.2& $-$1.67& 0.04& 0.946\\
5474 & 37.18 & 37.36 & 165 & 29 & 6 & 36 & 39.82& 37.26& 0.3& $-$1.18& 0.11& 0.614\\
5850 & 38.82 & 37.59 & 155 & 138 & 25.5 & 220 & 39.26& 37.30& 0.3& $-$1.69& 0.11& 0.960\\
5921 & 38.36 & 37.80 & 274 & 122 & 25.2 & 153 & 39.28& 37.50& 0.2& $-$1.38& 0.11& 0.944\\
5964 & 38.48 & 37.79 & 111 & 120 & 24.7 & 220 & 39.67& 37.50& 0.3& $-$1.96& 0.20& 0.995\\
6140 & 38.34 & 37.00 & 127 & 90 & 18.6 & 126 & 39.29& 37.38& 0.3& $-$1.50& 0.12& 0.970\\
6384 & 38.69 & 39.75 & 148 & 129 & 18.6 & 230 & 40.79& 39.48& 0.2& $-$1.08& 0.19& 0.830\\
6946 & 37.11 & 36.65 & 1528 & 27 & 5.5 & 40 & 39.38& 36.36& 0.2& $-$1.44& 0.03& 0.974\\
7727 & 38.32 & PQ & 24 & 113 & 23.3 & 196 & 38.16& 37.61& 0.7&\multicolumn{3}{c}{Too few \hii\ regions}\\
7741 & 37.96 & 38.5 & 246 & 60 & 12.3 & 87 & 39.51& 36.76& 0.3& $-$1.35& 0.06& 0.823\\
\hline
\end{tabular}
\caption{({\bf Online only}) General properties of the sample
galaxies, \ha\ images and resulting \hii\ region
catalogues. Calibration constant, in log of $L$ (erg\,s$^{-1}$) per
count in the \ha\ image (col.~2), lower completeness limit of the LF
(log\,$L$; col.~3), number of \hii\ regions catalogued (col.~4), image
scale in pc per arcsec (col.~5), distance to the galaxy in Mpc (from
Paper~II; col. 6), and resolution (seeing, FWHM) in the \ha\ image in
pc (col.~7). N/A in col.~2 means no calibration available, in col.~3,
PQ means poor quality LF and NDR means no disk \hii\
regions. Cols.~8-9 tabulate the upper and lower limits (in $\log L$)
of the luminosity function; col.~10 the bin width used for the plotting
and fitting of the LF, ranging from 0.2 to 0.7\,dex as a function of
the number of \hii\ regions; col.~11 the value of the slope, $a$, of a
single power law fitted to the data; col.~12 the combined error (in
$\log L$), including background effects and calibration errors, and
col.~13 the coefficient of determination, $r^2$, where values closer
to unity indicate better fits.}
\label{table1}
\end{table*}

\begin{table}
\centering
\begin{tabular}{lcccc}
\hline
NGC & 
Type & 
Upper Slope & 
Lower Slope & 
Break $\log L$\\
\hline
628&  Sc    & $-$1.26$\pm $0.03& -2.05$\pm $0.10& N/A \\
3184& Sc    & $-$1.35$\pm $0.10& -2.14$\pm $0.11& 38.7$\pm $0.1 \\
3631& Sbc   & $-$1.21$\pm $0.03& $-$1.81$\pm $0.16& 38.6$\pm $0.1 \\
4123& SBbc  & $-$1.32$\pm $0.09& -2.32$\pm $0.28& 38.7$\pm $0.1 \\
4151& Sab   & $-$1.17$\pm $0.09& -2.10$\pm $0.22& 38.3$\pm $0.1 \\
4254& Sc    & $-$1.22$\pm $0.08& -2.12$\pm $0.13&38.7$\pm $0.1 \\
4321& Sc    & $-$1.72$\pm $0.10& $-$1.97$\pm $0.41 & 38.6$\pm $0.1 \\
6946& Sc    & $-$1.44$\pm $0.03& -2.81$\pm $0.27& 38.2$\pm $0.1 \\
\hline
\end{tabular}
\caption{({\bf Online only}) Double fits to LFs for those galaxies
 whose LFs allowed such fits.}
\label{table2}
\end{table}

\setcounter{figure}{0}
\begin{figure}
\psfig{figure=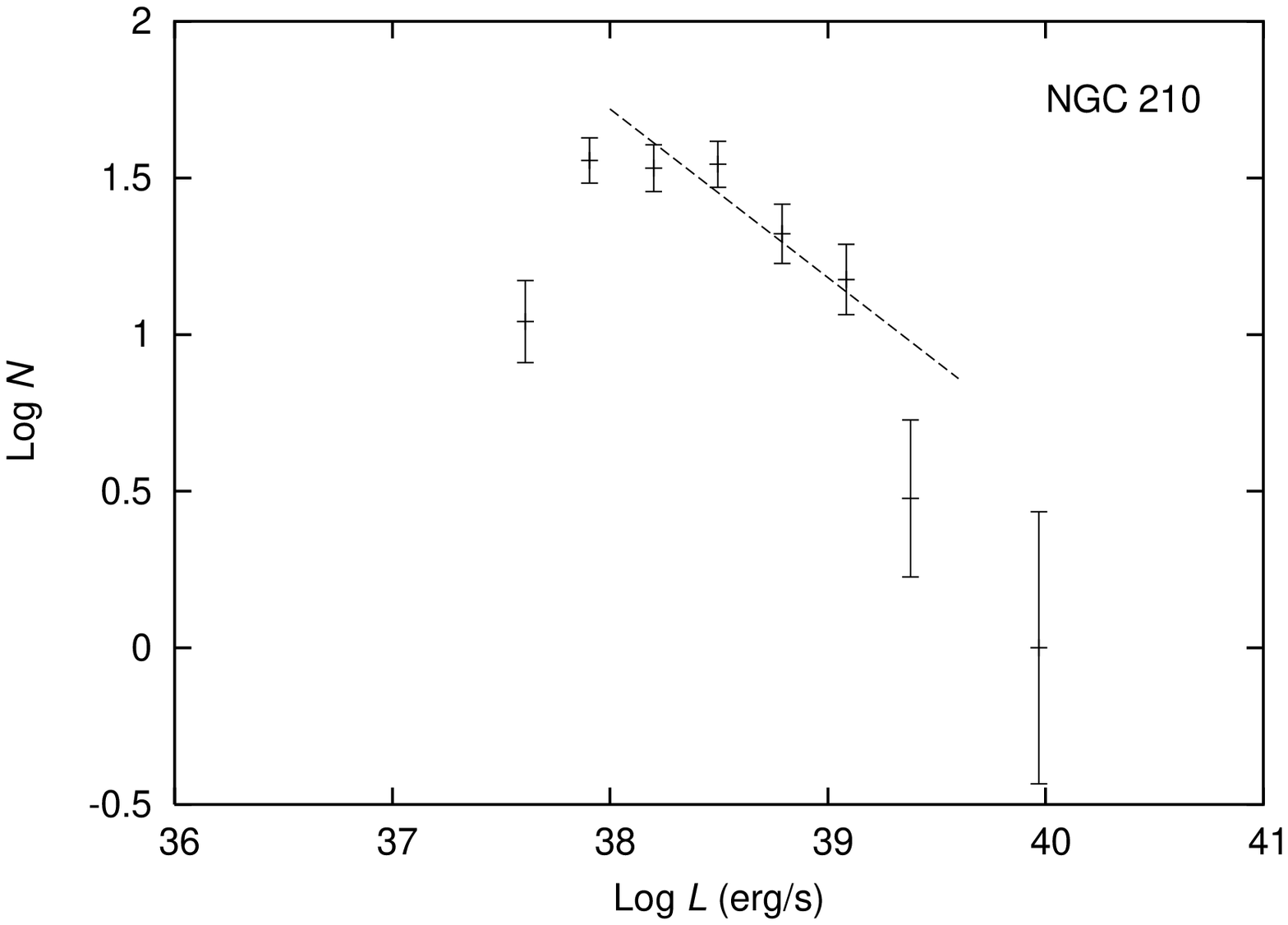,angle=-90,width=9cm}
\psfig{figure=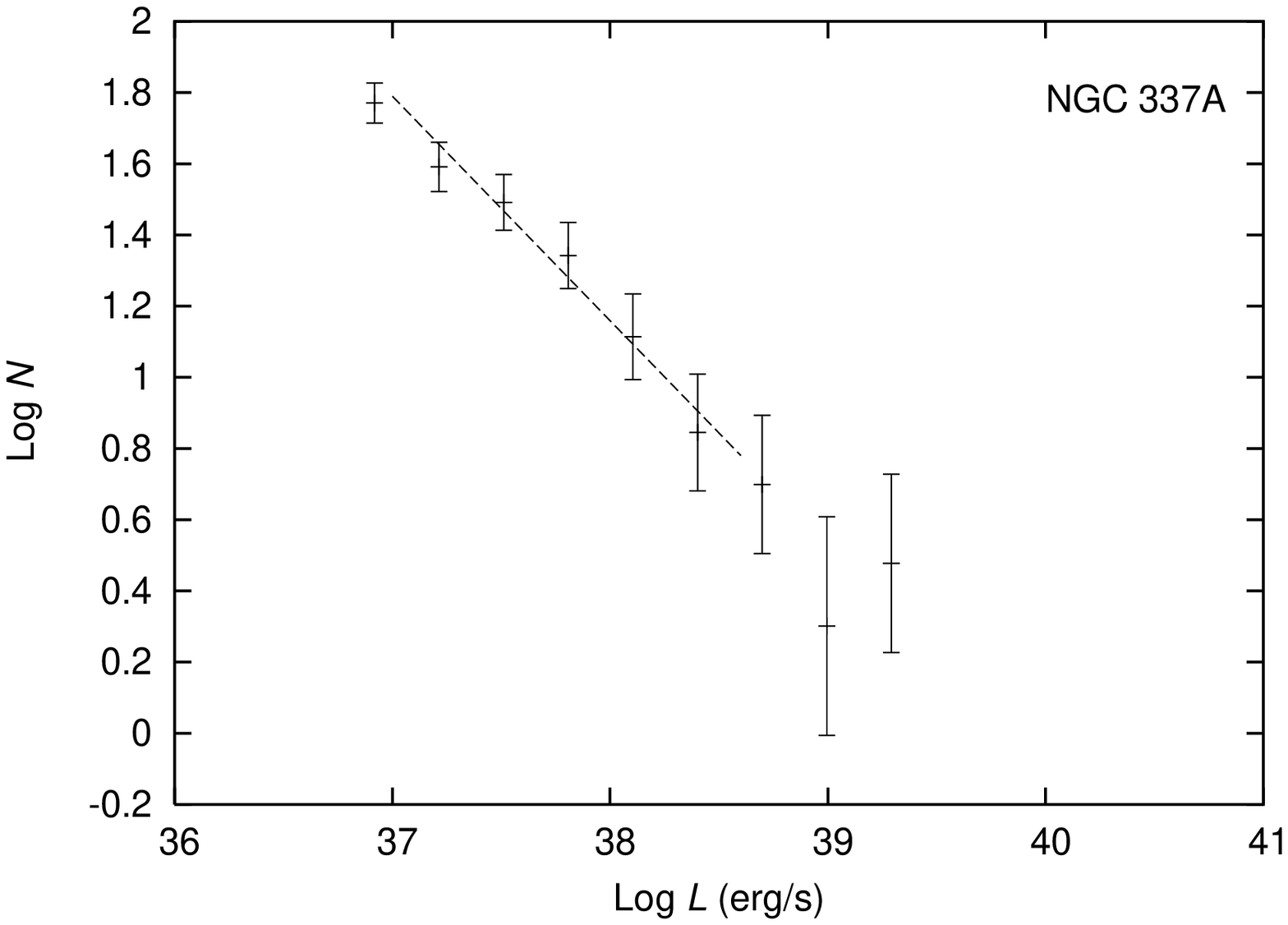,angle=-90,width=9cm}
\psfig{figure=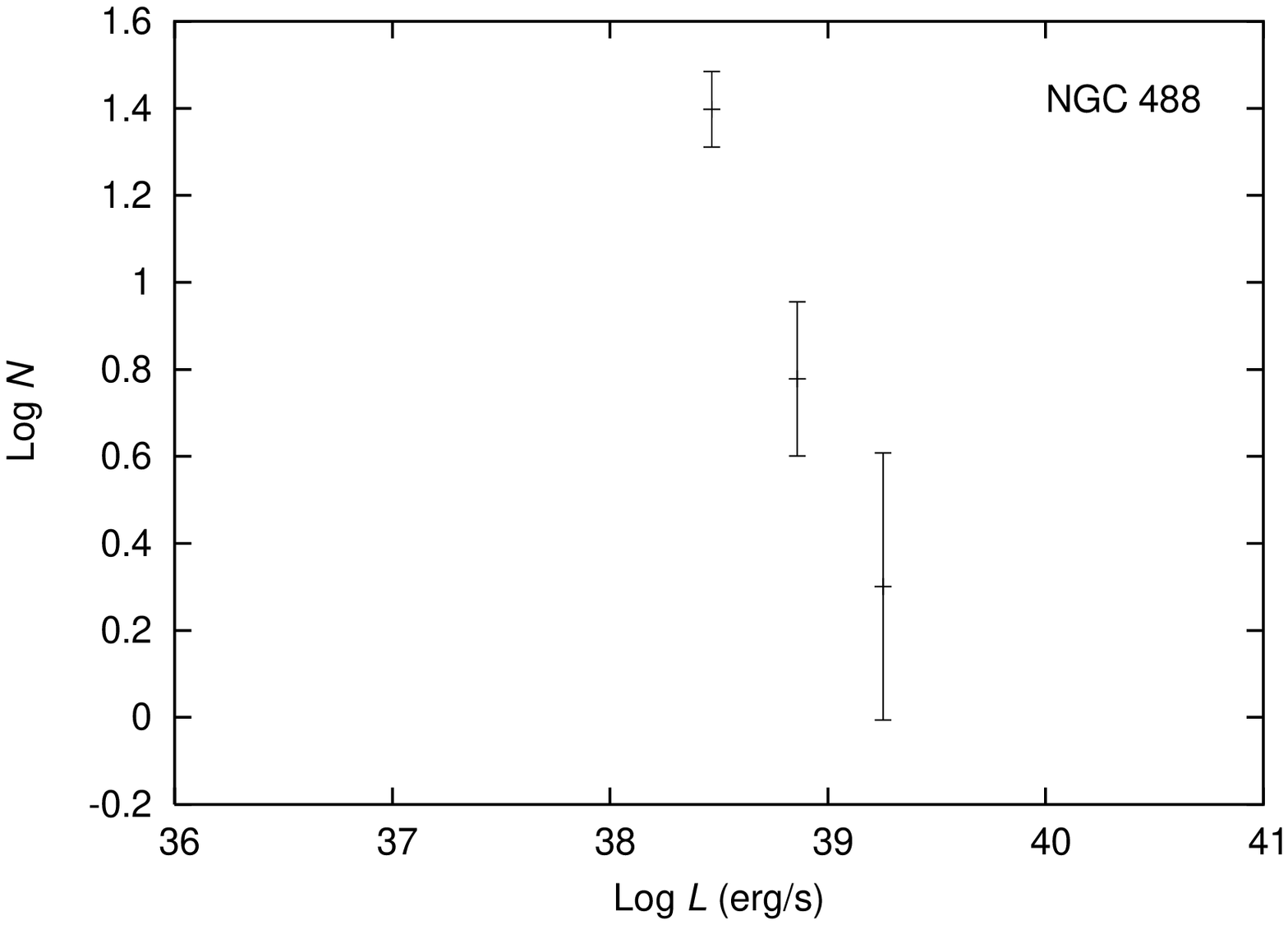,angle=-90,width=9cm}
\psfig{figure=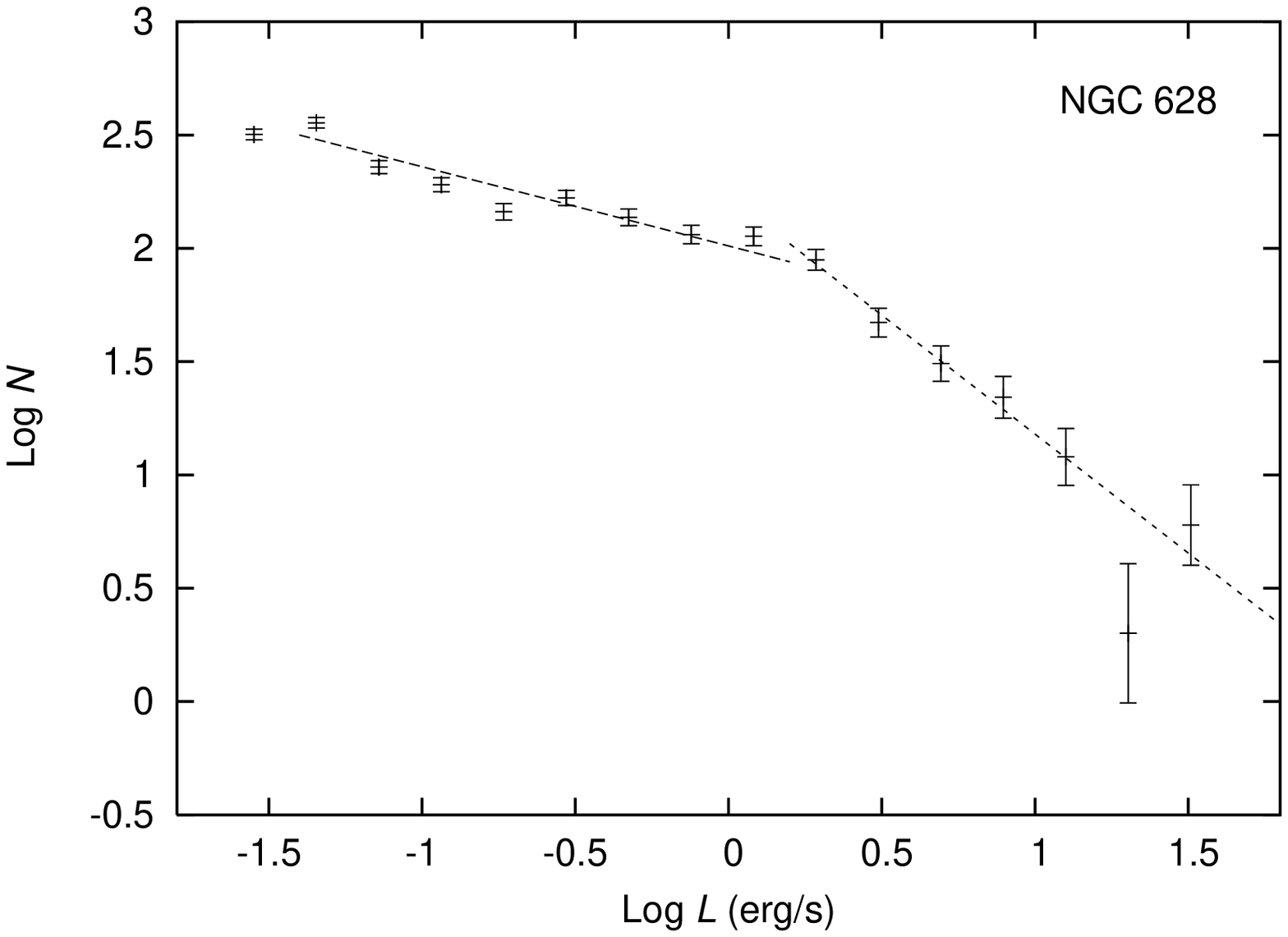,angle=-90,width=9cm}
\caption{{\bf Online only} LFs of 56 of our sample galaxies, with
  best power law fits overlaid. Bin sizes for plotting and for LF
  fitting vary depending on the number of HII regions, as given in
  Table~\ref{table1}. See text for further details.}
\label{lfs}
\end{figure}
\setcounter{figure}{0}
\begin{figure}
\psfig{figure=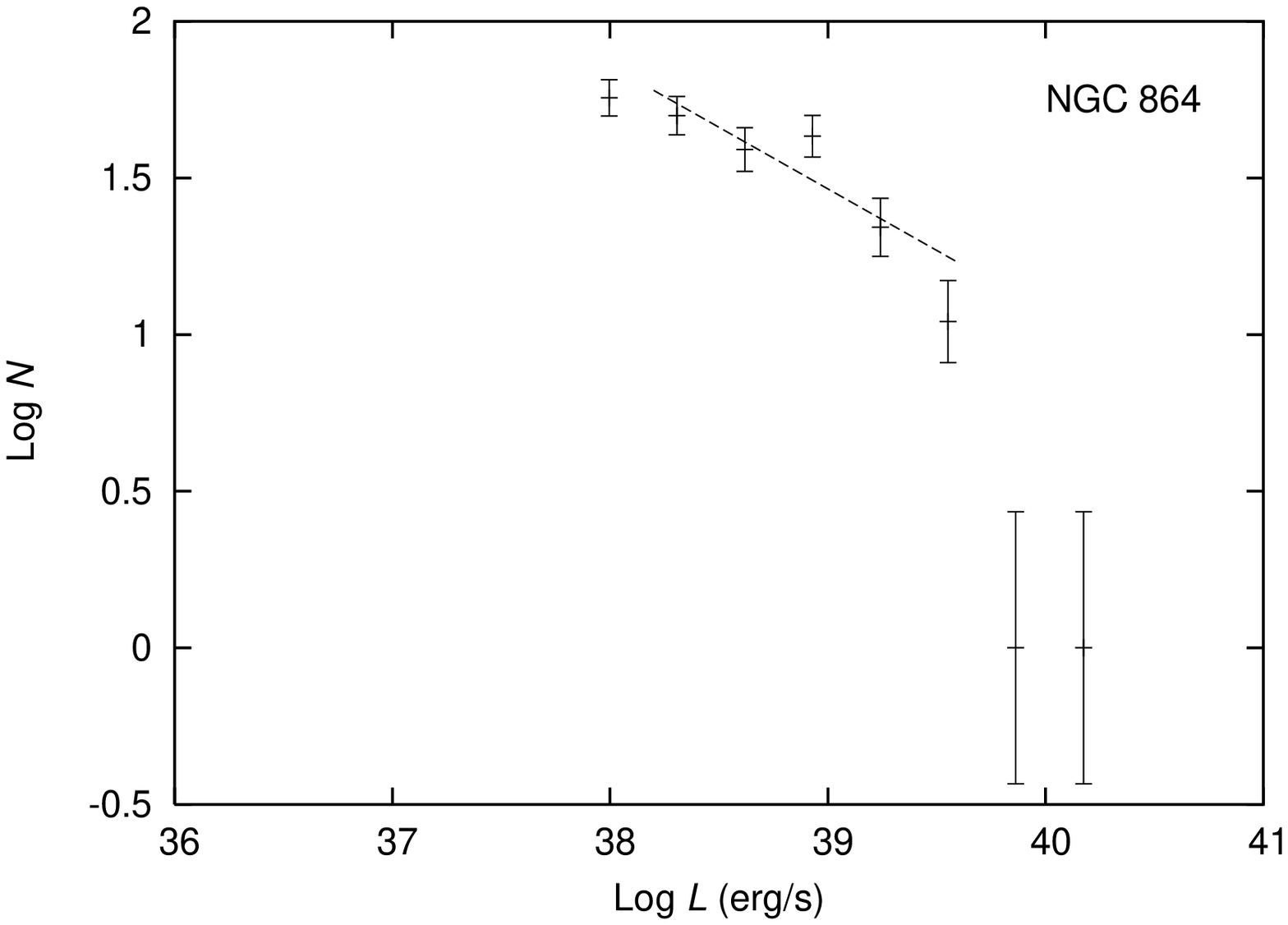,angle=-90,width=9cm}
\psfig{figure=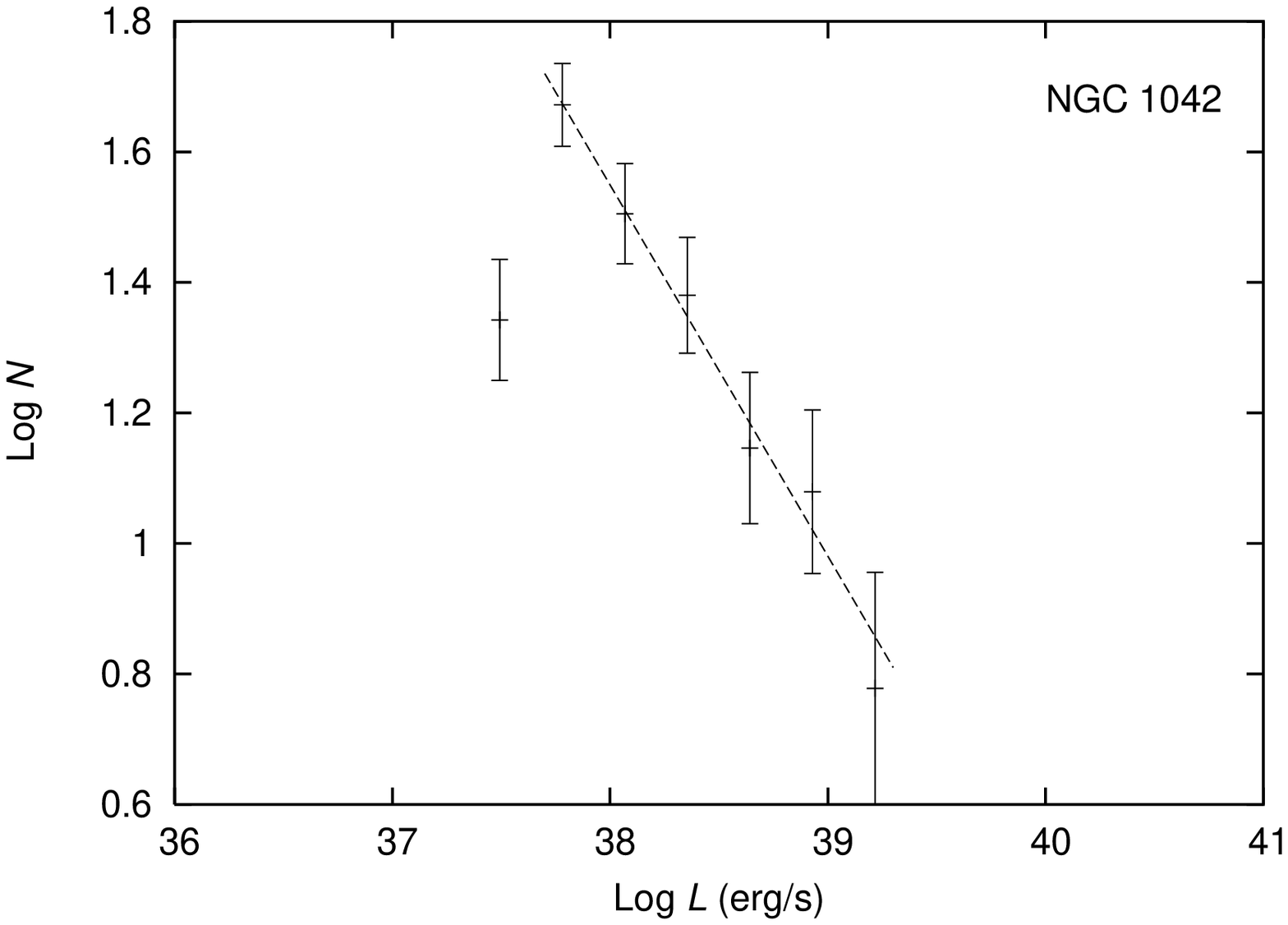,angle=-90,width=9cm}
\psfig{figure=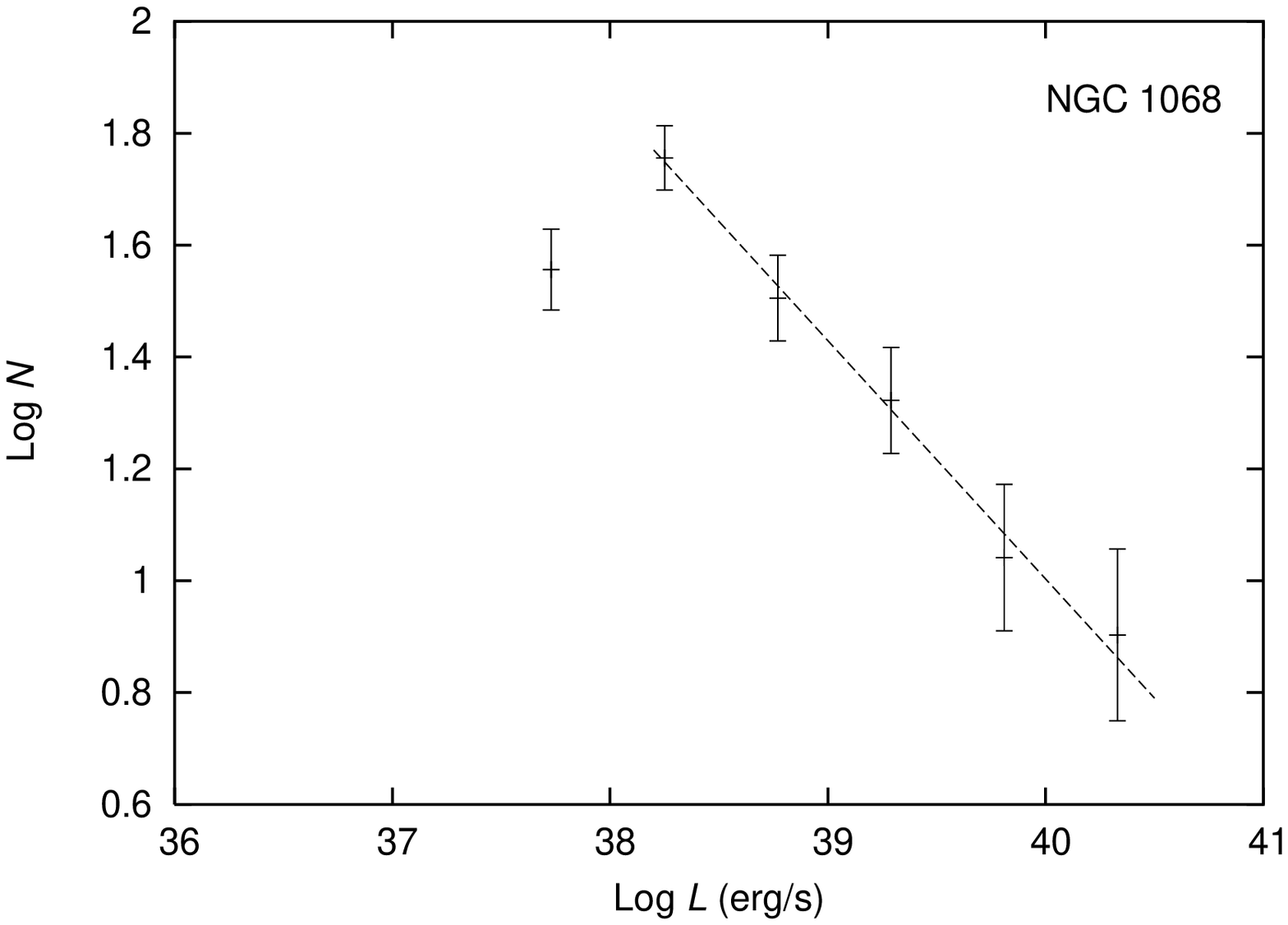,angle=-90,width=9cm}
\psfig{figure=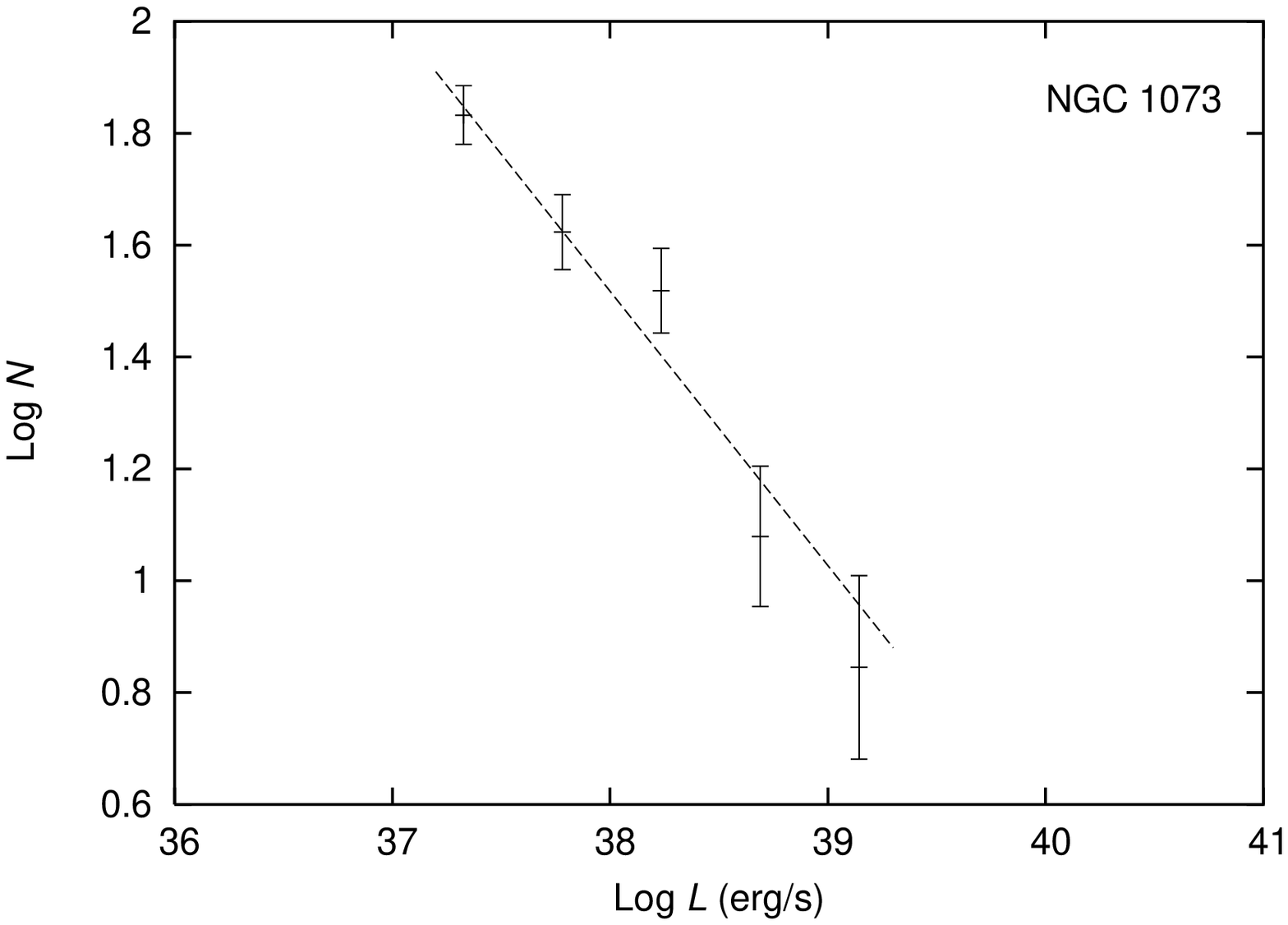,angle=-90,width=9cm}
\caption{(Continued)}
\end{figure}
\setcounter{figure}{0}
\begin{figure}
\psfig{figure=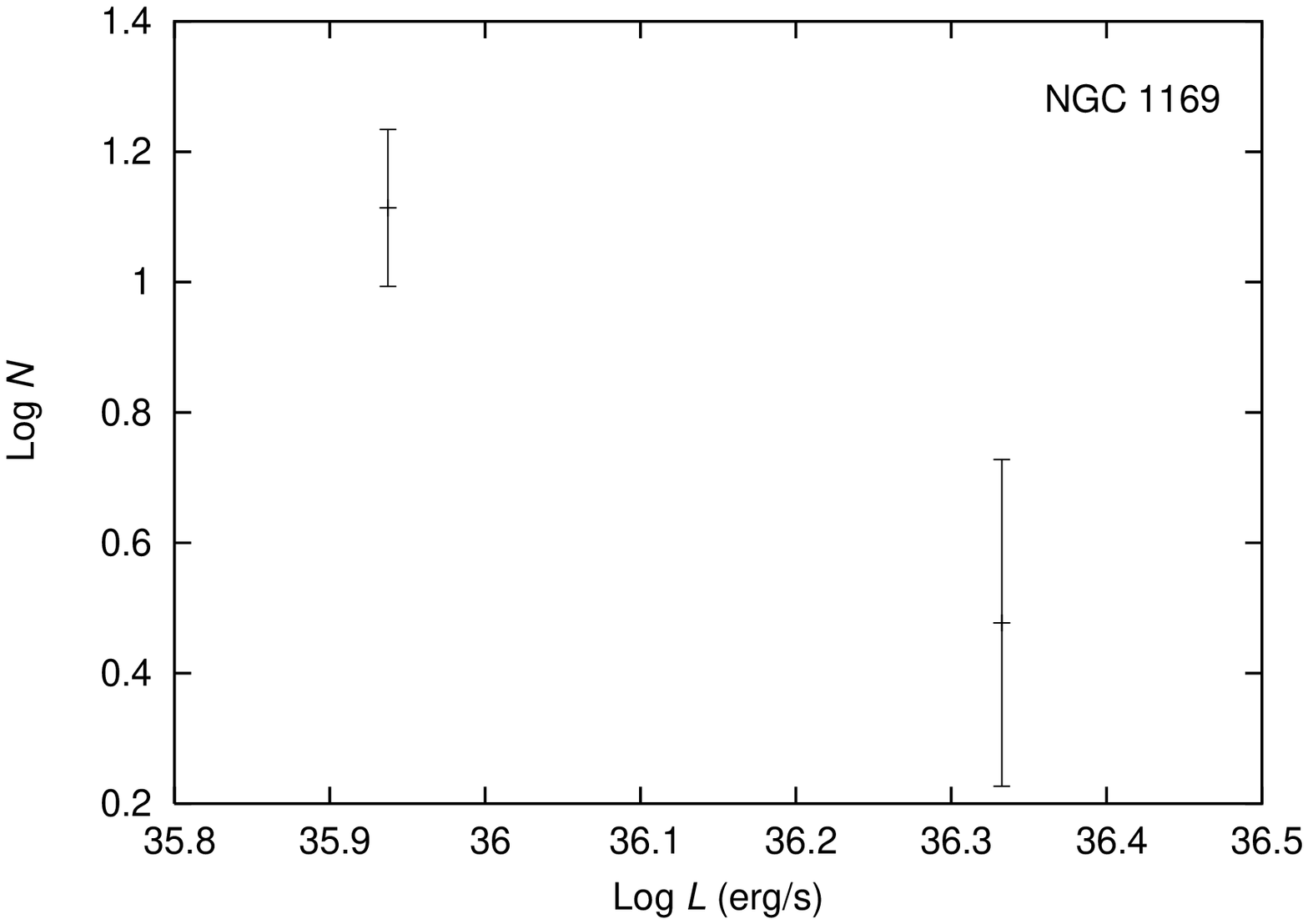,angle=-90,width=9cm}
\psfig{figure=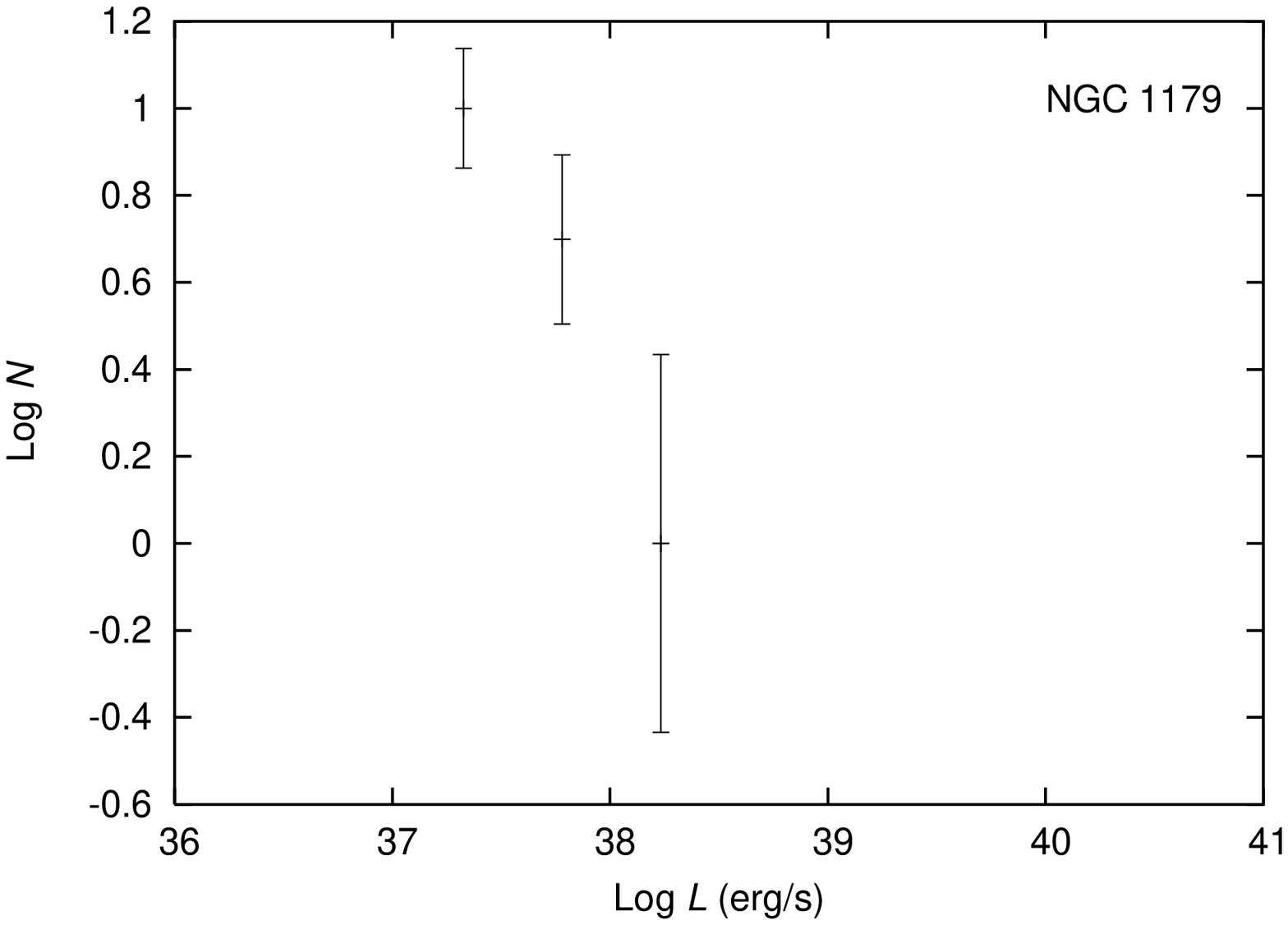,angle=-90,width=9cm}
\psfig{figure=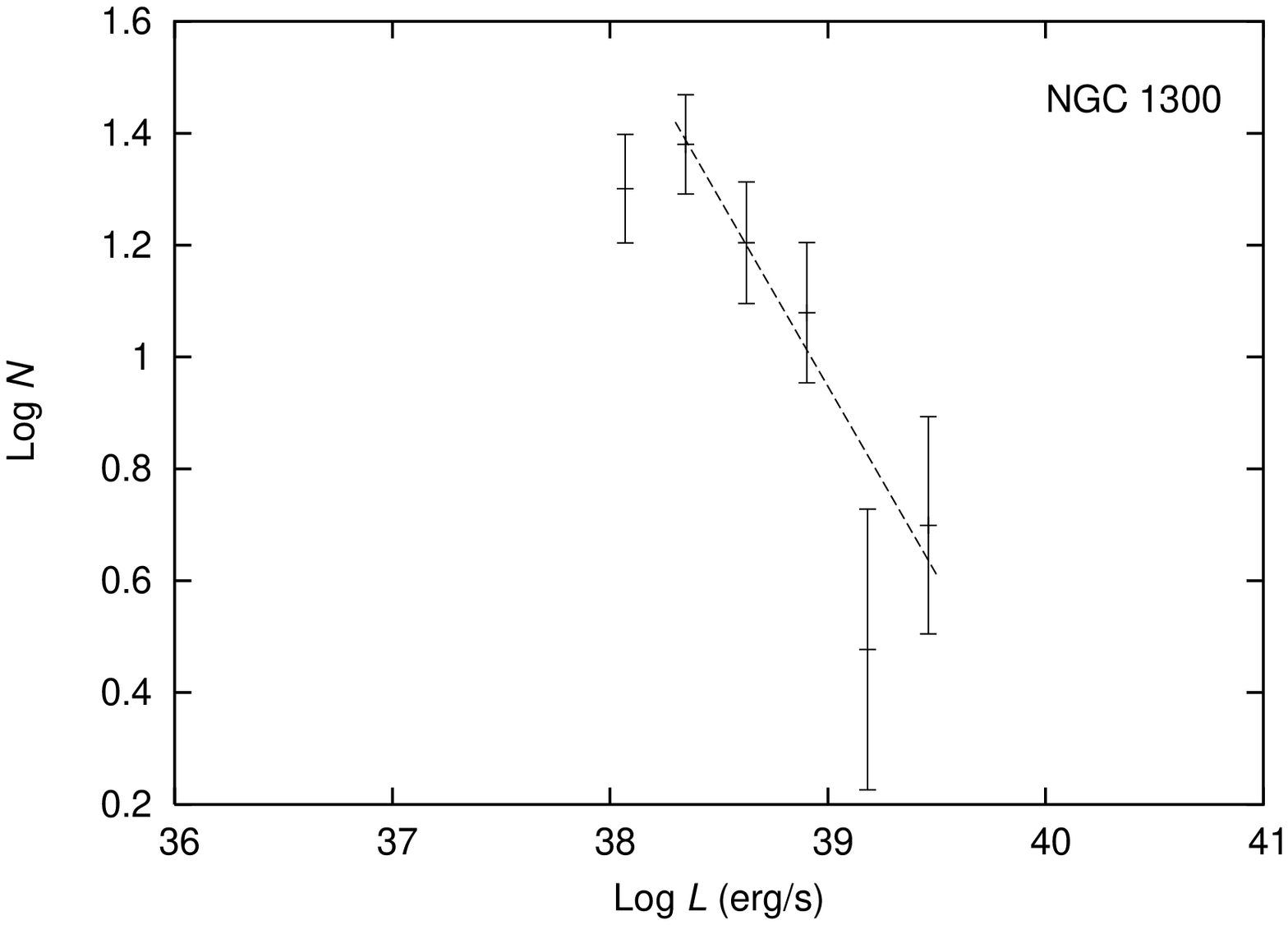,angle=-90,width=9cm}
\psfig{figure=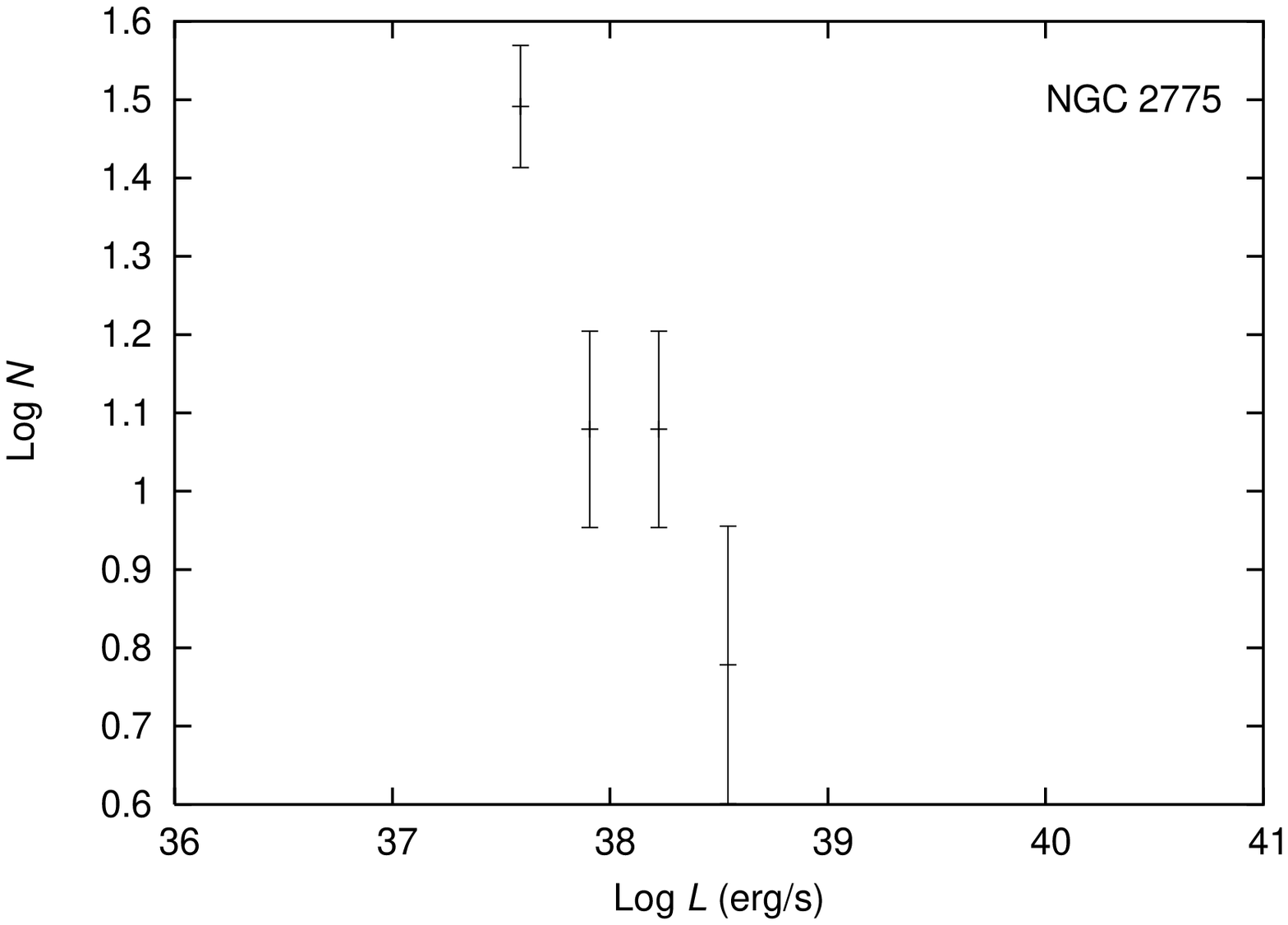,angle=-90,width=9cm}
\caption{(Continued)}
\end{figure}
\setcounter{figure}{0}
\begin{figure}
\psfig{figure=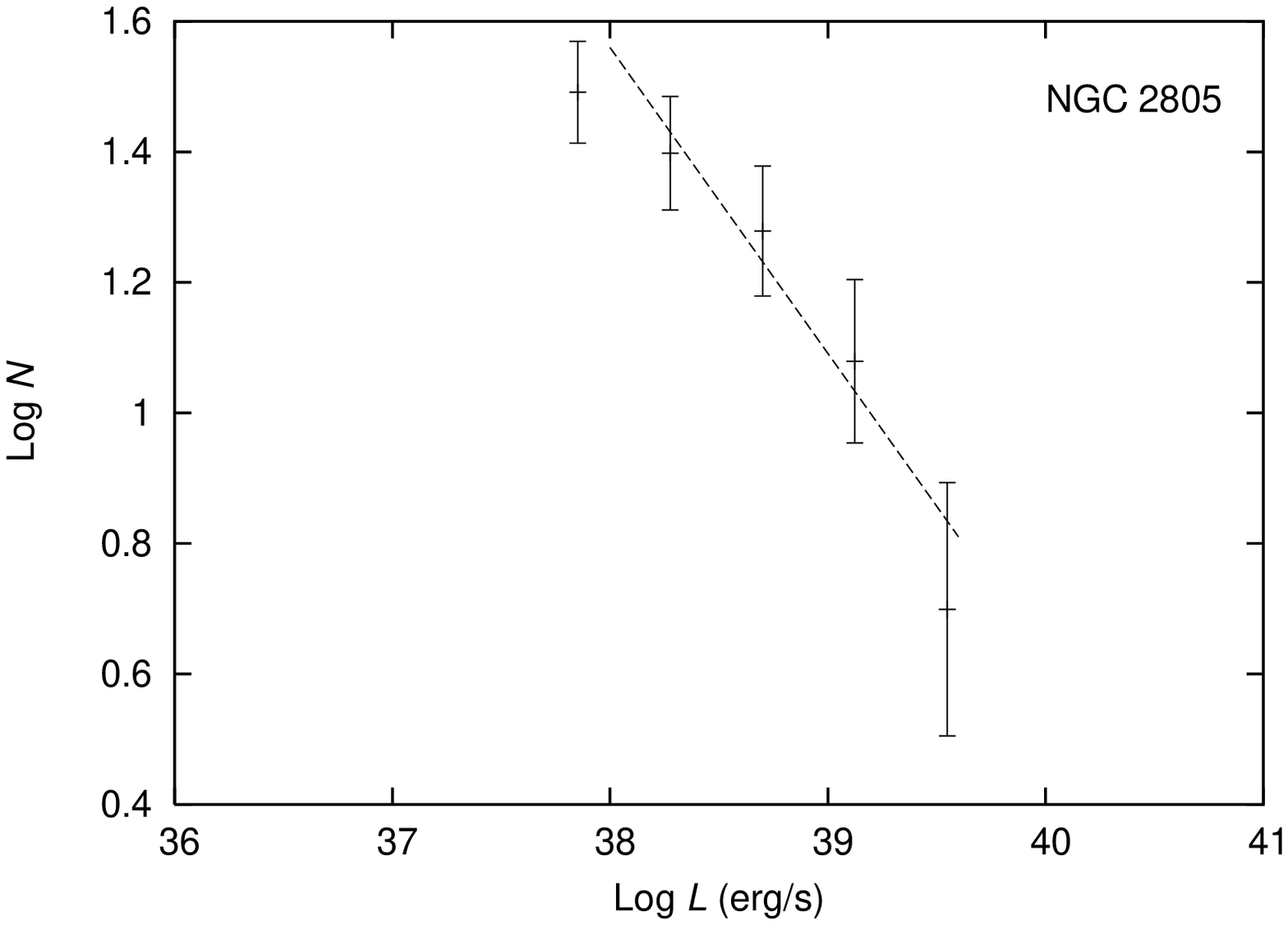,angle=-90,width=9cm}
\psfig{figure=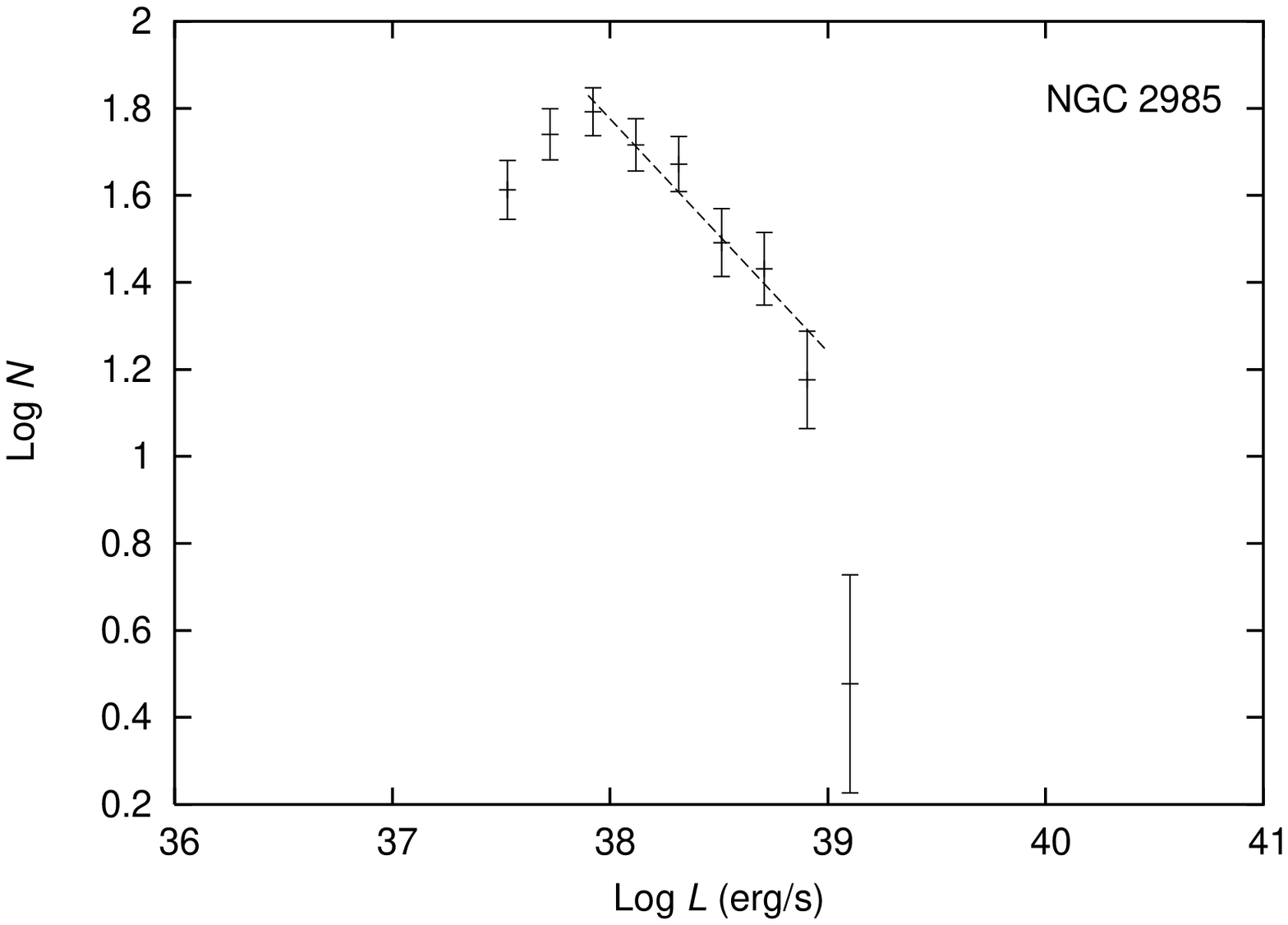,angle=-90,width=9cm}
\psfig{figure=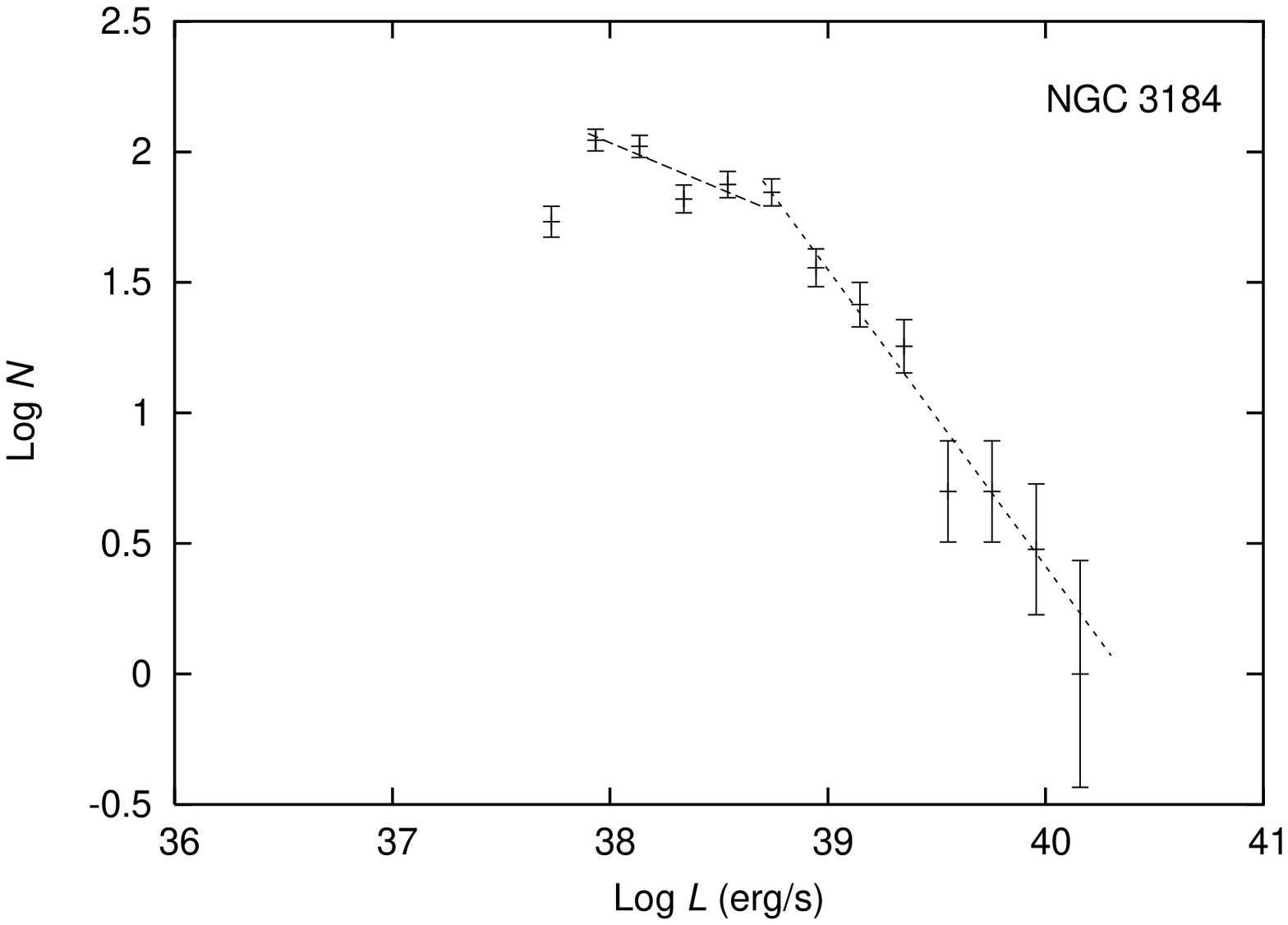,angle=-90,width=9cm}
\psfig{figure=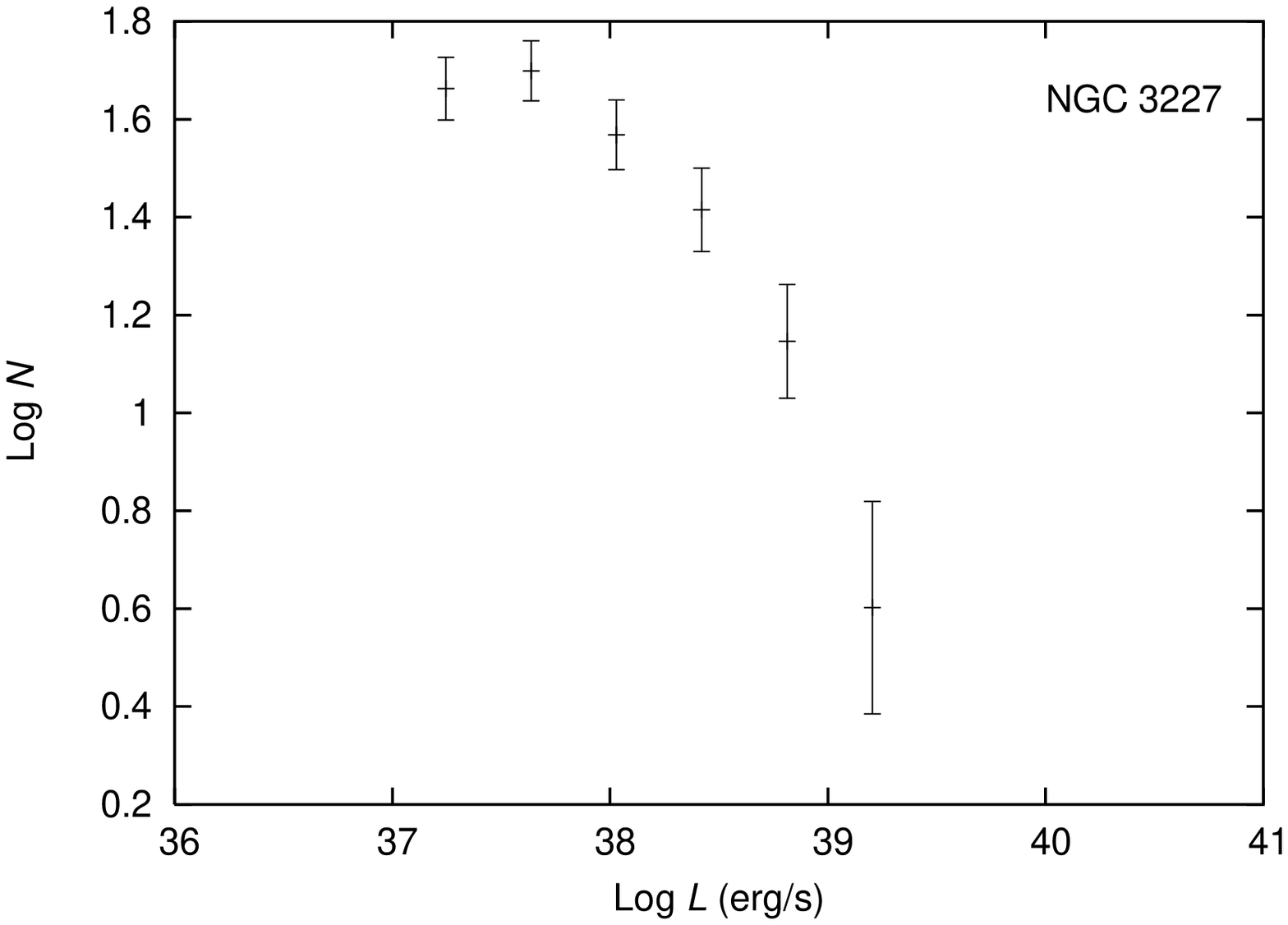,angle=-90,width=9cm}
\caption{(Continued)}
\end{figure}
\setcounter{figure}{0}
\begin{figure}
\psfig{figure=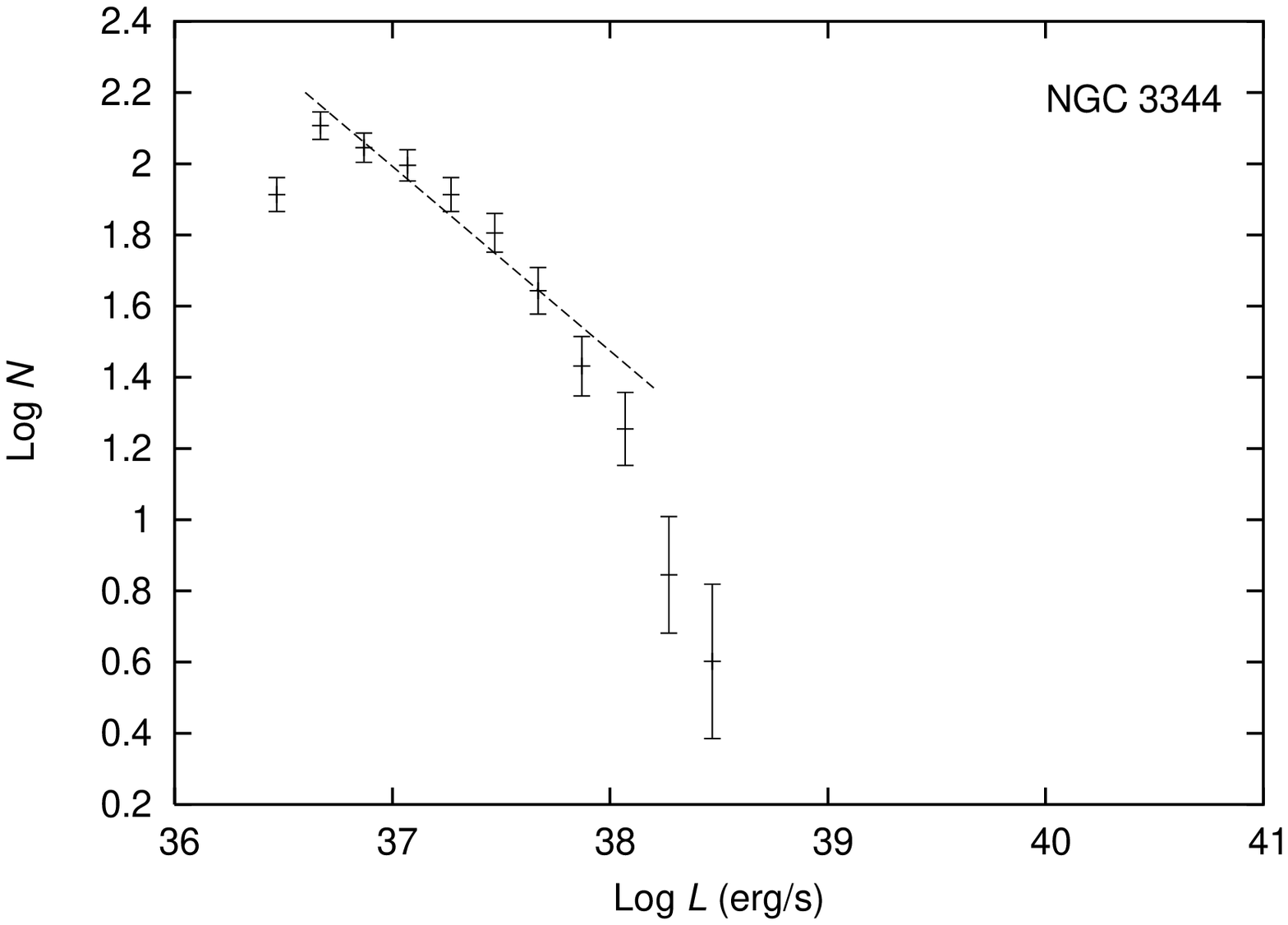,angle=-90,width=9cm}
\psfig{figure=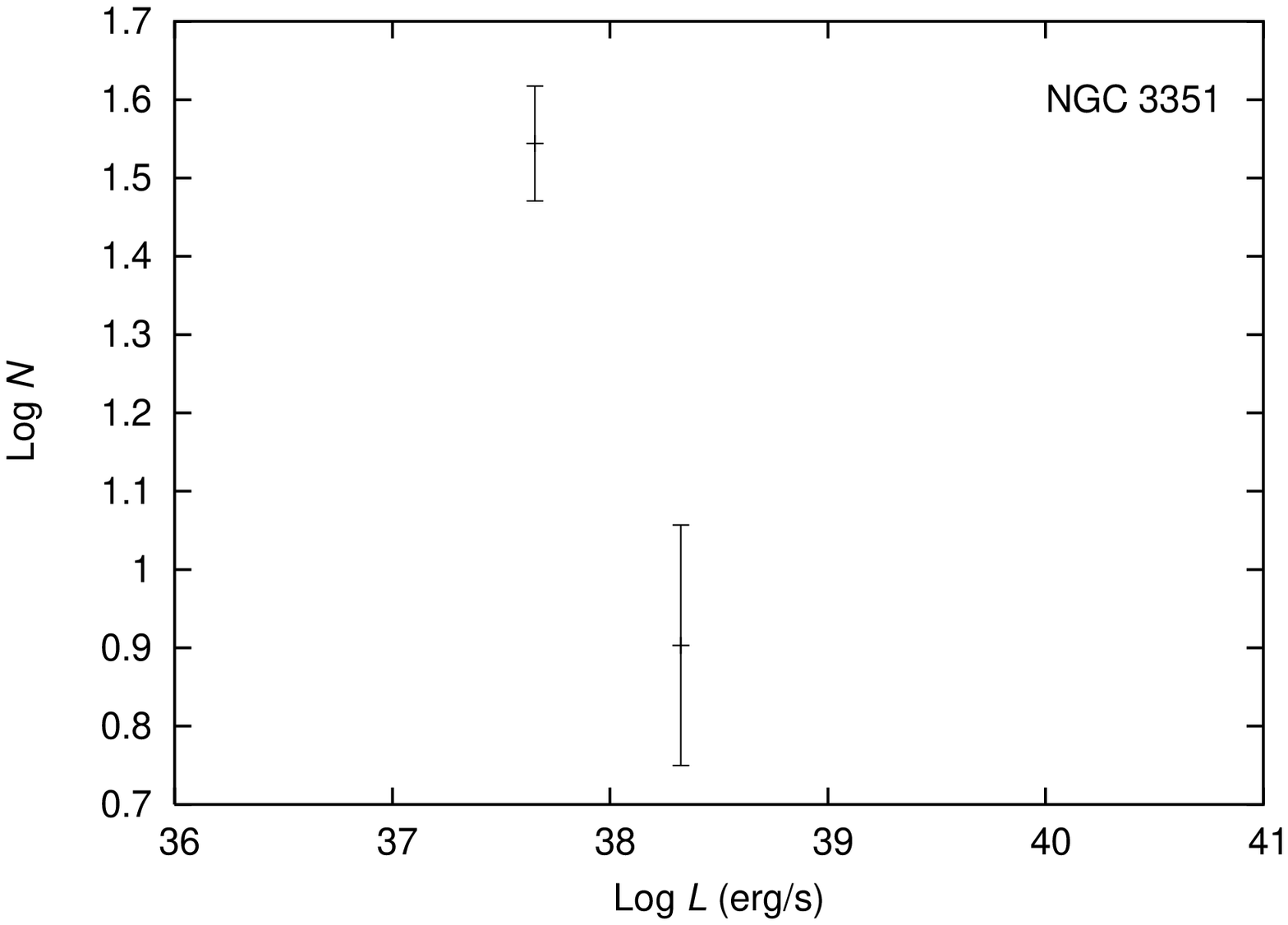,angle=-90,width=9cm}
\psfig{figure=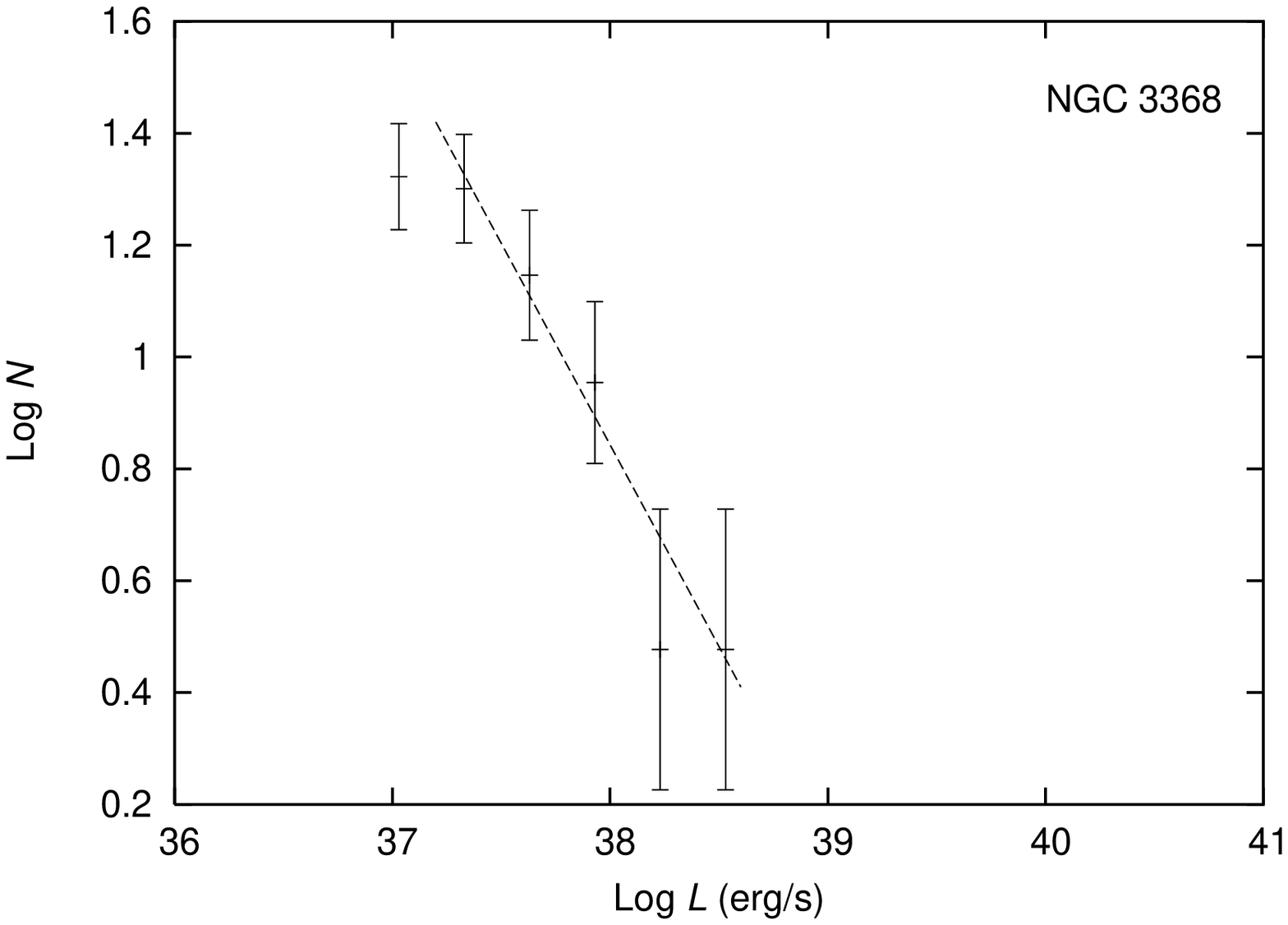,angle=-90,width=9cm}
\psfig{figure=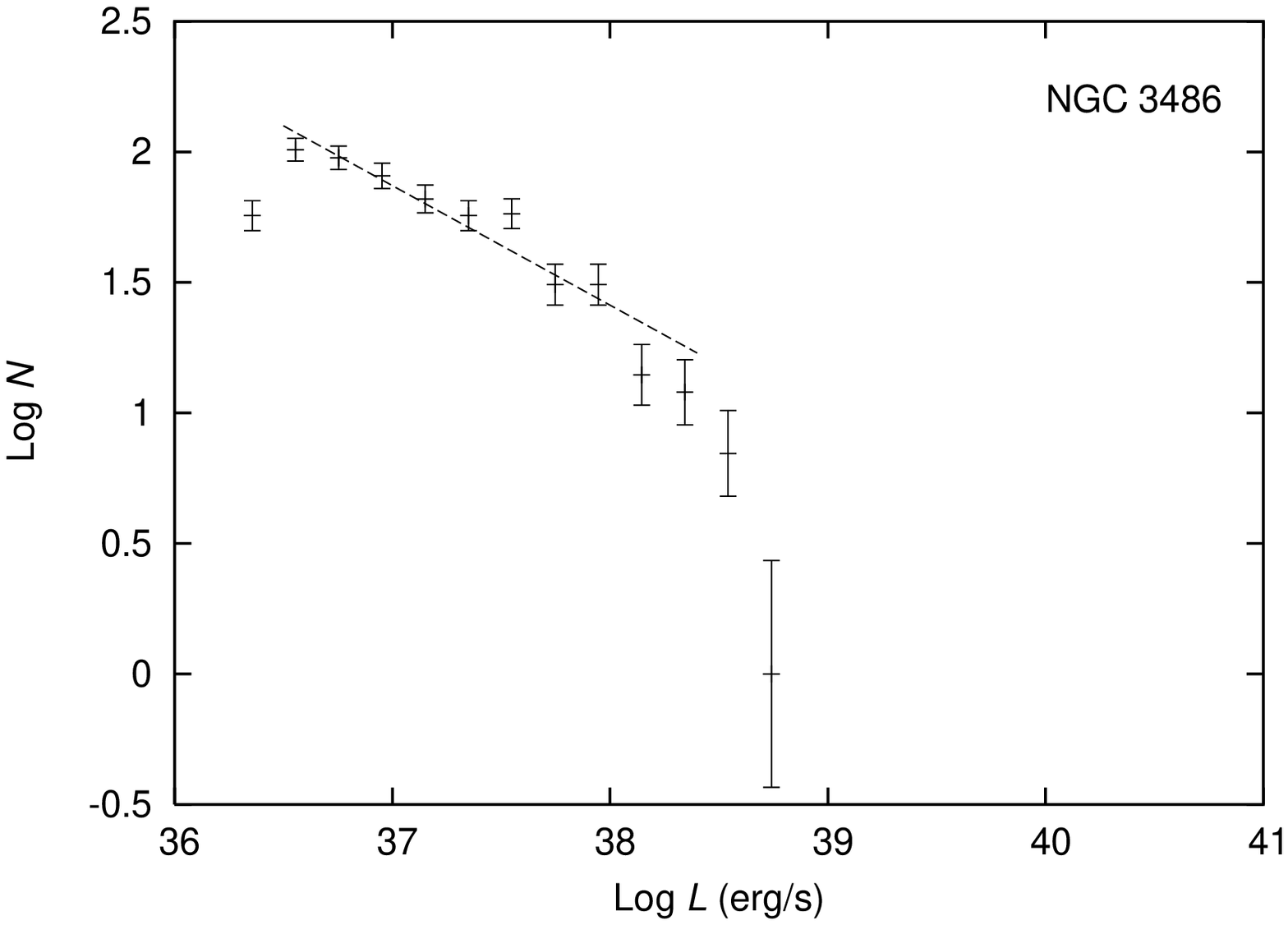,angle=-90,width=9cm}
\caption{(Continued)}
\end{figure}
\setcounter{figure}{0}
\begin{figure}
\psfig{figure=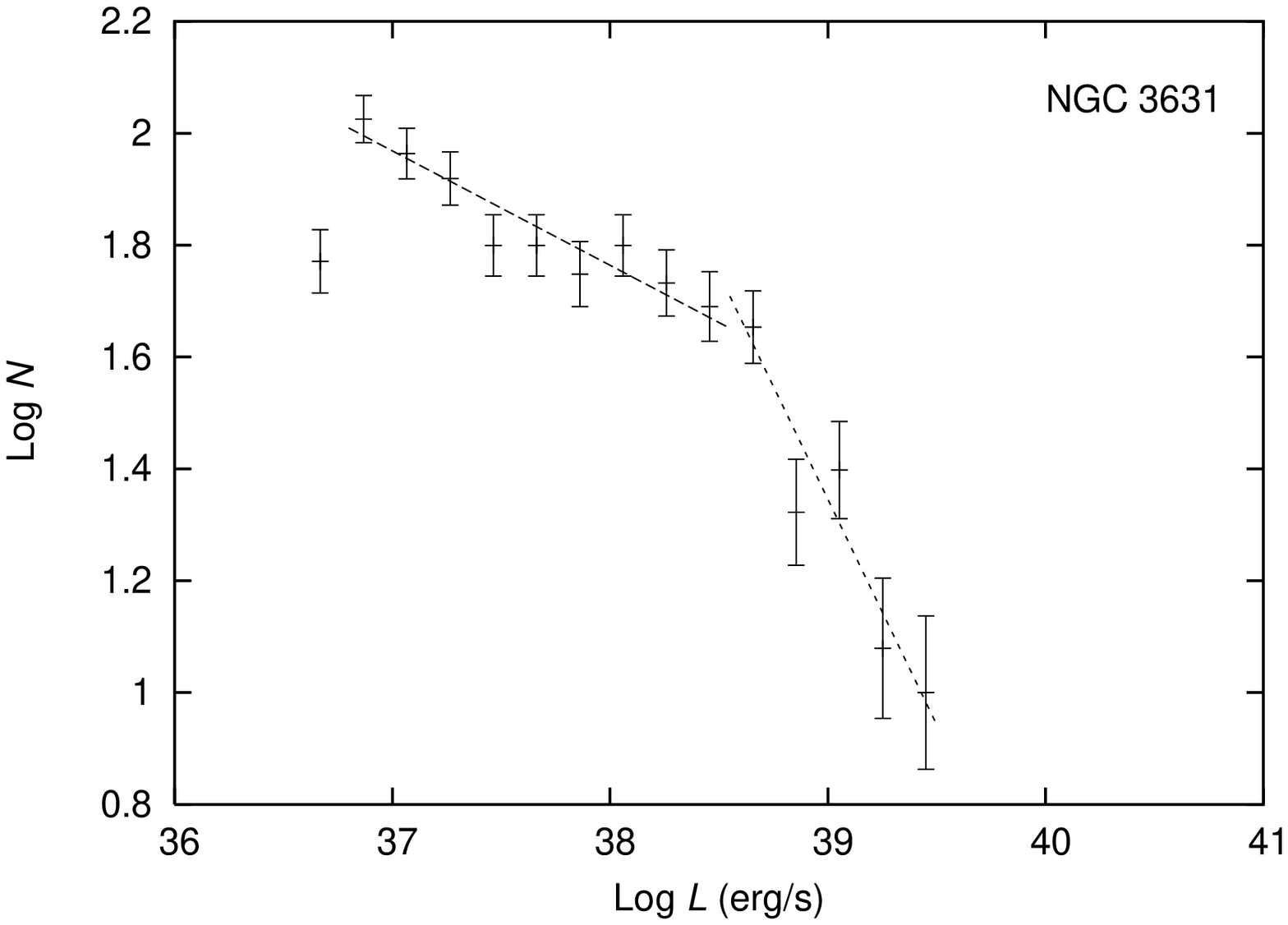,angle=-90,width=9cm}
\psfig{figure=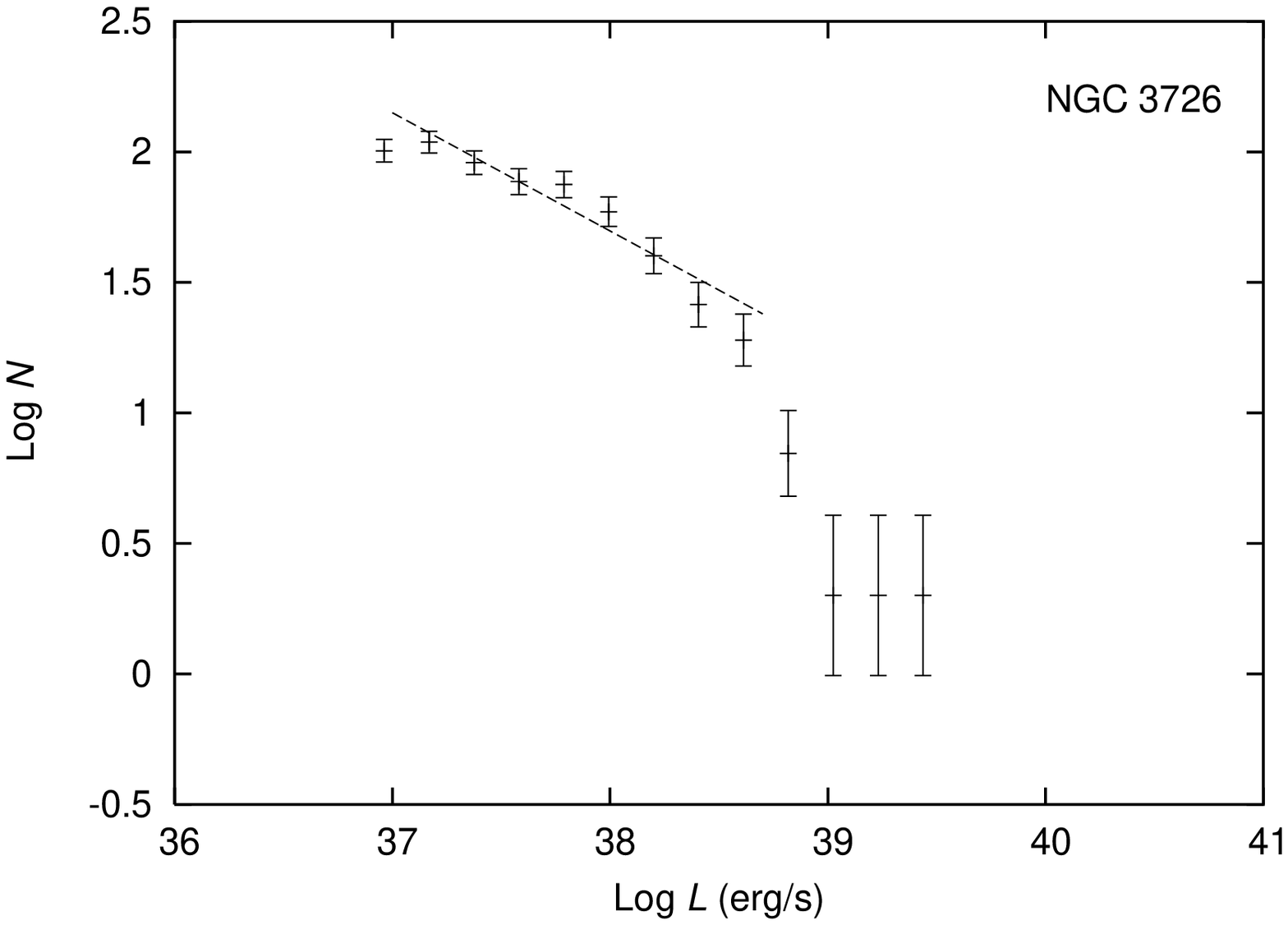,angle=-90,width=9cm}
\psfig{figure=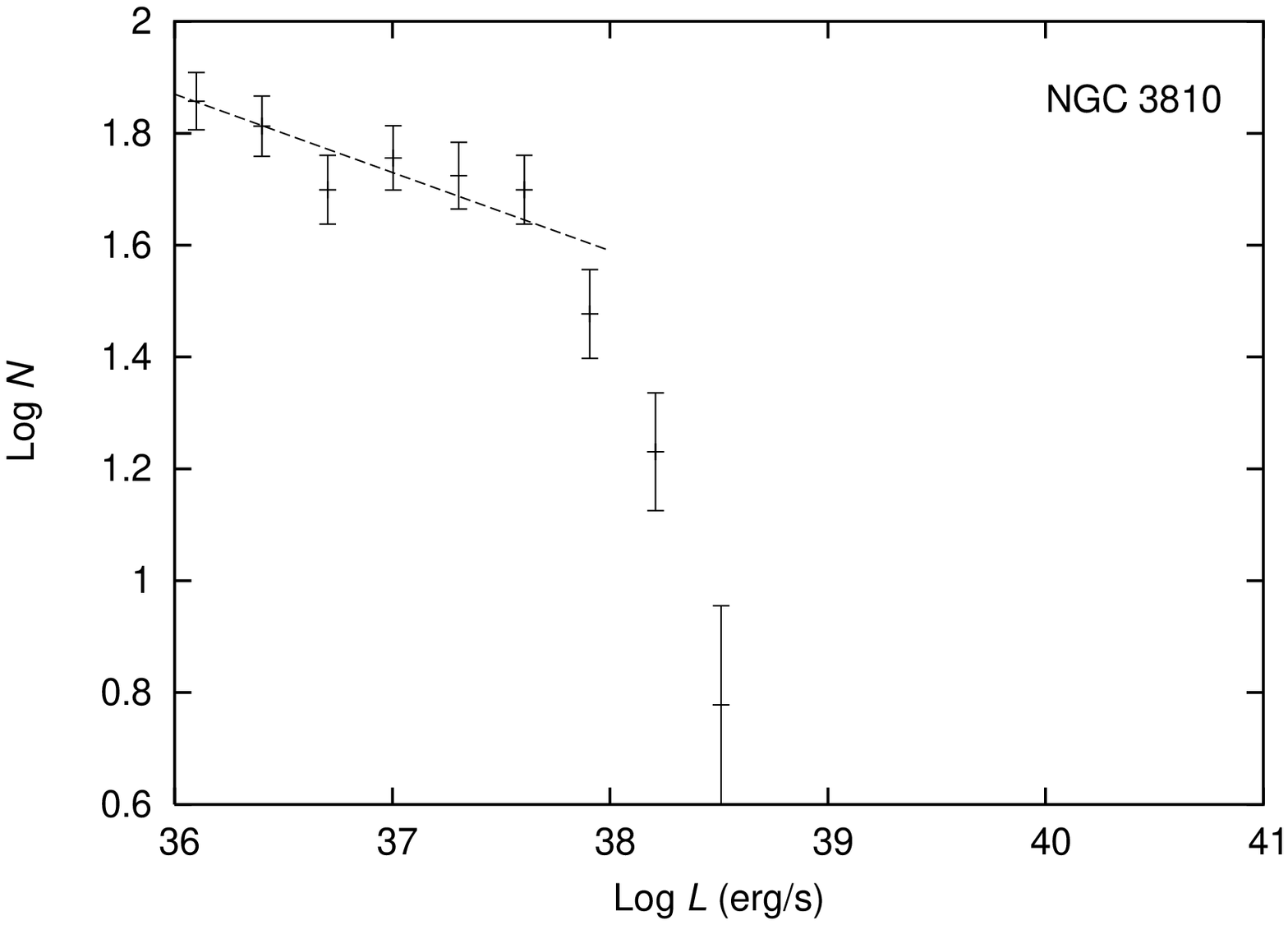,angle=-90,width=9cm}
\psfig{figure=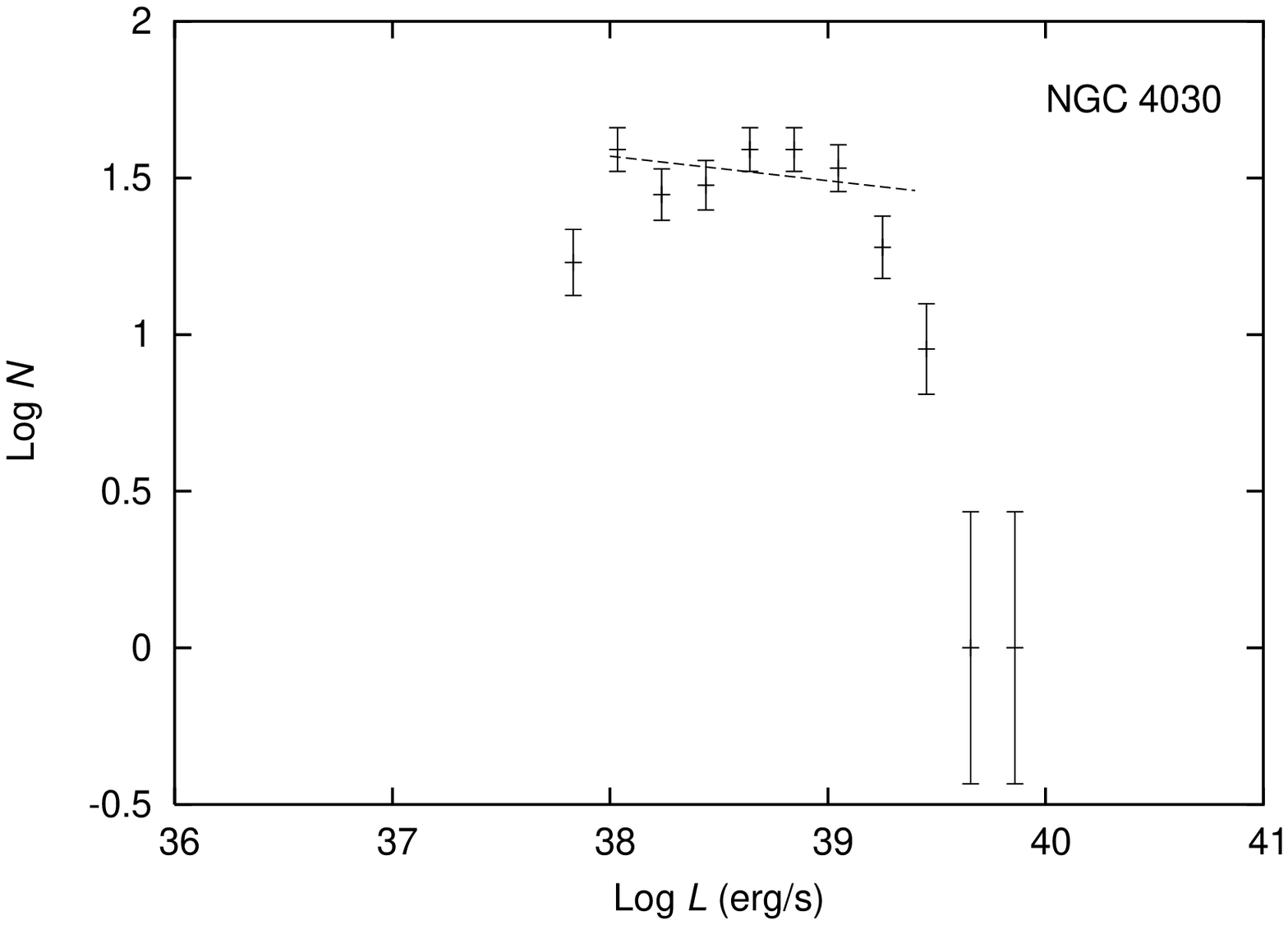,angle=-90,width=9cm}
\caption{(Continued)}
\end{figure}
\setcounter{figure}{0}
\begin{figure}
\psfig{figure=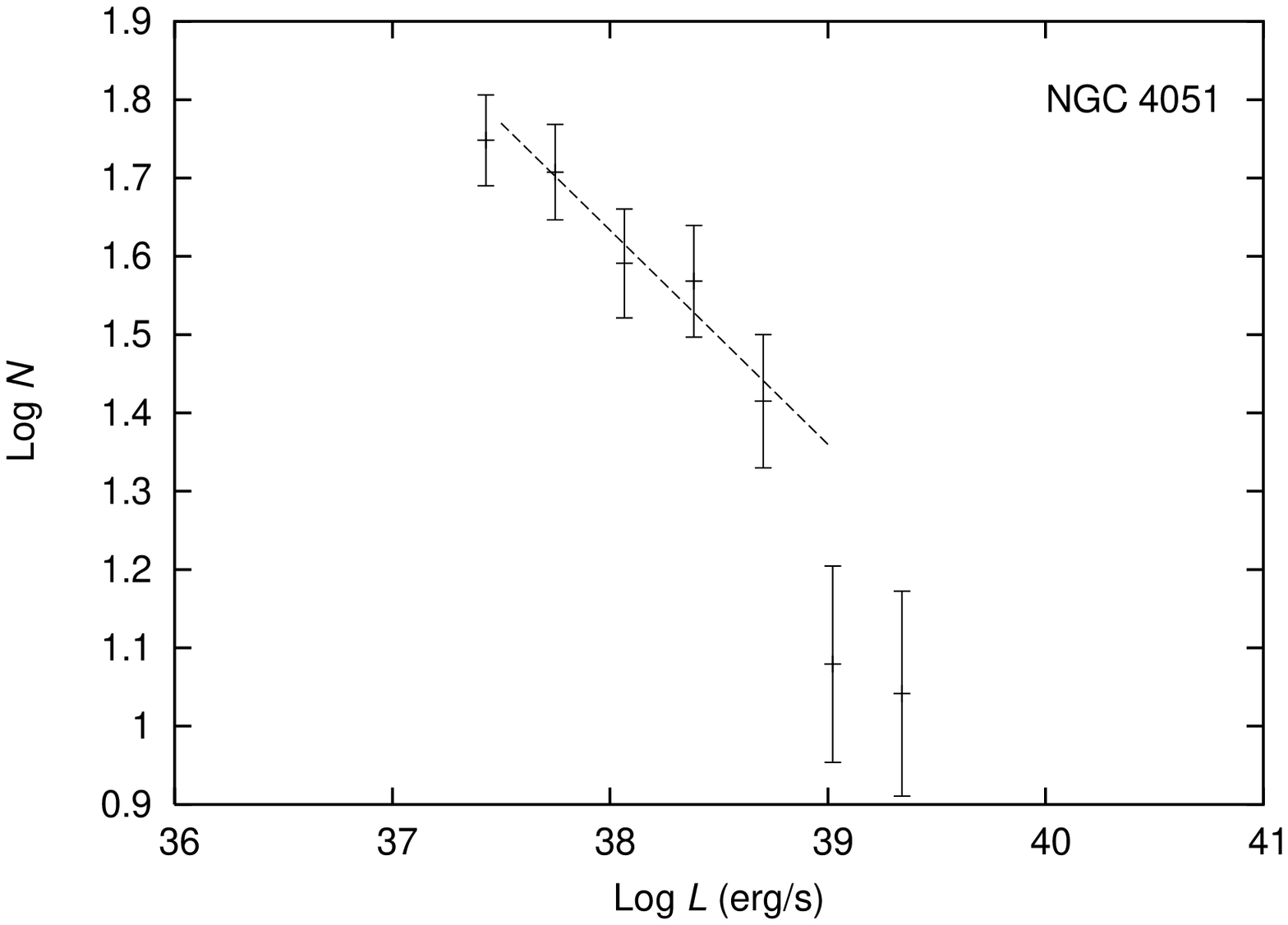,angle=-90,width=9cm}
\psfig{figure=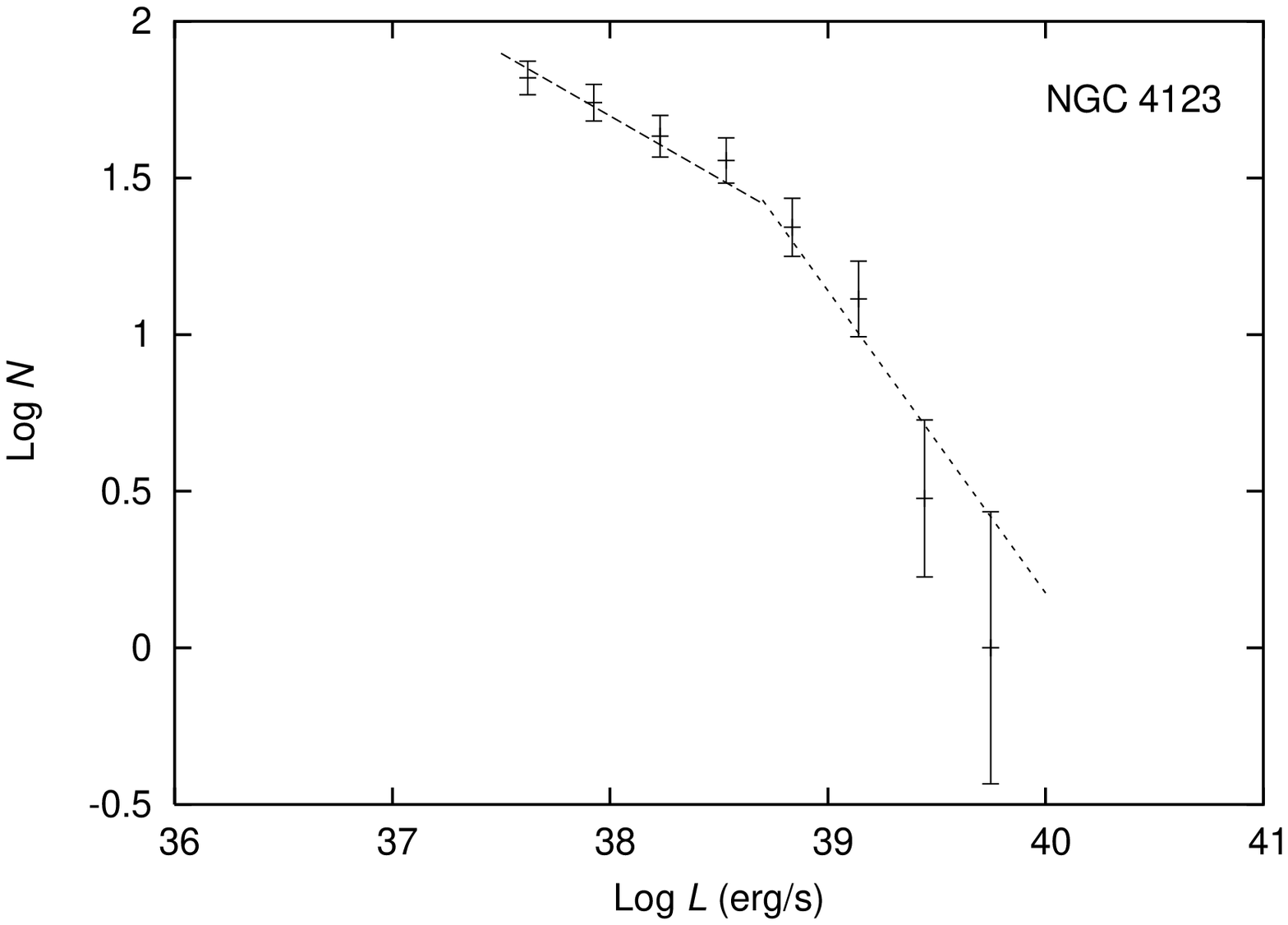,angle=-90,width=9cm}
\psfig{figure=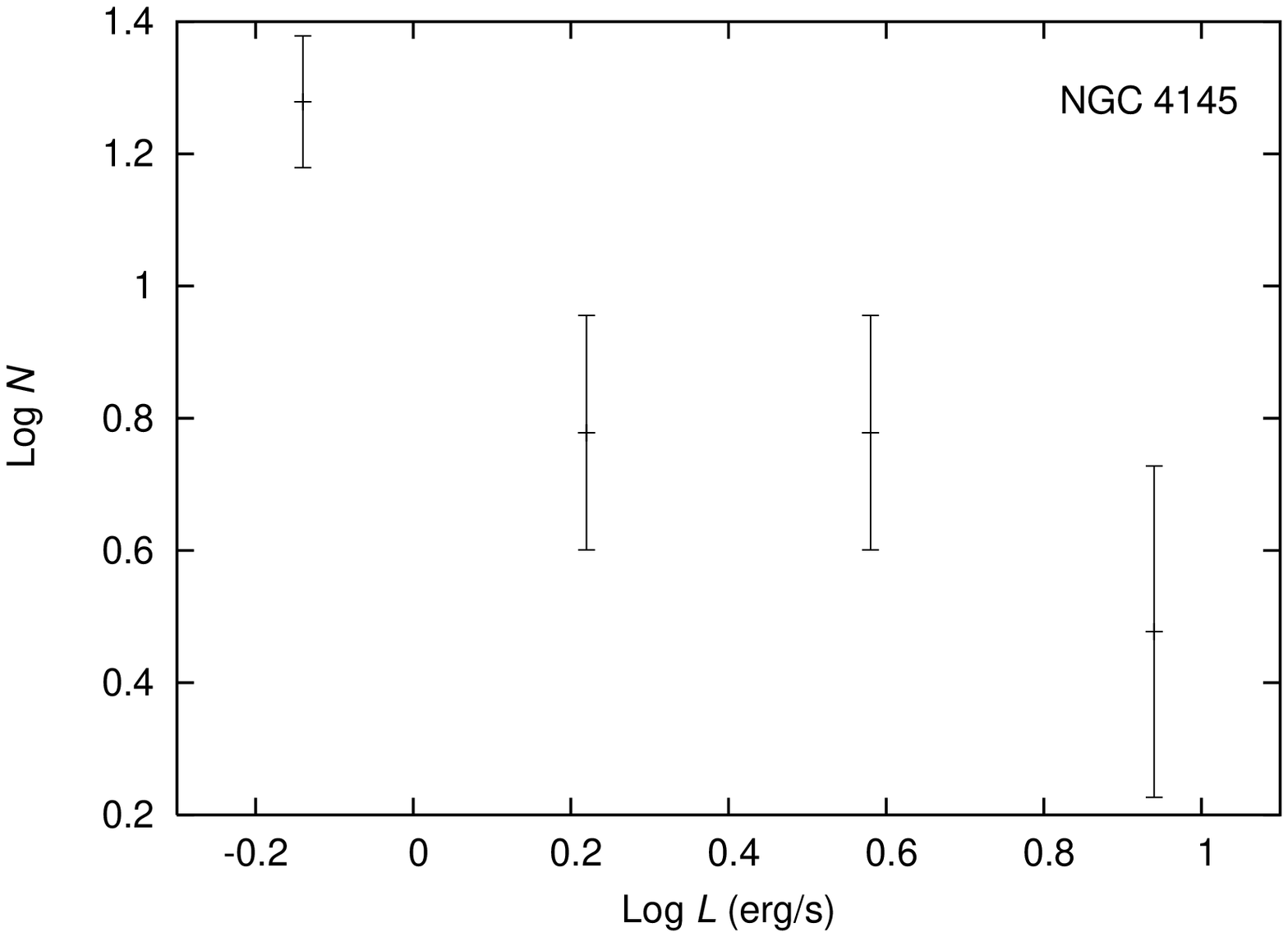,angle=-90,width=9cm}
\psfig{figure=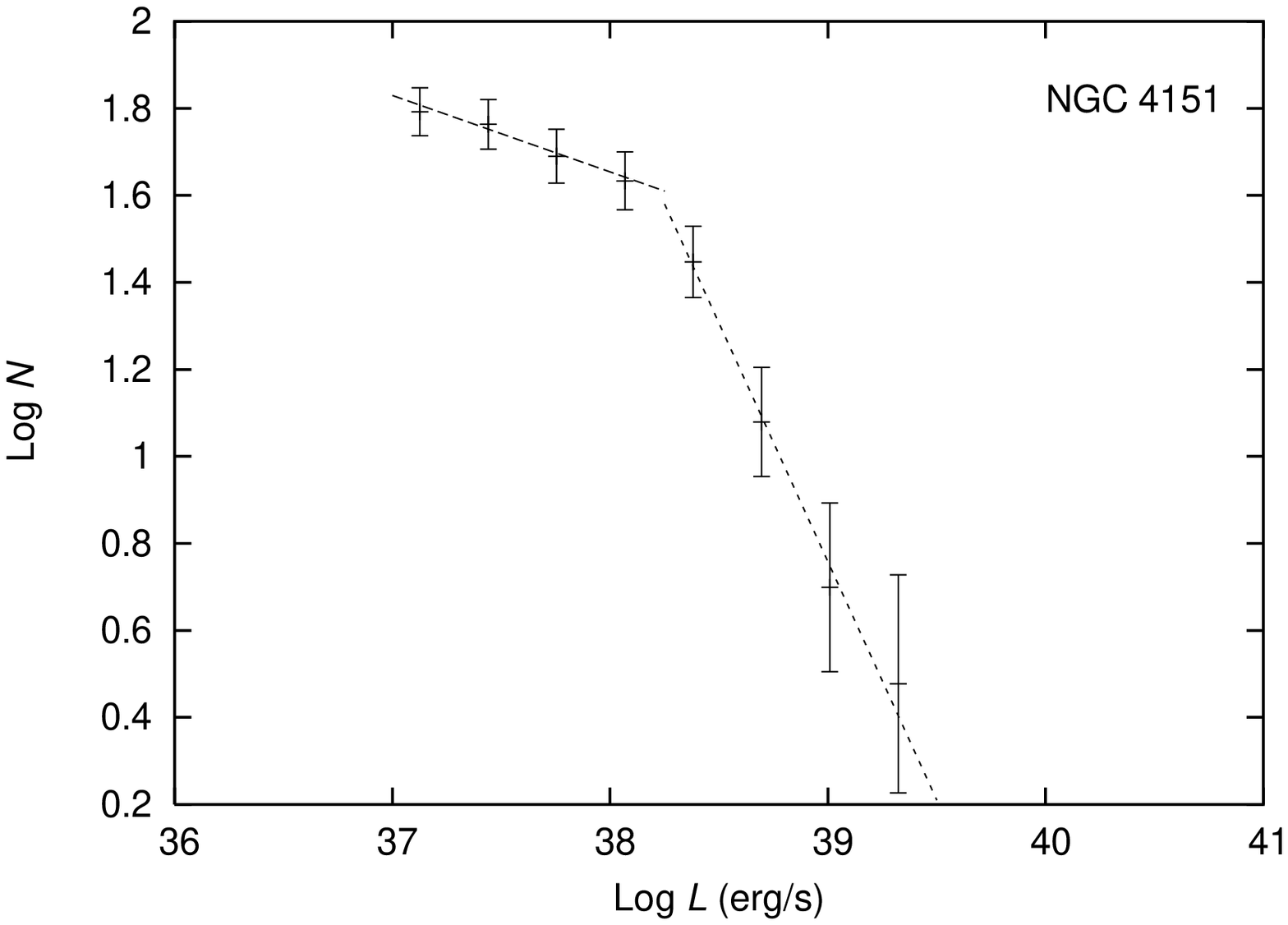,angle=-90,width=9cm}
\caption{(Continued)}
\end{figure}
\setcounter{figure}{0}
\begin{figure}
\psfig{figure=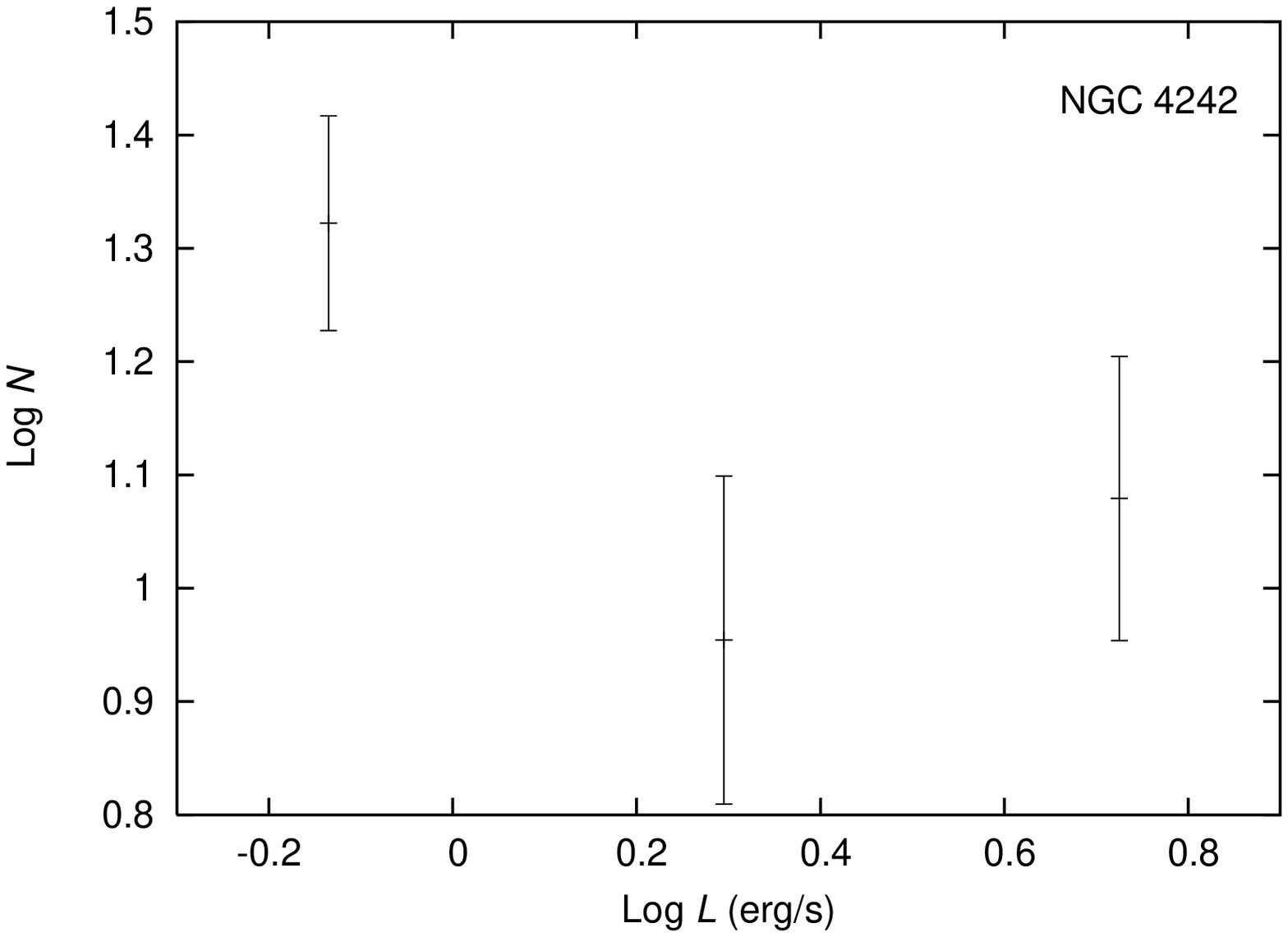,angle=-90,width=9cm}
\psfig{figure=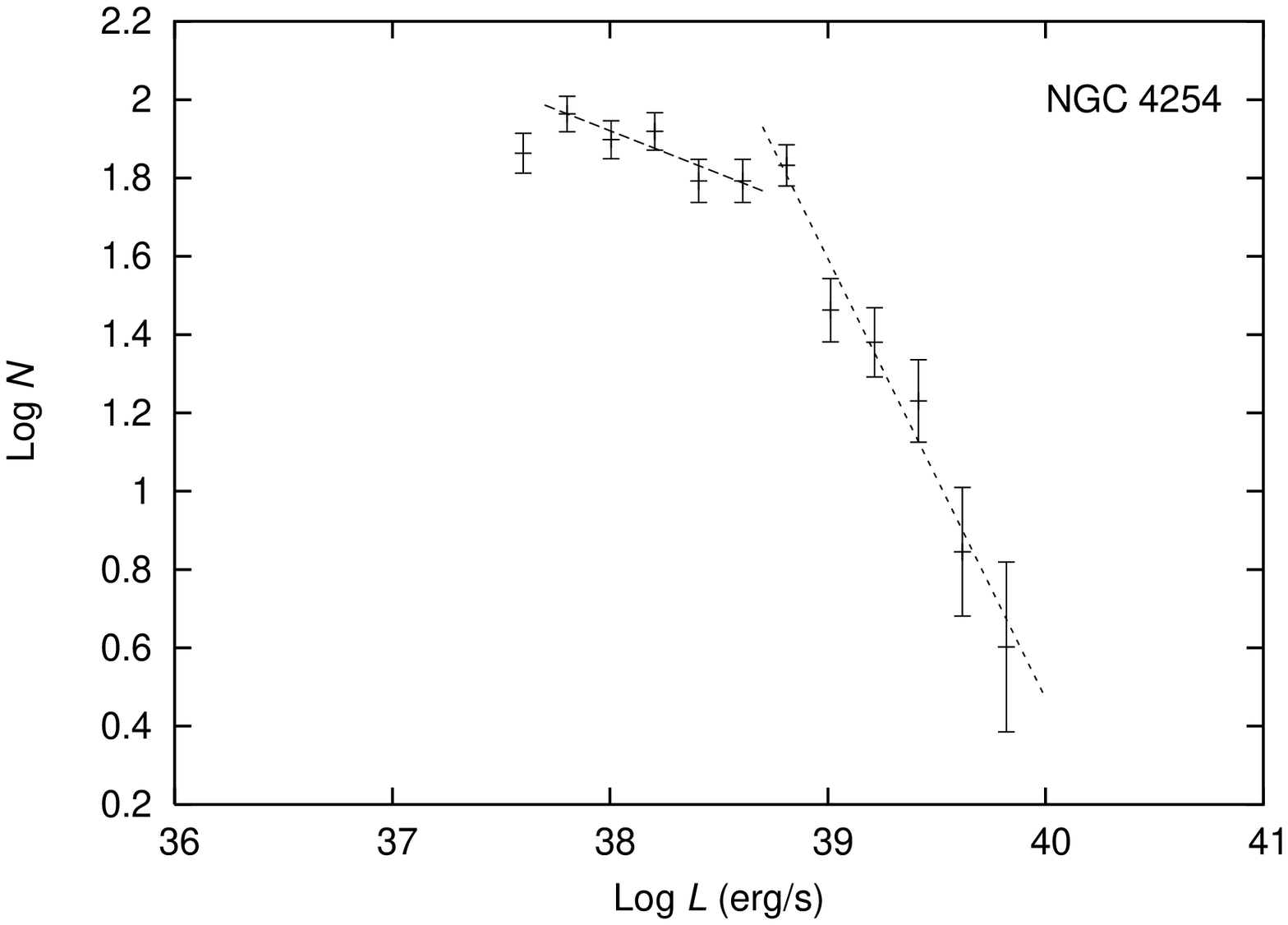,angle=-90,width=9cm}
\psfig{figure=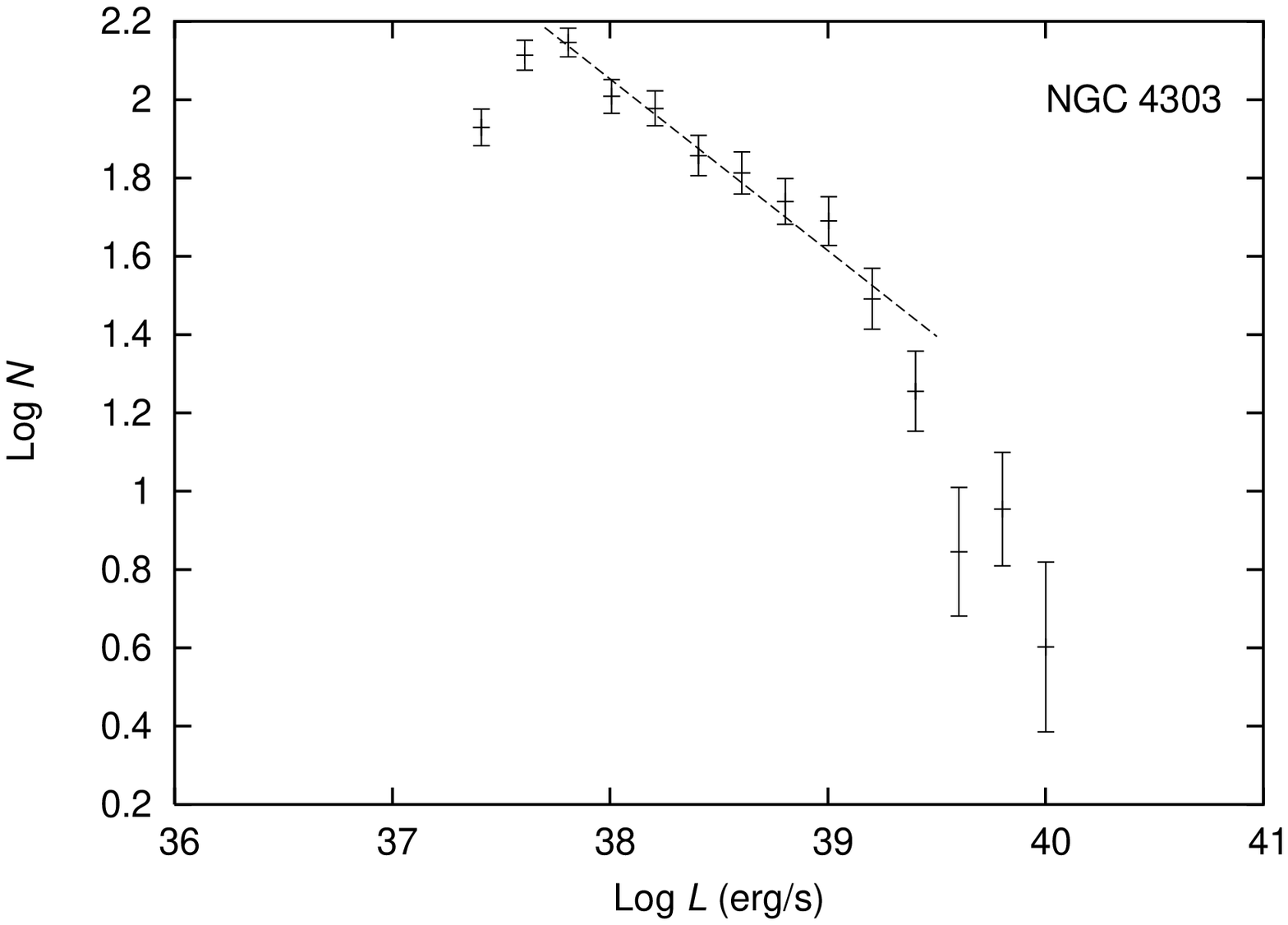,angle=-90,width=9cm}
\psfig{figure=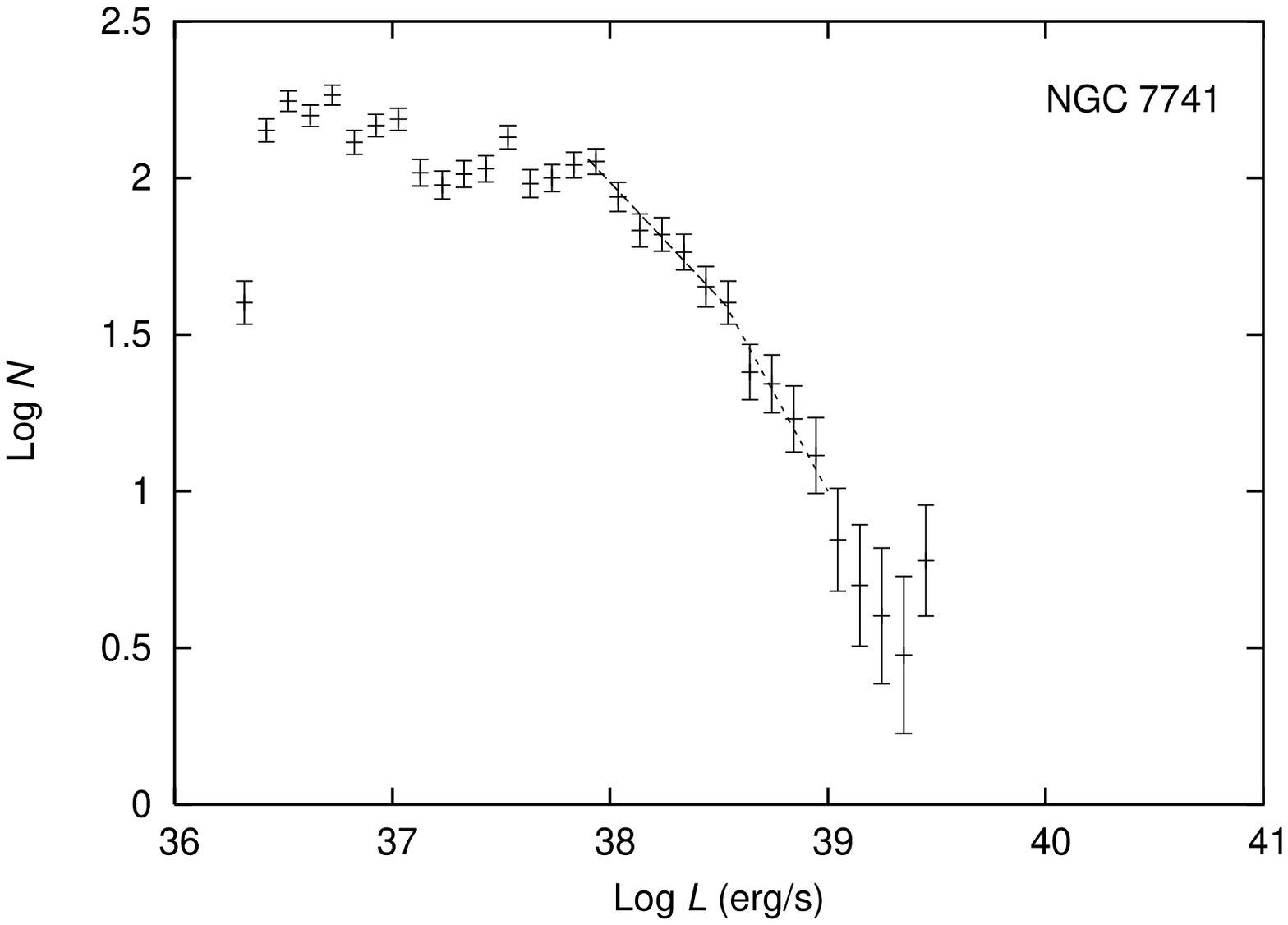,angle=-90,width=9cm}
\caption{(Continued)}
\end{figure}
\setcounter{figure}{0}
\begin{figure}
\psfig{figure=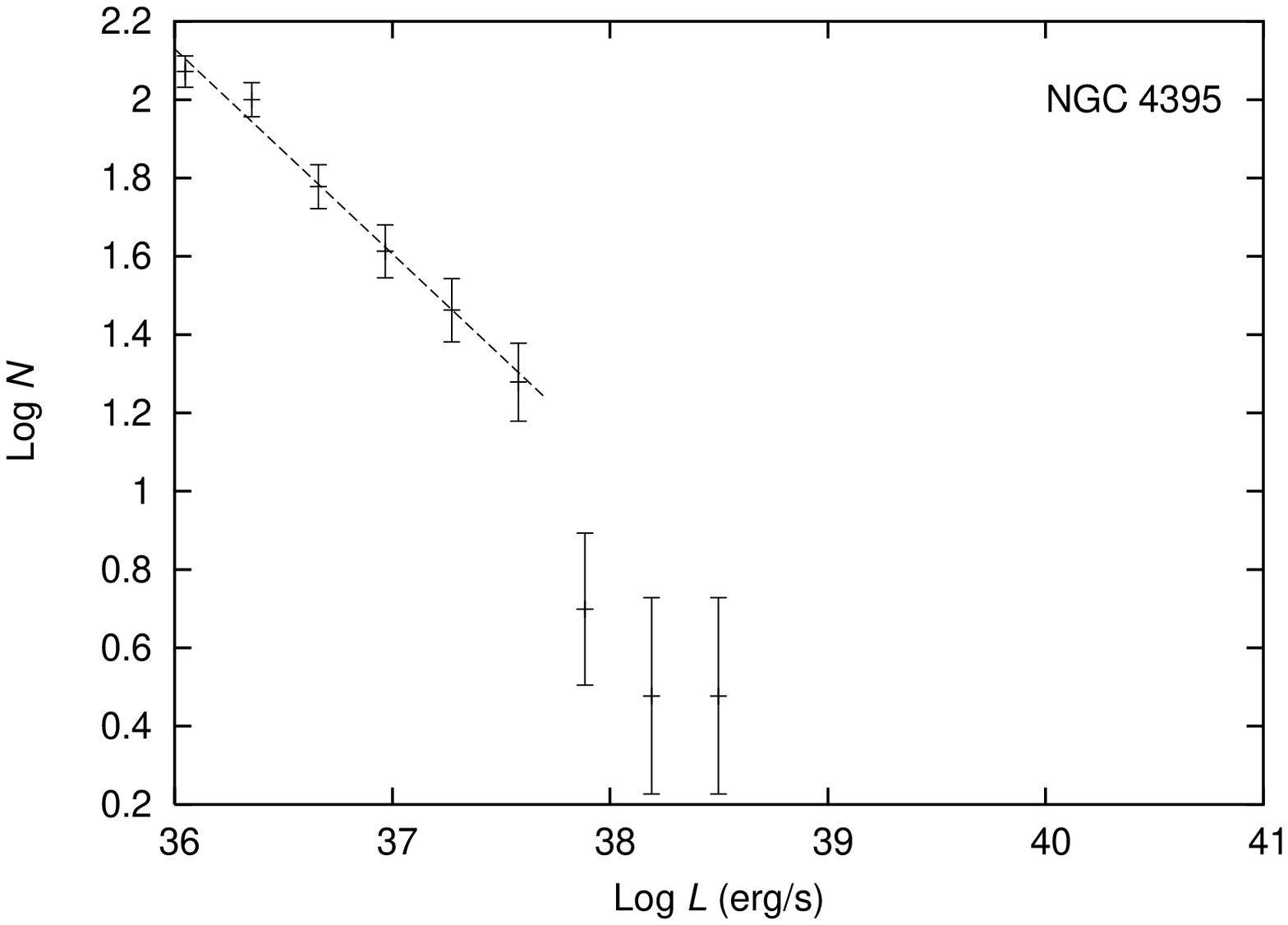,angle=-90,width=9cm}
\psfig{figure=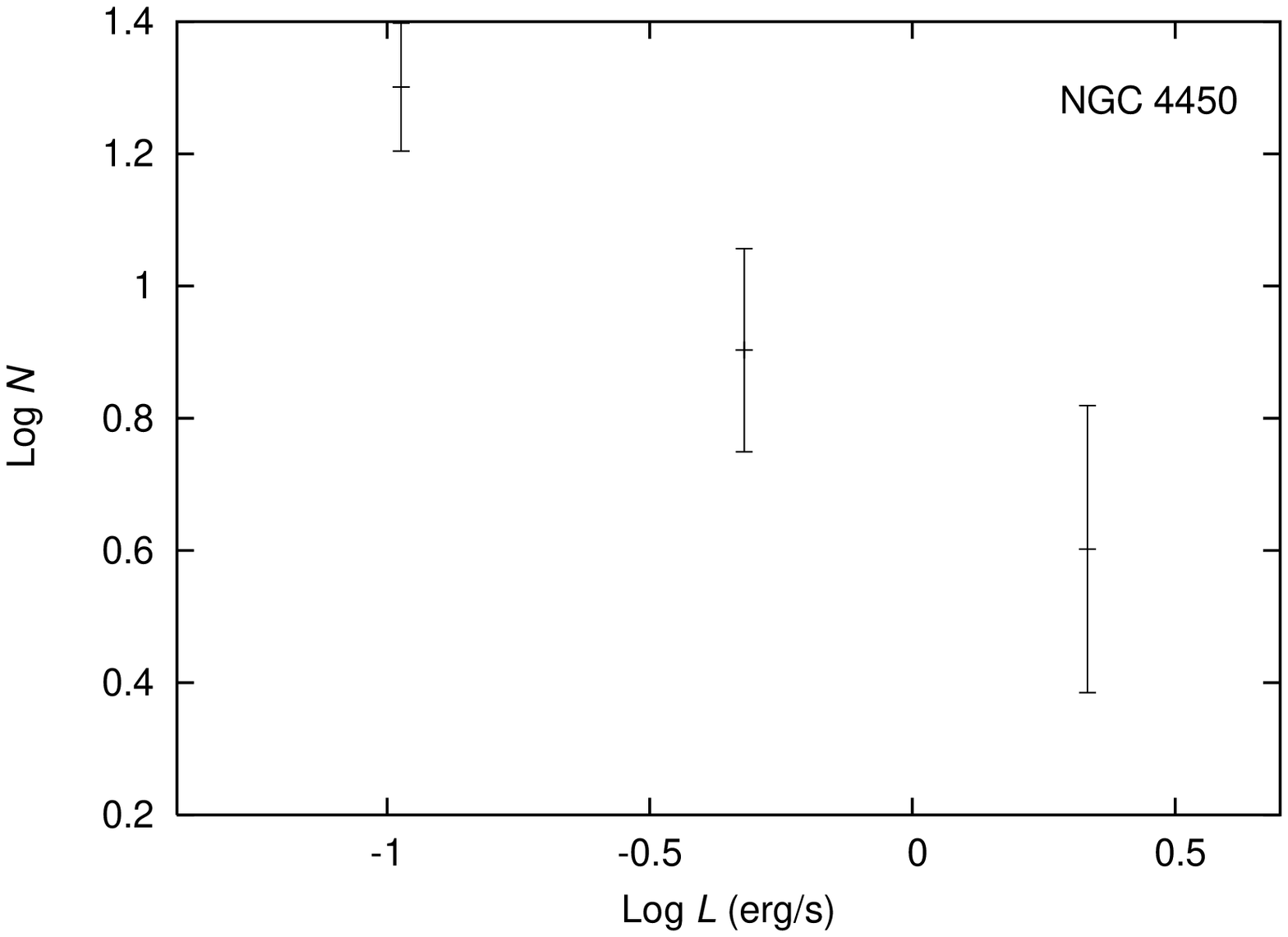,angle=-90,width=9cm}
\psfig{figure=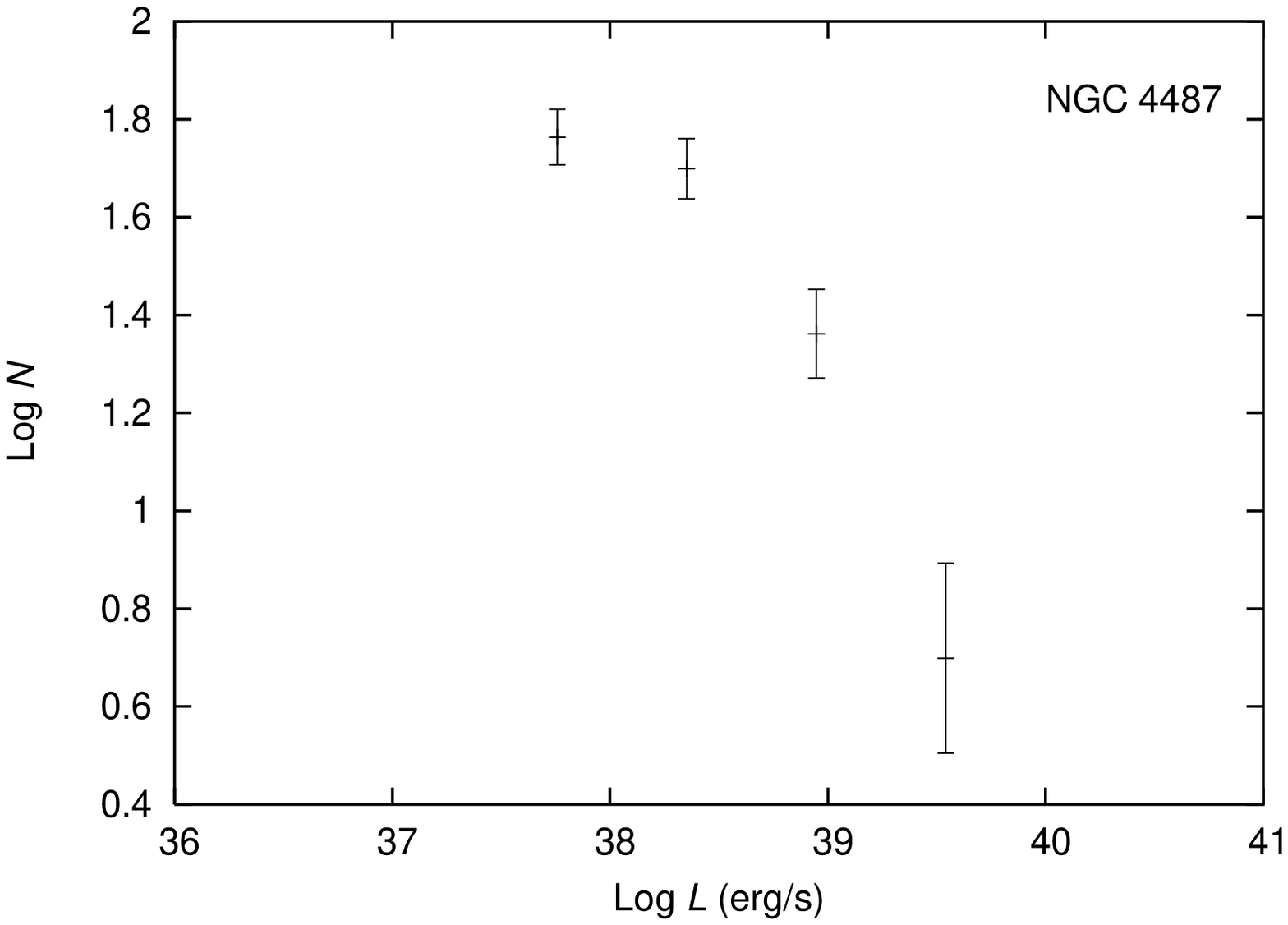,angle=-90,width=9cm}
\psfig{figure=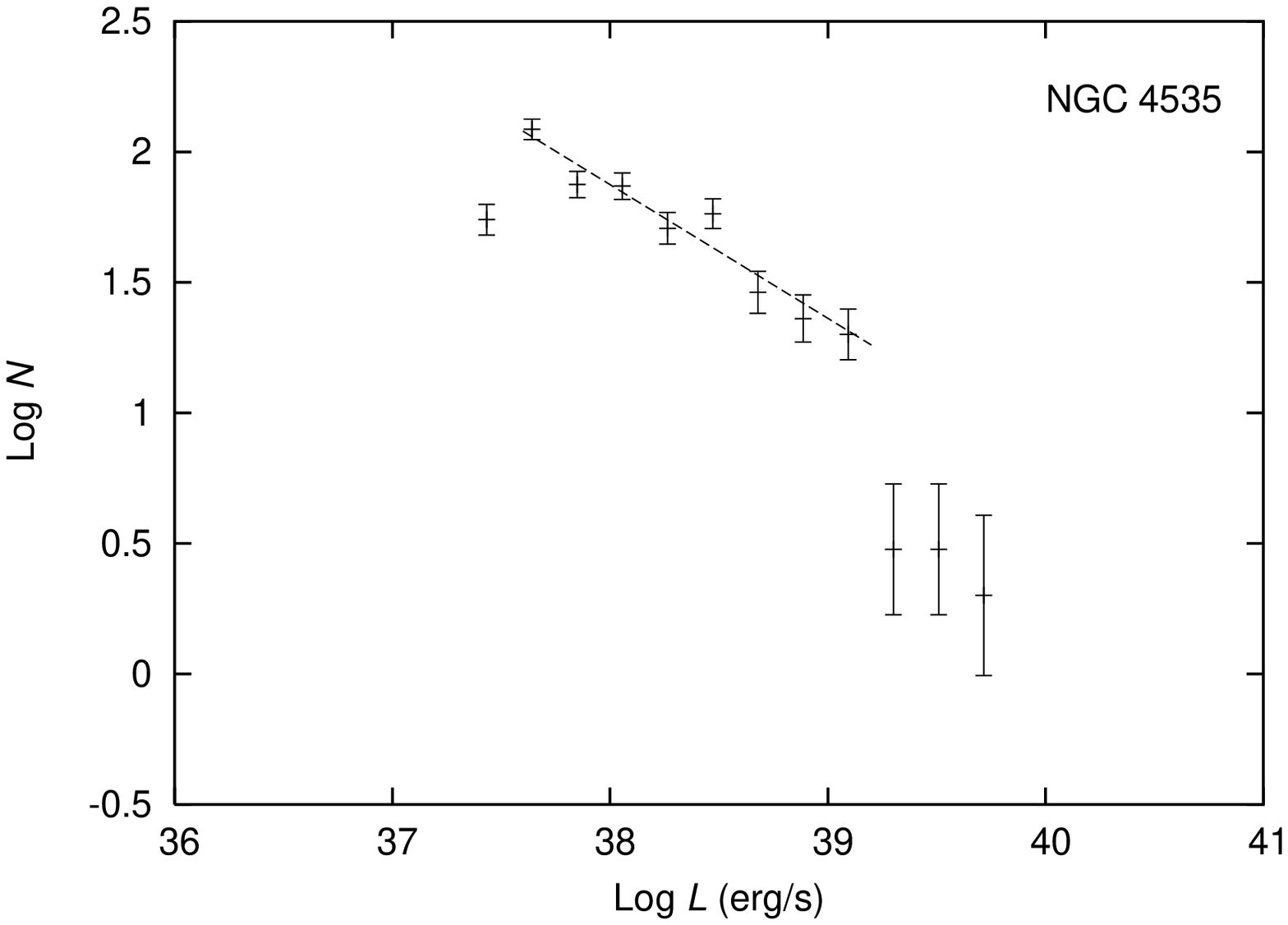,angle=-90,width=9cm}
\caption{(Continued)}
\end{figure}
\setcounter{figure}{0}
\begin{figure}
\psfig{figure=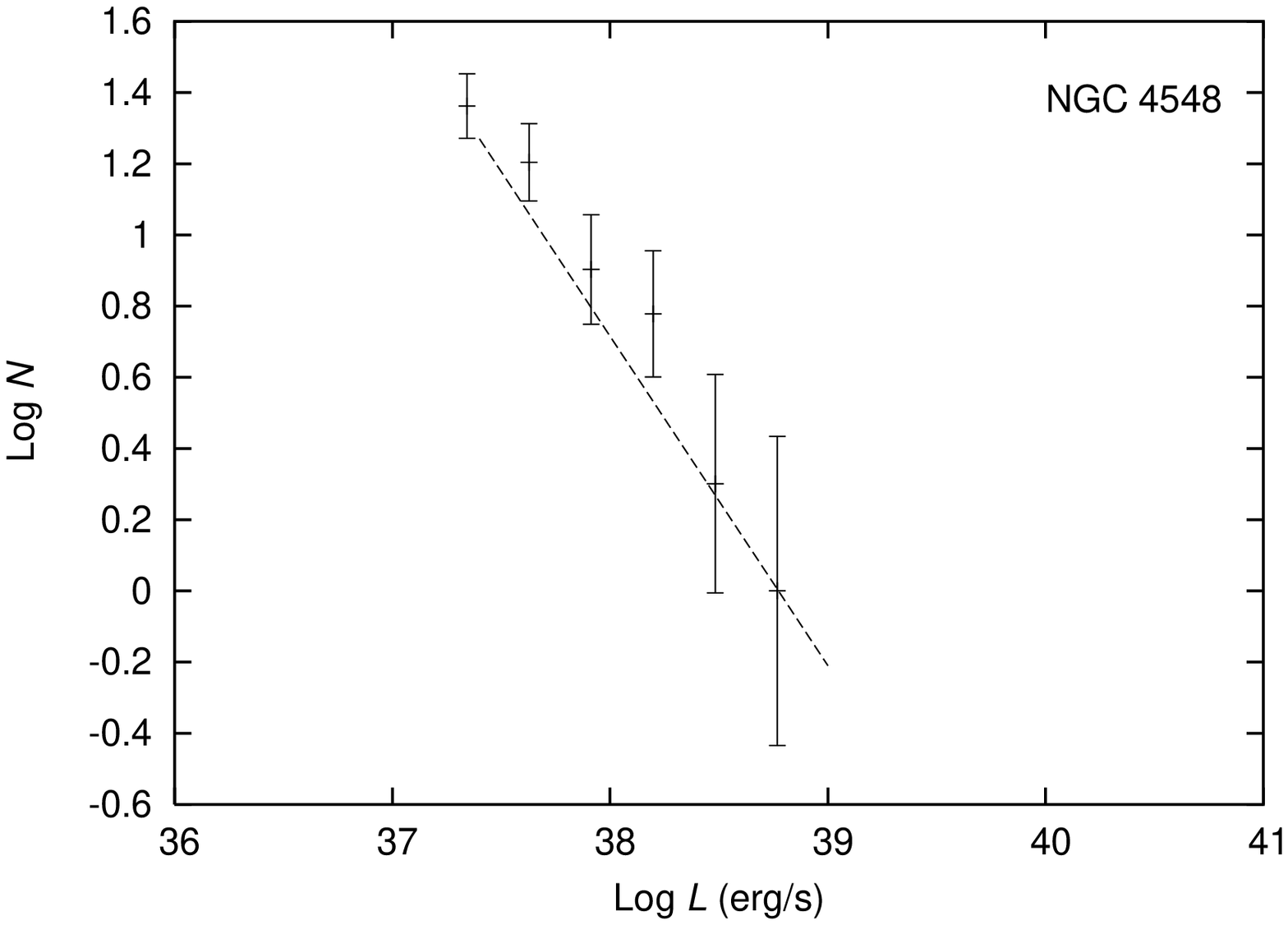,angle=-90,width=9cm}
\psfig{figure=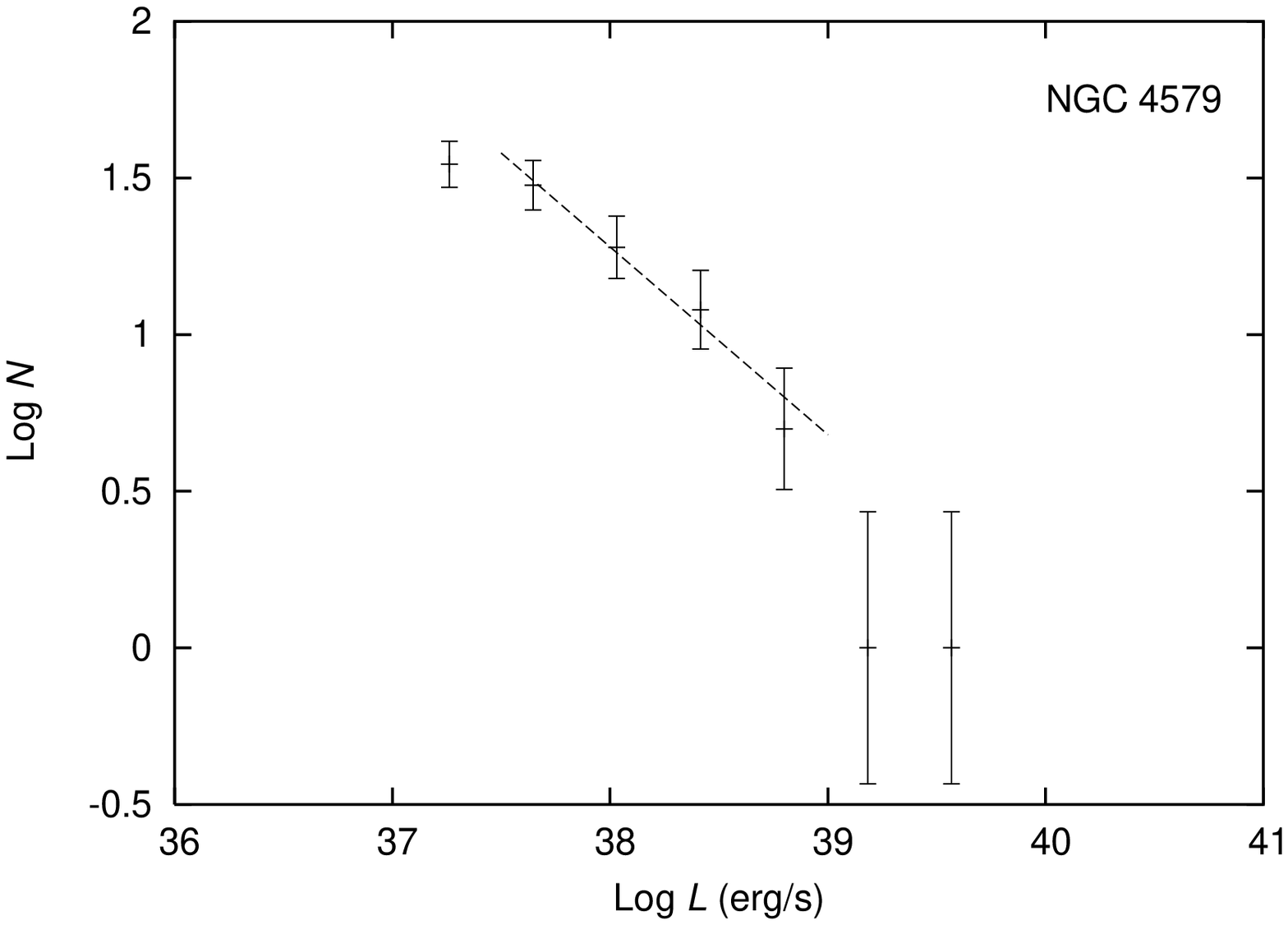,angle=-90,width=9cm}
\psfig{figure=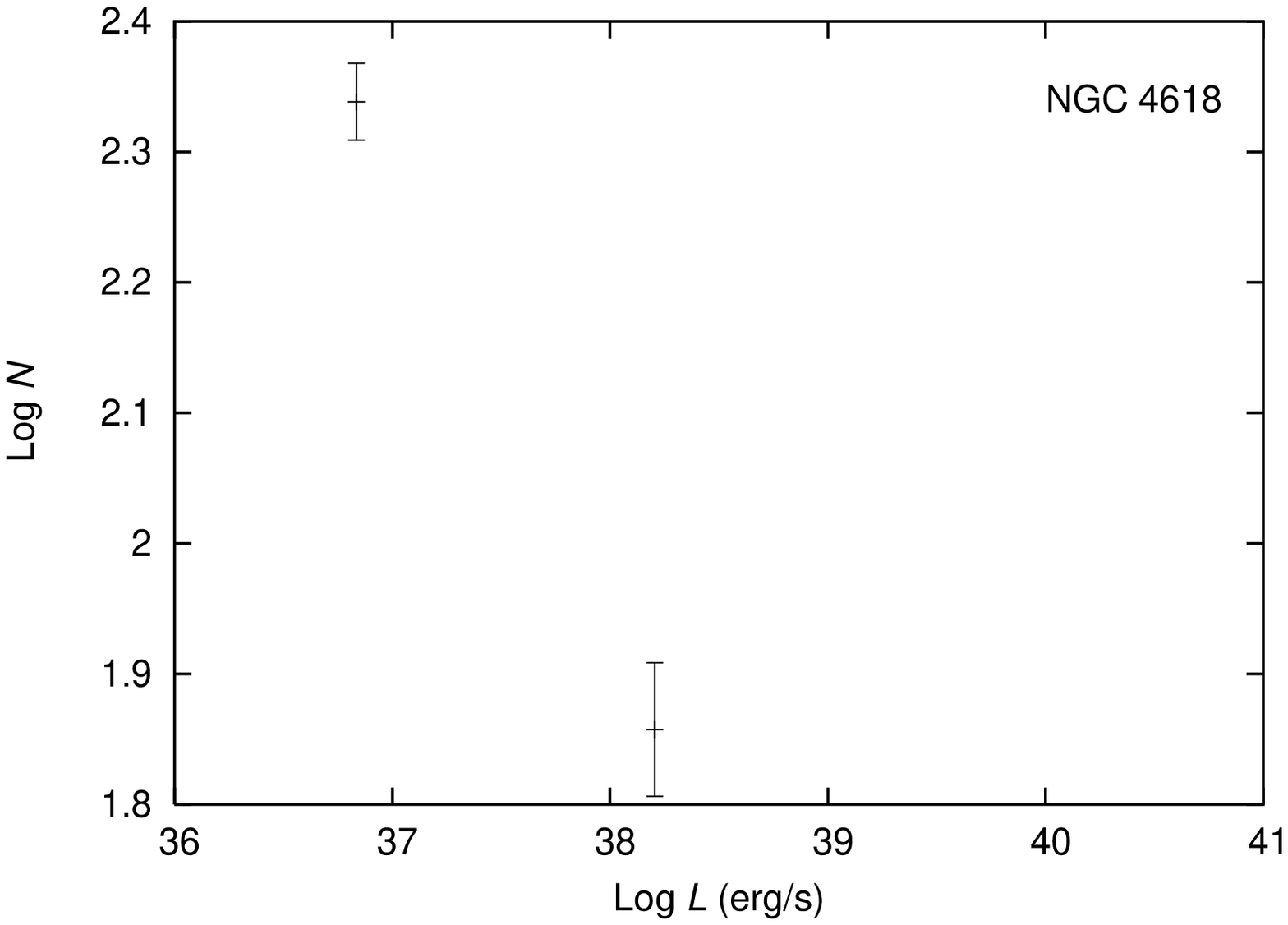,angle=-90,width=9cm}
\psfig{figure=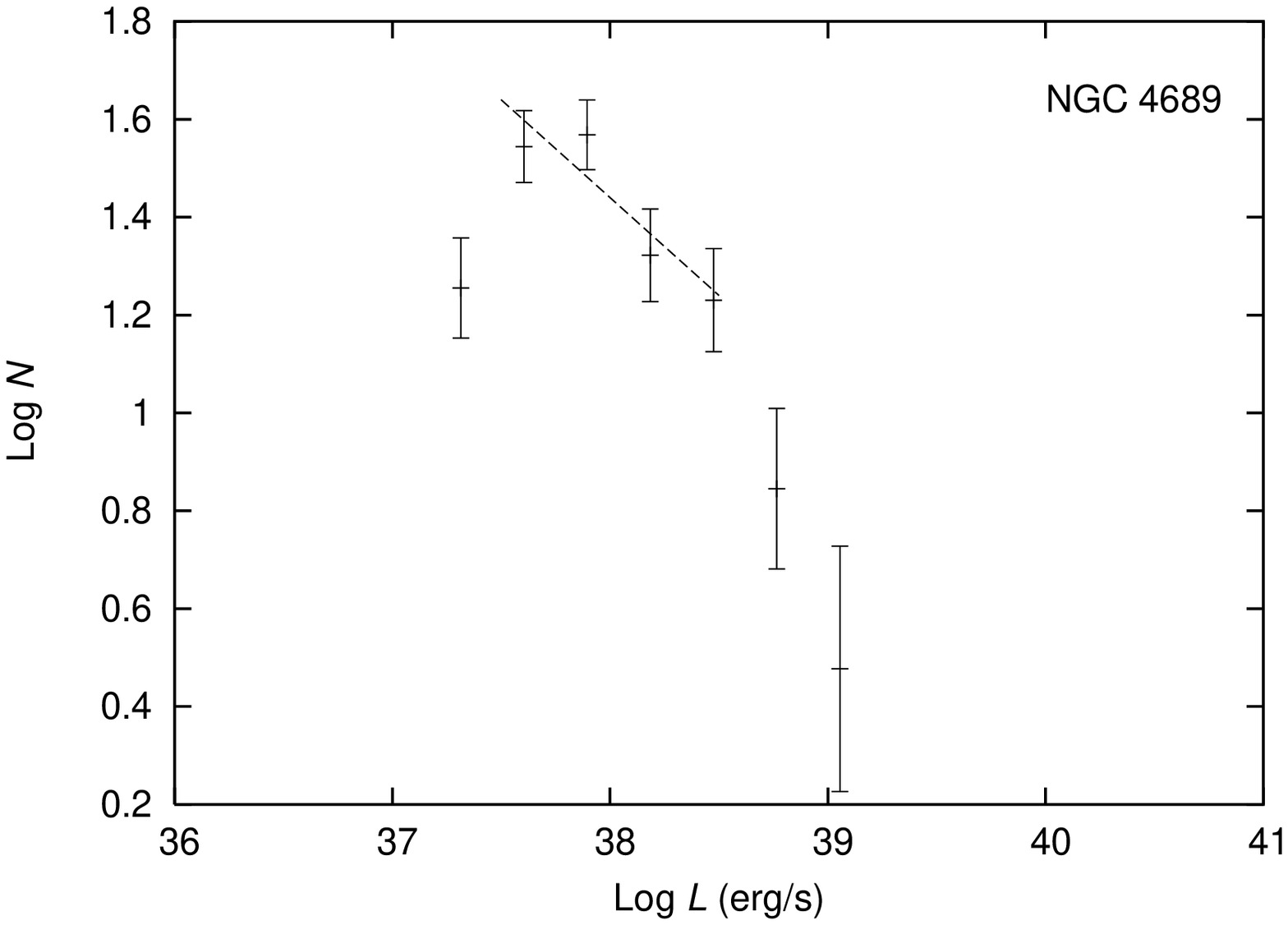,angle=-90,width=9cm}
\caption{(Continued)}
\end{figure}
\setcounter{figure}{0}
\begin{figure}
\psfig{figure=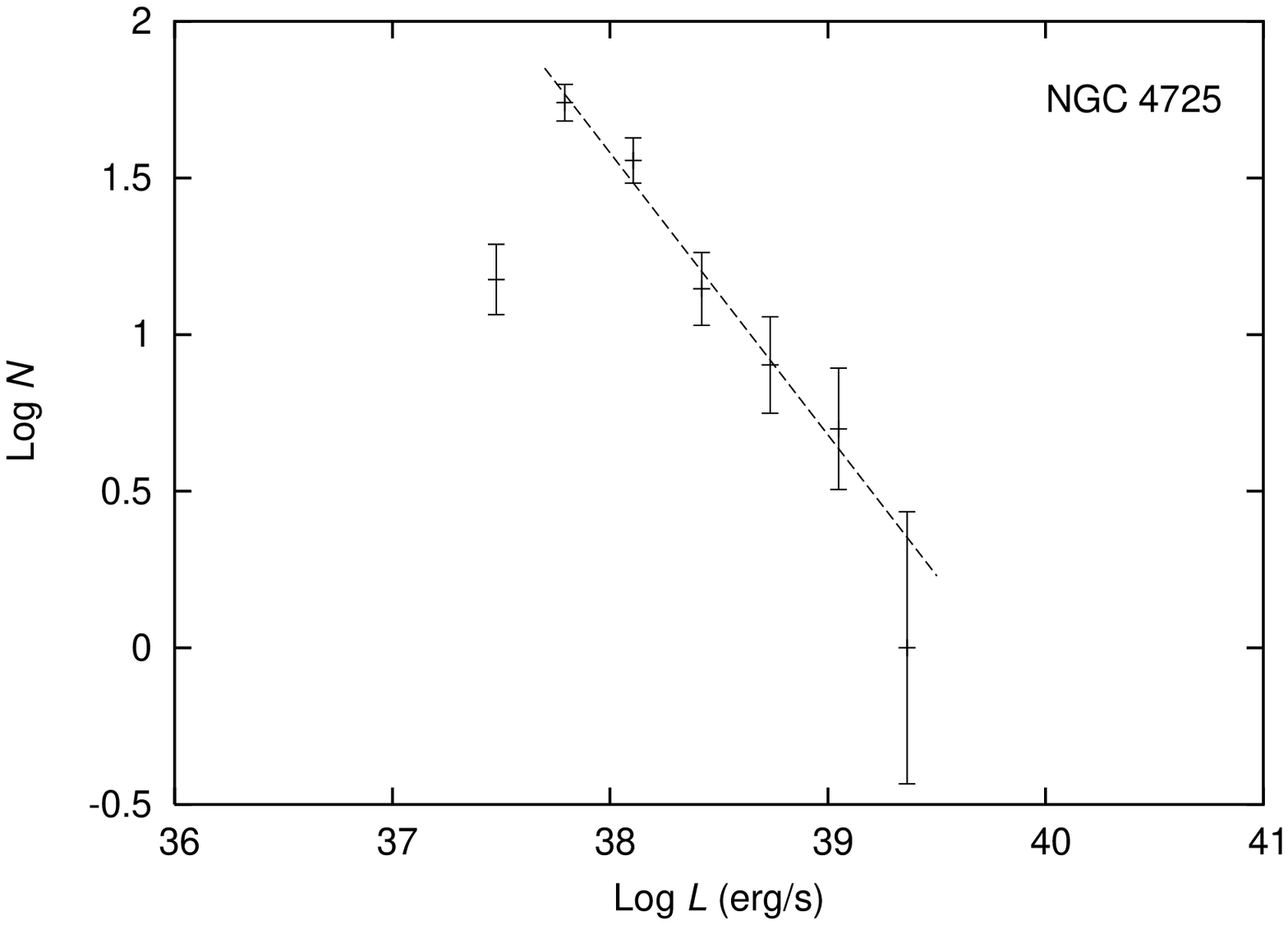,angle=-90,width=9cm}
\psfig{figure=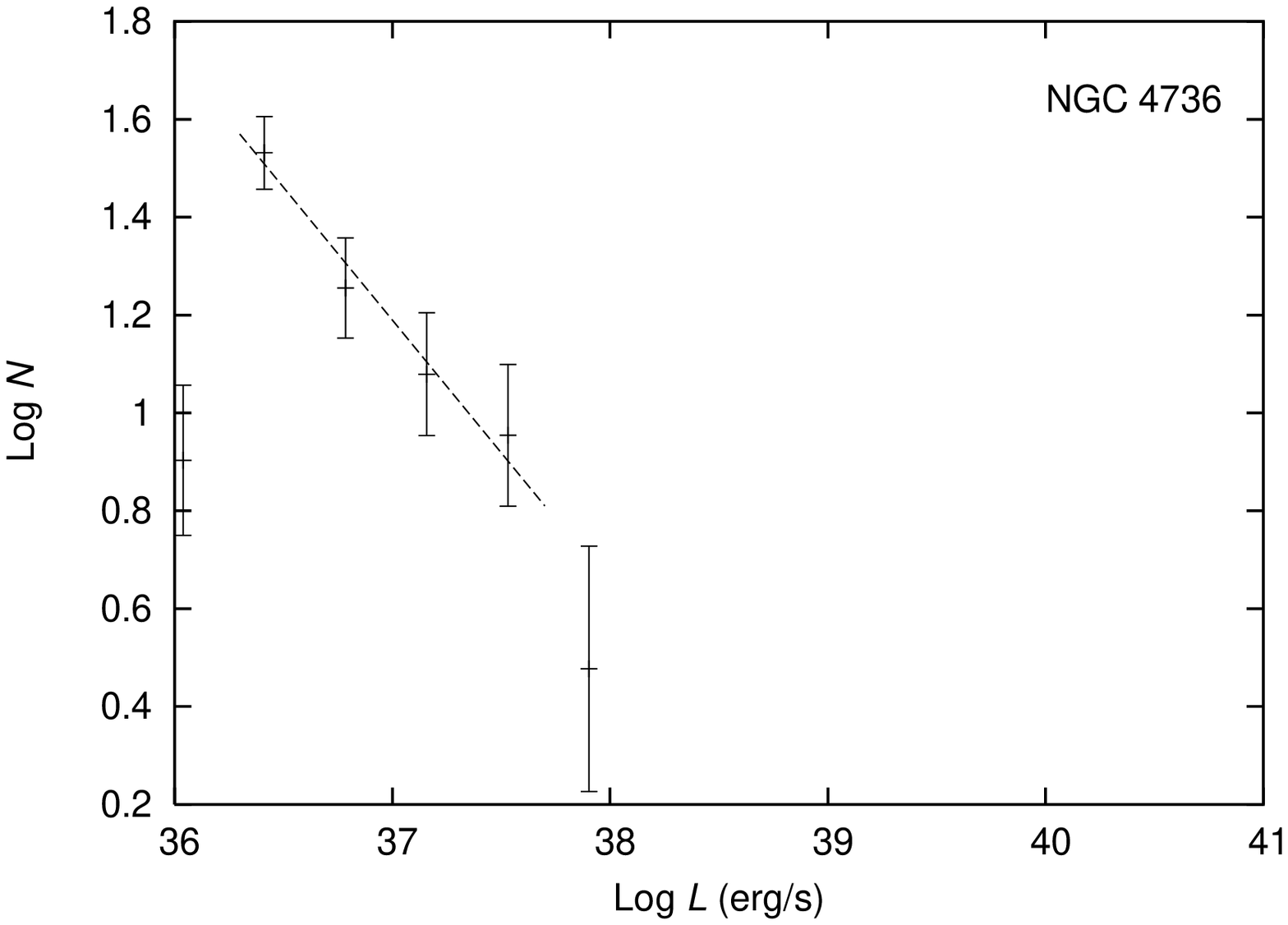,angle=-90,width=9cm}
\psfig{figure=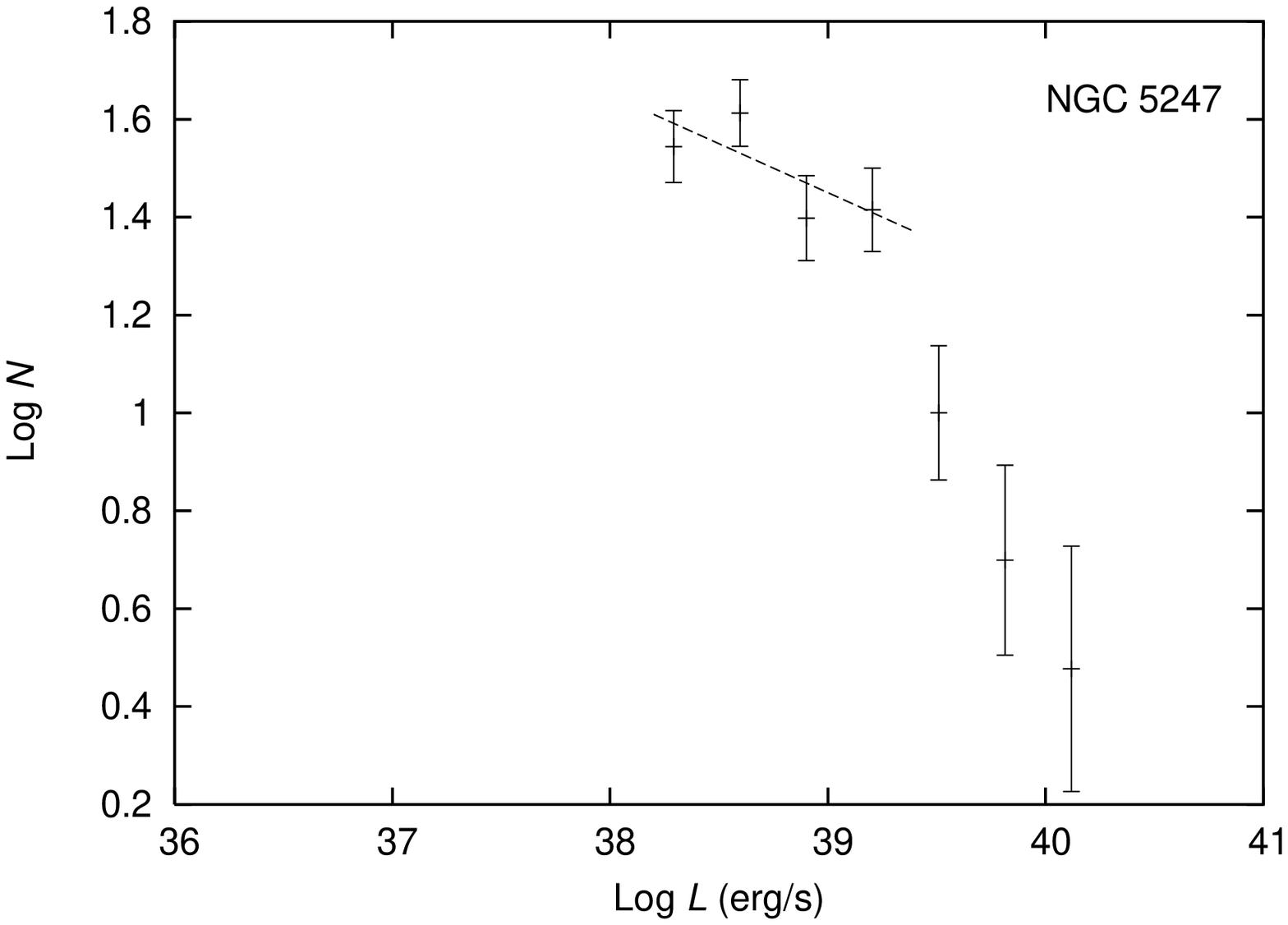,angle=-90,width=9cm}
\psfig{figure=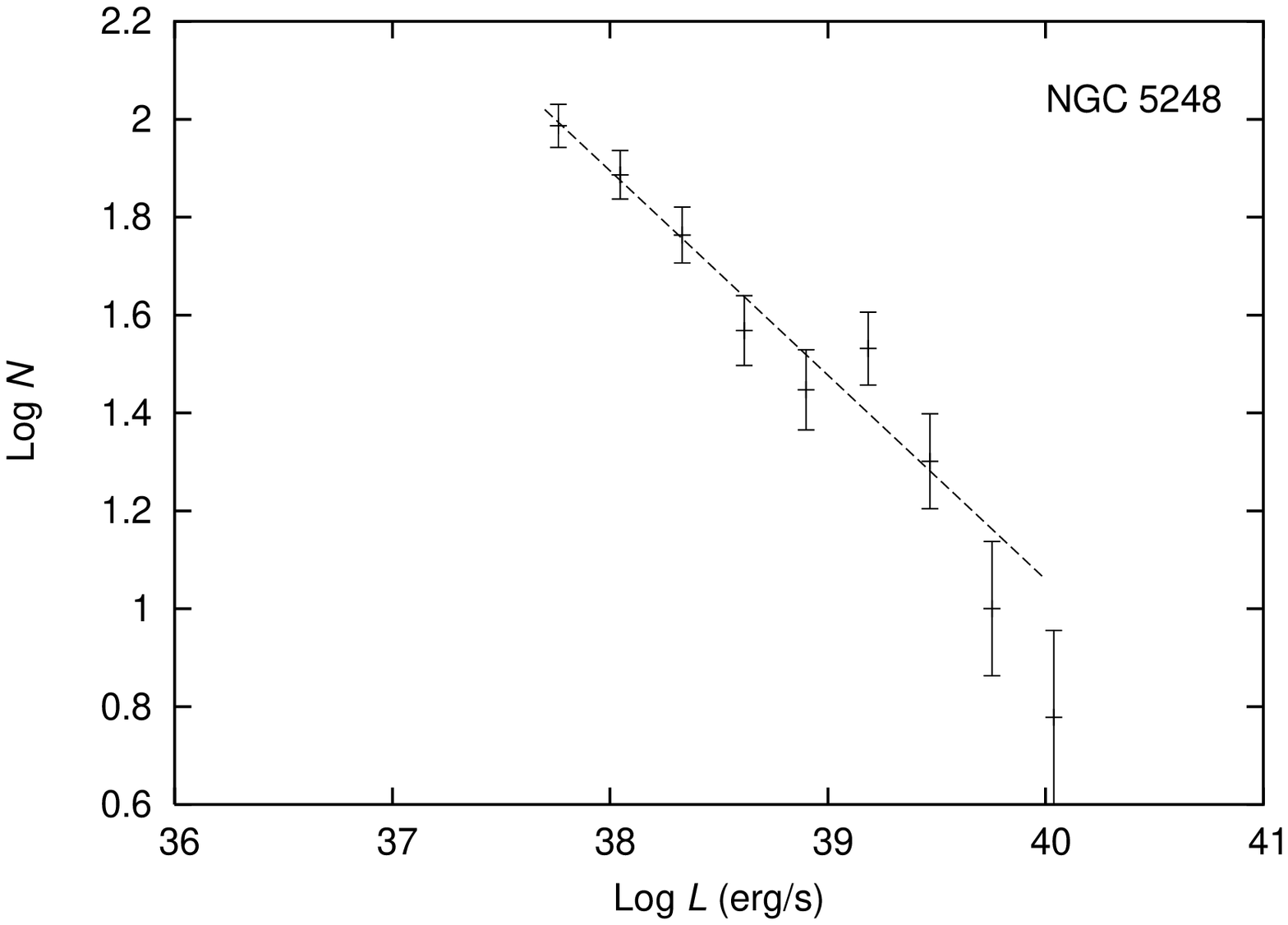,angle=-90,width=9cm}
\caption{(Continued)}
\end{figure}
\setcounter{figure}{0}
\begin{figure}
\psfig{figure=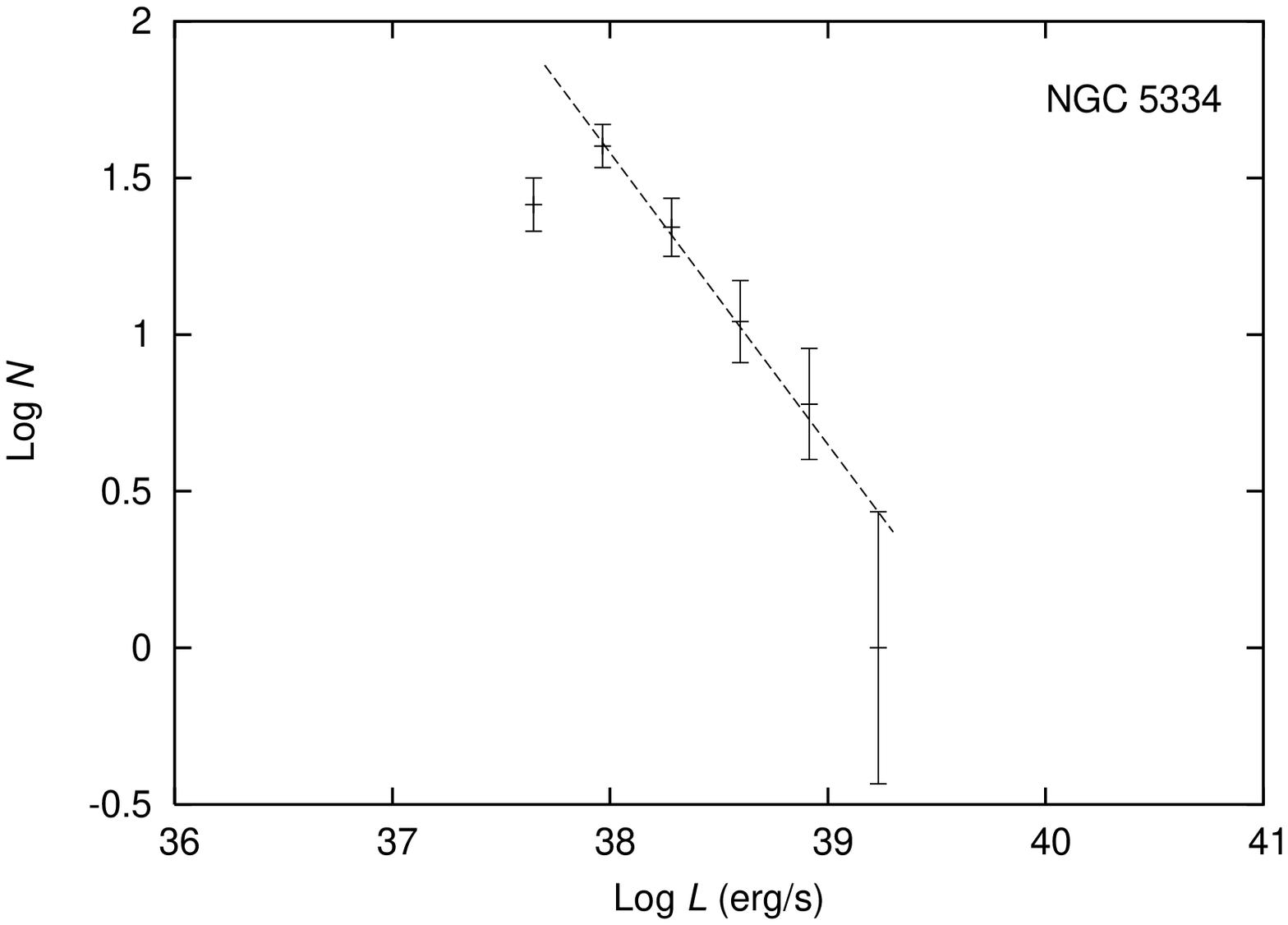,angle=-90,width=9cm}
\psfig{figure=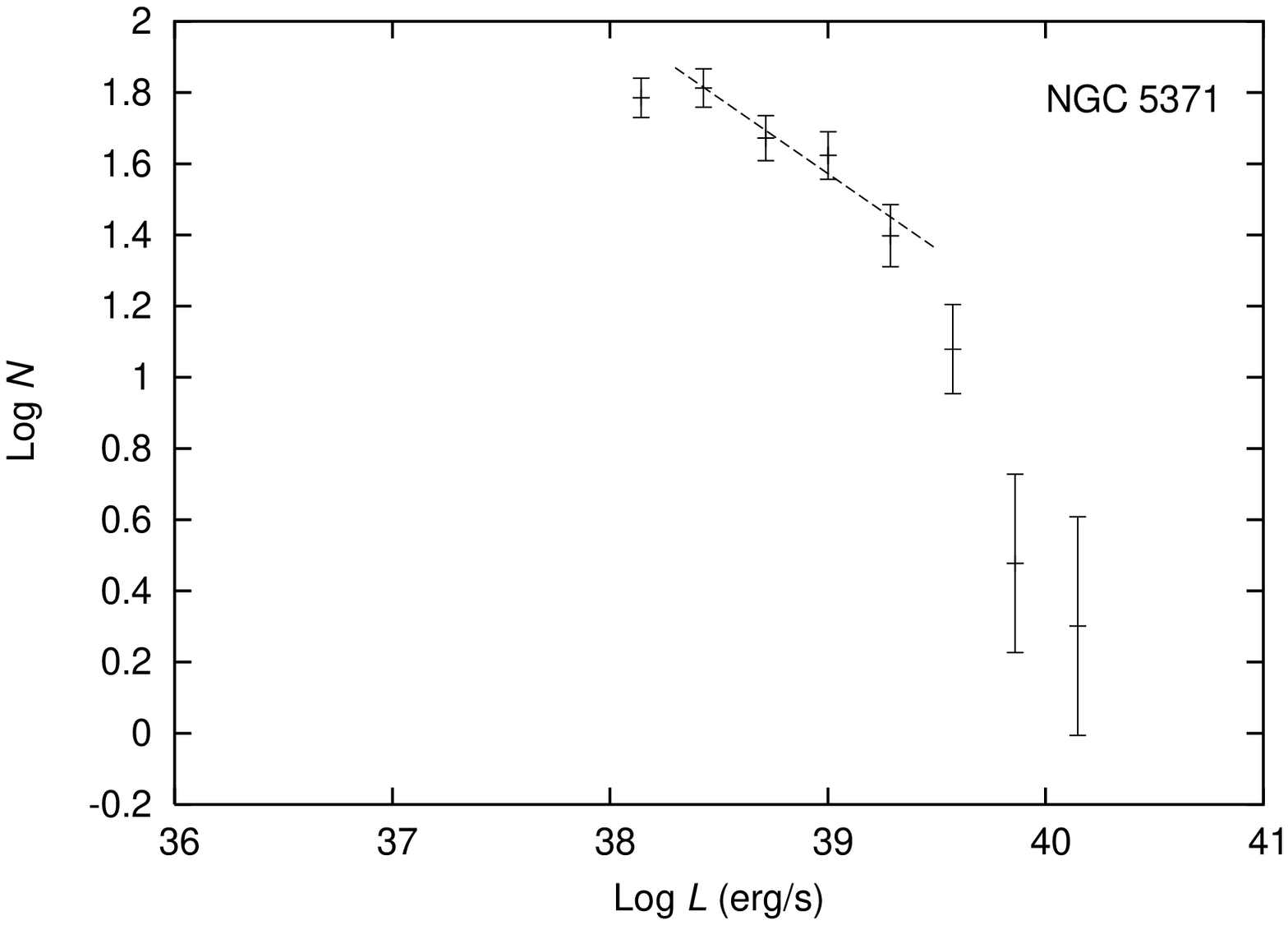,angle=-90,width=9cm}
\psfig{figure=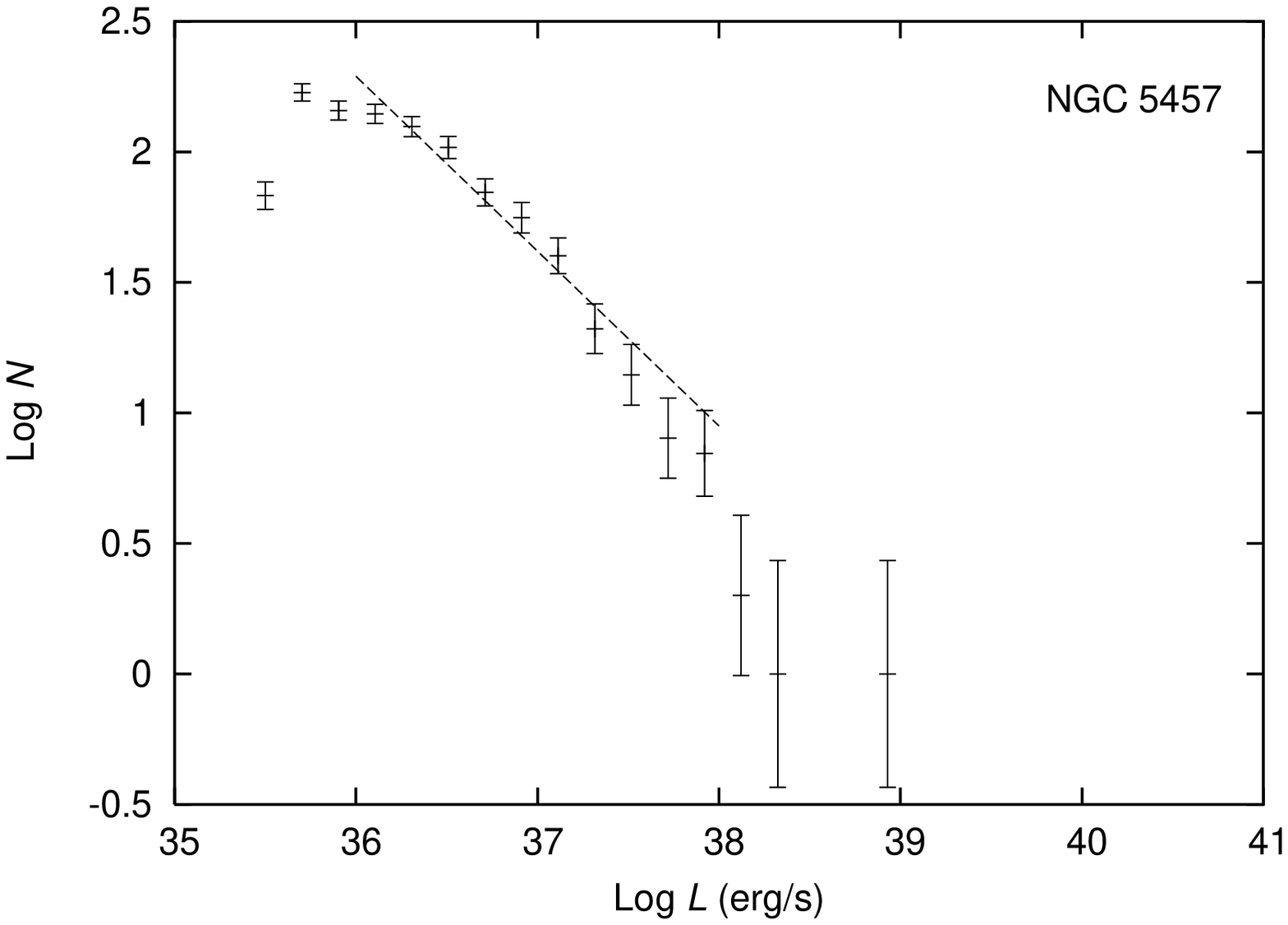,angle=-90,width=9cm}
\psfig{figure=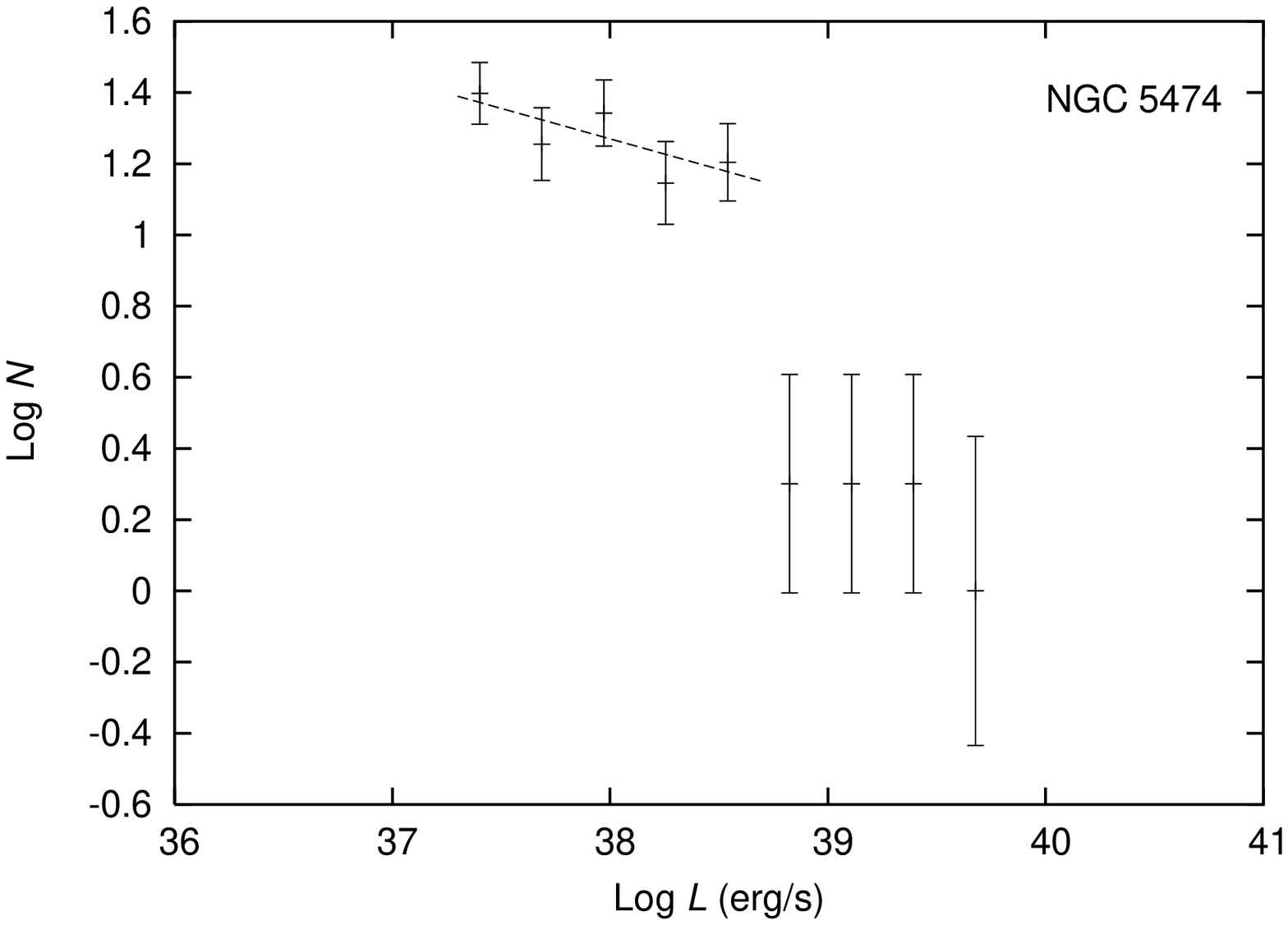,angle=-90,width=9cm}
\caption{(Continued)}
\end{figure}
\setcounter{figure}{0}
\begin{figure}
\psfig{figure=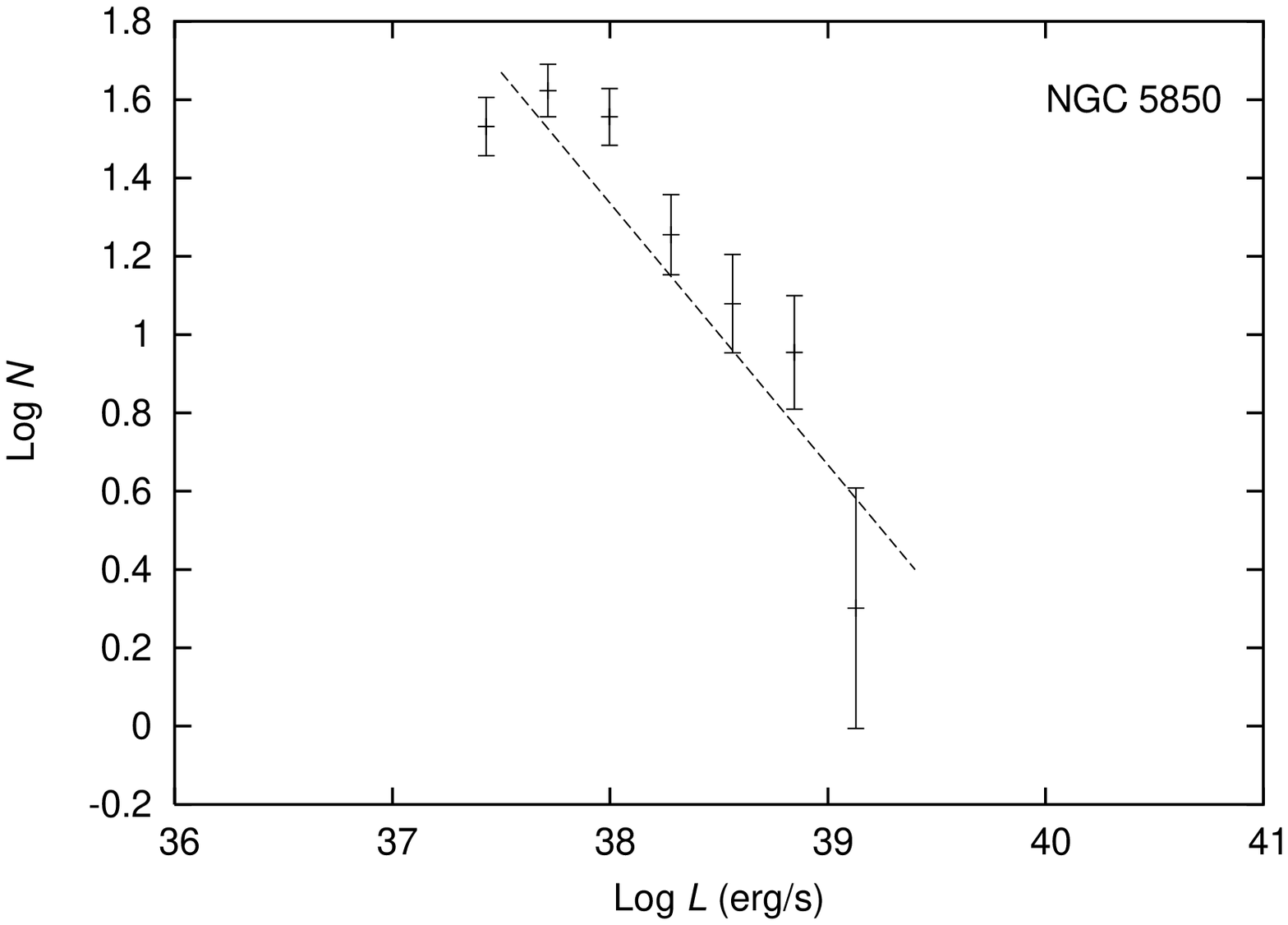,angle=-90,width=9cm}
\psfig{figure=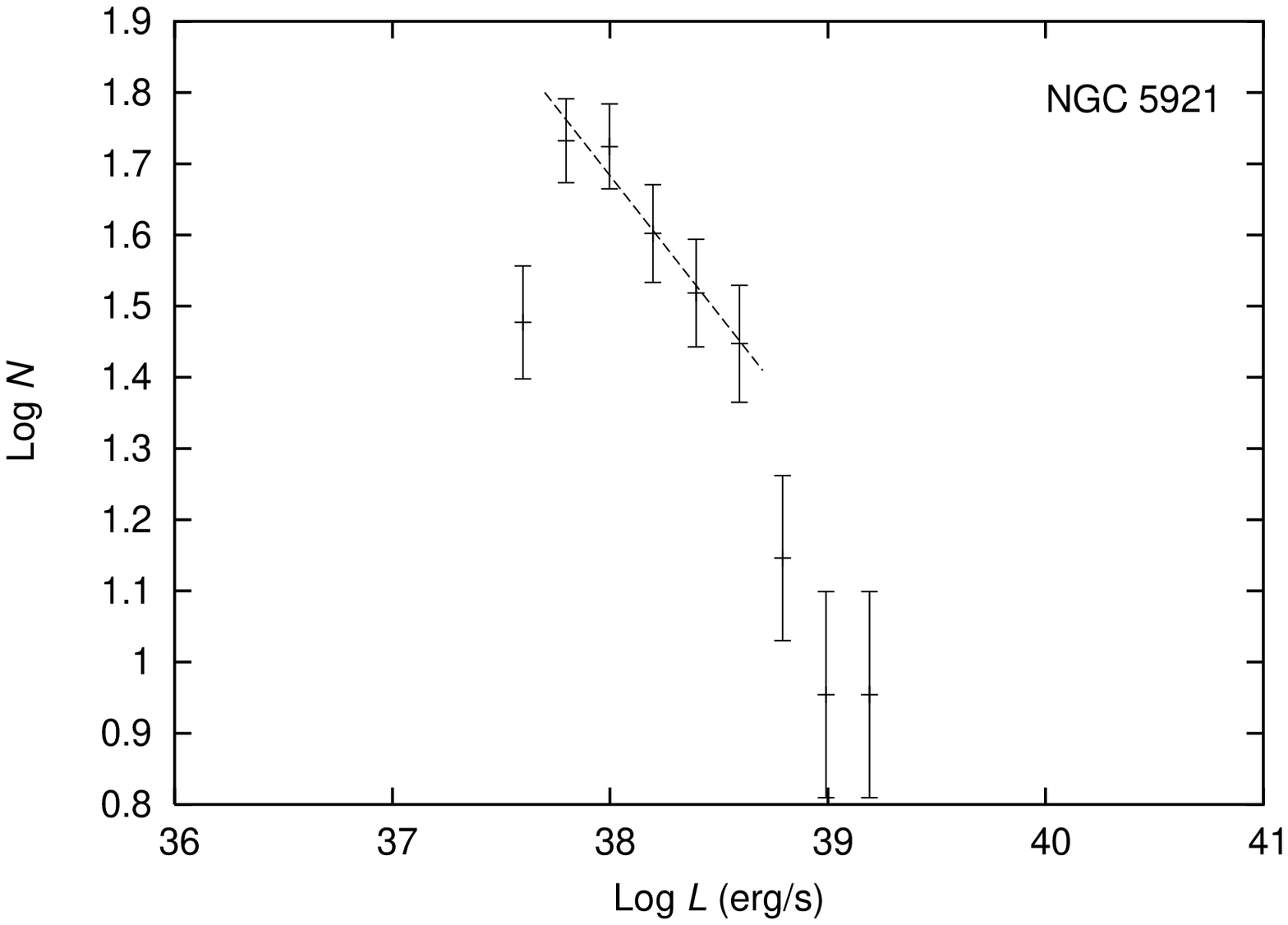,angle=-90,width=9cm}
\psfig{figure=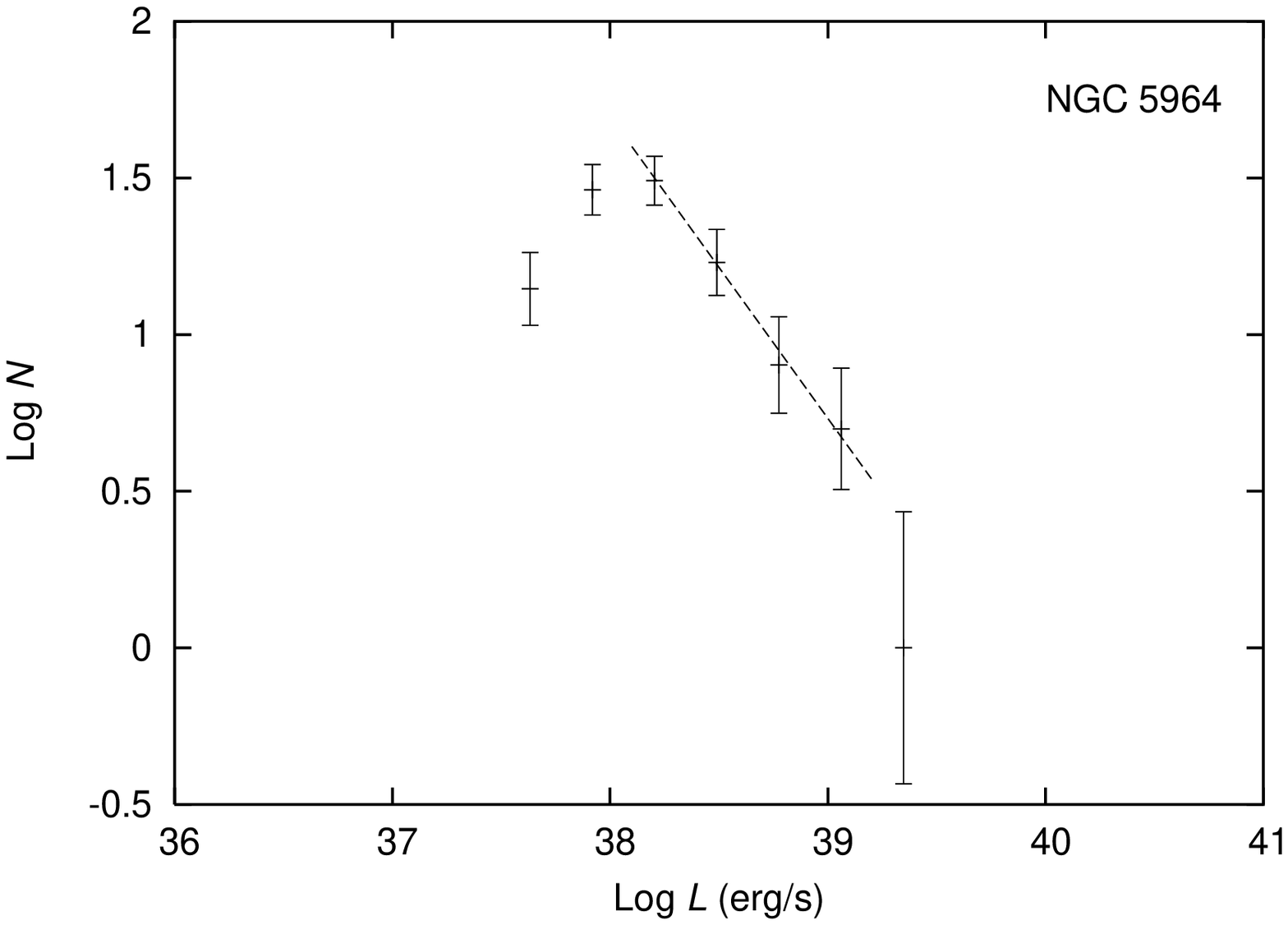,angle=-90,width=9cm}
\psfig{figure=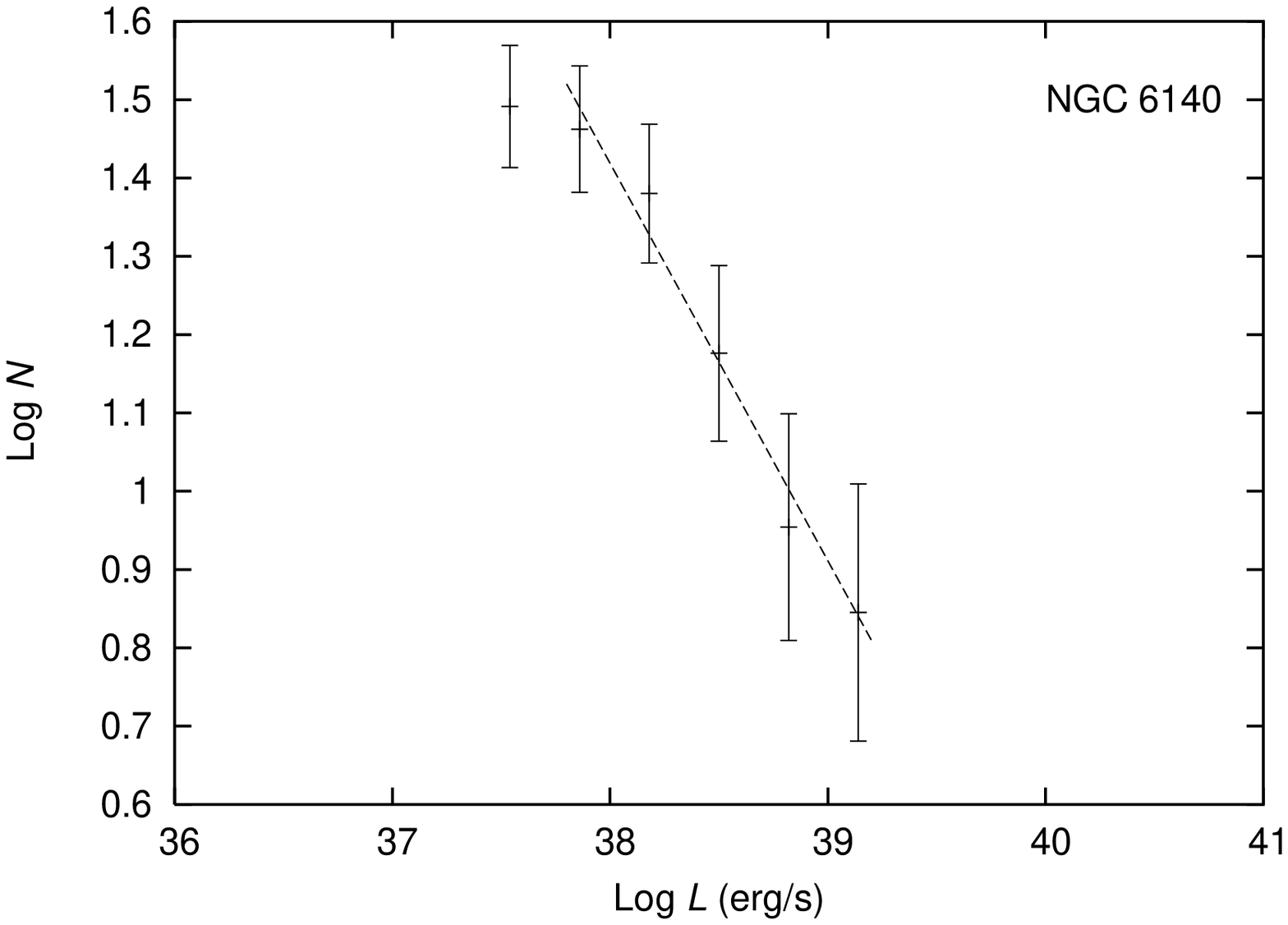,angle=-90,width=9cm}
\caption{(Continued)}
\end{figure}
\setcounter{figure}{0}
\begin{figure}
\psfig{figure=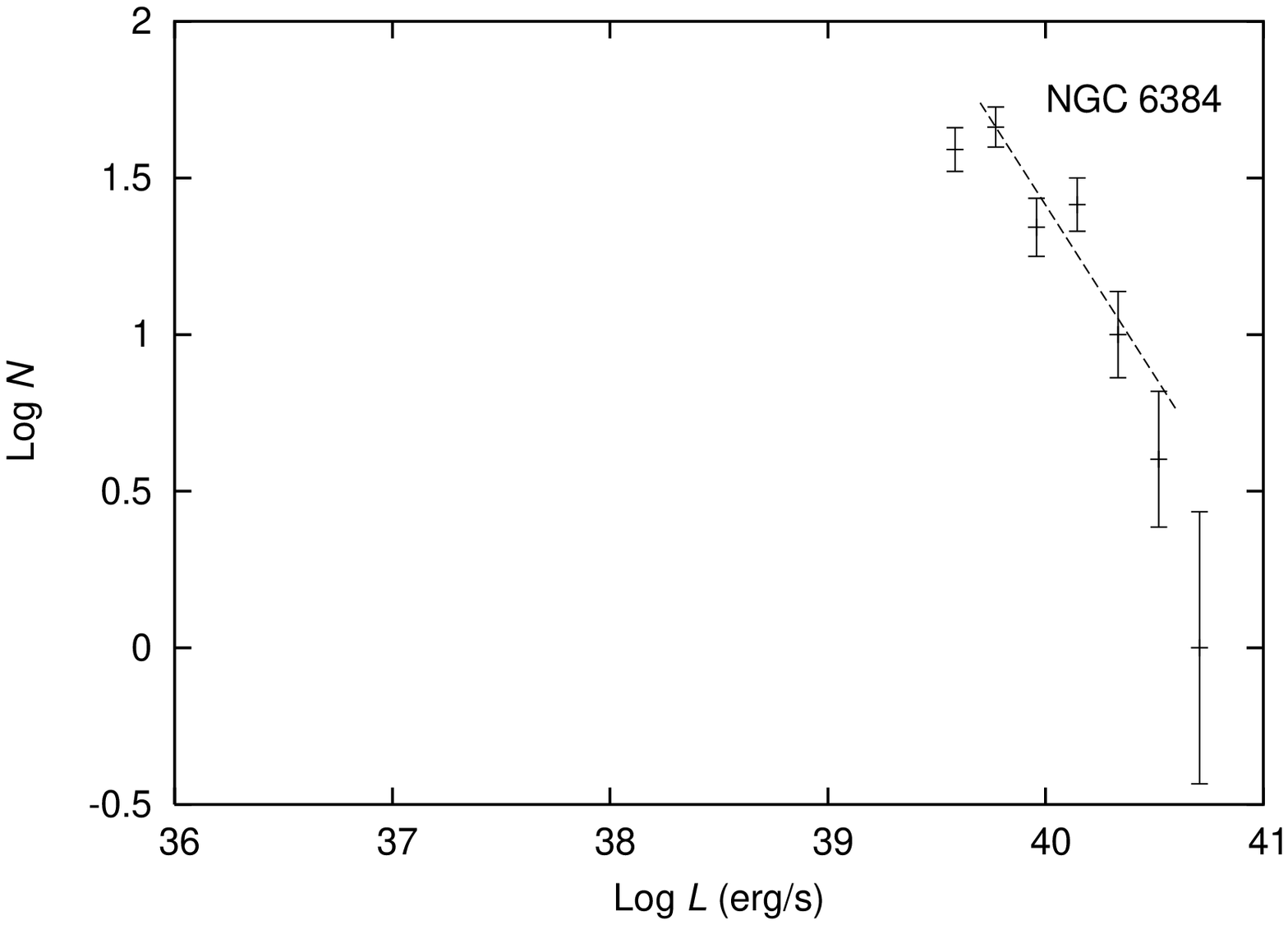,angle=-90,width=9cm}
\psfig{figure=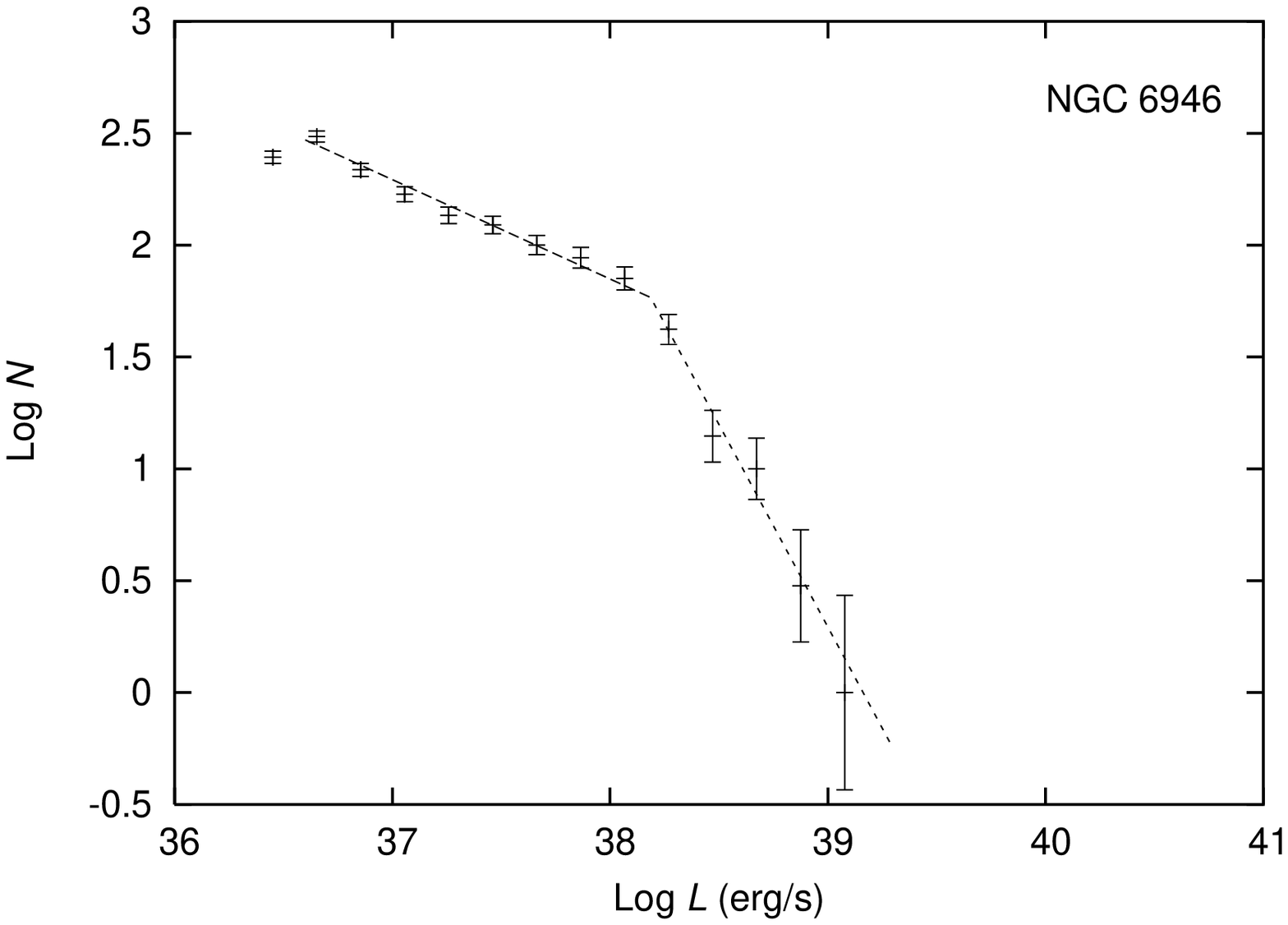,angle=-90,width=9cm}
\psfig{figure=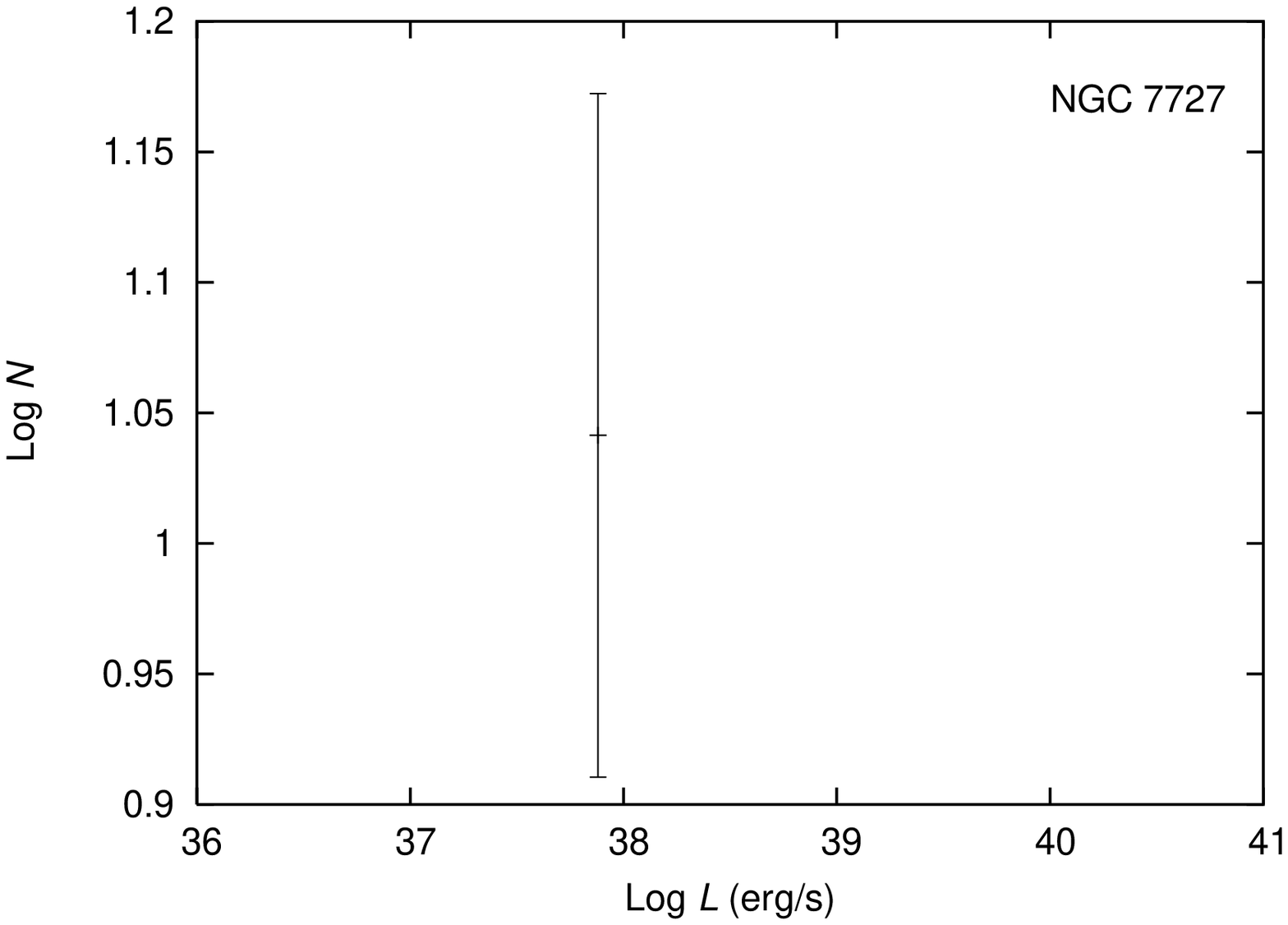,angle=-90,width=9cm}
\psfig{figure=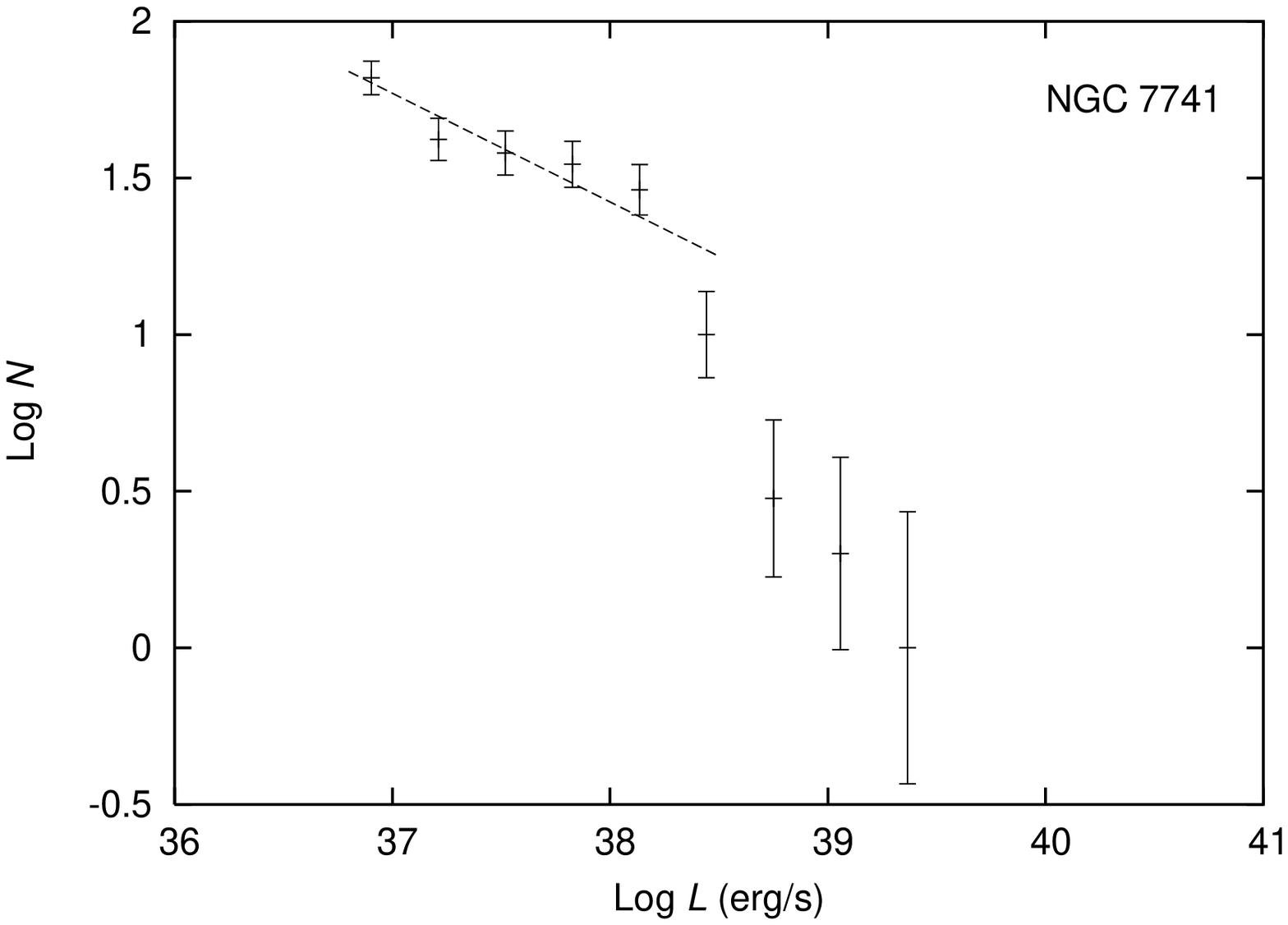,angle=-90,width=9cm}
\caption{(Continued)}
\end{figure}

\label{lastpage}

\end{document}